\documentclass[10pt]{article}
\usepackage{fullpage,epsfig,graphics,cancel,bbm,slashed}
\usepackage{amsthm,amssymb,amsmath,amsfonts,amscd,amsbsy,upgreek}
\usepackage{lmodern}
\usepackage{slantsc}
\usepackage{psfrag,hyperref,color}
\usepackage[]{youngtab}
\usepackage{graphicx}
\usepackage{wasysym}
\usepackage{mathdots}
\usepackage{mathrsfs}

\usepackage{tikz}
\usetikzlibrary{arrows,chains,matrix,positioning,scopes,decorations.pathreplacing,calc}
\usepackage{tikz-cd}

\makeatletter
\tikzset{join/.code=\tikzset{after node path={%
\ifx\tikzchainprevious\pgfutil@empty\else(\tikzchainprevious)%
edge[every join]#1(\tikzchaincurrent)\fi}}}
\makeatother

\tikzset{>=stealth',every on chain/.append style={join},
         every join/.style={->}}
\tikzset{
    >=stealth',
    punkt/.style={
           rectangle,
           rounded corners,
           draw=black, very thick,
           text width=6.5em,
           minimum height=2em,
           text centered},
    pil/.style={
           ->,
           thick,
           shorten <=2pt,
           shorten >=2pt,}
}

\def\centerarc[#1](#2)(#3:#4:#5){\draw[#1] ($(#2)+({#5*cos(#3)},{#5*sin(#3)})$) arc (#3:#4:#5); }

\newcommand{\BB}{\mathbb}
\newcommand{\SF}{\mathsf}
\newcommand{\FR}{\mathfrak}

\newcommand{\bea}{\begin{eqnarray}}
\newcommand{\eea}{\end{eqnarray}}
\newcommand{\nn}{\nonumber}
\newcommand{\Tr}{\operatorname{Tr}}

\newcommand{\sbullet}{\vcenter{\hbox{\tiny$\bullet$}}}

\newcommand{\bra}{\langle}
\newcommand{\ket}{\rangle}

\newcommand{\reeb}{\textsl{\textsc{r}}}

\newcommand{\opn}{\operatorname}

\def\ga{\alpha}

\def\Gc{\Gamma}
\def\gd{\delta}
\def\gdh{\textrm{\dh}}
\def\Gd{\Delta}
\def\ep{\epsilon}
\def\gt{\theta}

\def\gs{\sigma}
\def\Gs{\Sigma}
\def\gk{\kappa}
\def\gl{\lambda}
\def\Gl{\Lambda}
\def\Go{\Omega}
\def\go{\omega}

\DeclareMathAlphabet{\mathpzc}{OT1}{pzc}{m}{it}

\newtheorem{theorem}{Theorem}[section]

\theoremstyle{definition}
\newtheorem{example}[theorem]{Example}
\newtheorem{remark}[theorem]{Remark}

\numberwithin{equation}{section}
\begin{document}
\begin{flushright} \small
UUITP-47/20
 \end{flushright}
\smallskip
\begin{center} \Large
{\bf Rozansky-Witten theory, Localised then Tilted}
 \\[12mm] \normalsize
{\bf Jian Qiu${}^{a,b}$} \\[8mm]
 {\small\it
    ${}^a$Department of Physics and Astronomy, Uppsala University,\\
        \vspace{.3cm}
      ${}^b$ Mathematics Institute,  Uppsala University, \\
      \vspace{.3cm}
   Uppsala, Sweden\\  }
\end{center}
\vspace{7mm}
\begin{abstract}
 \noindent The paper has two parts, in the first part, we apply the localisation technique to the Rozansky-Witten theory on compact HyperK\"ahler targets. We do so via first reformulating the theory as some supersymmetric sigma-model. We obtain the exact formula for the partition function with Wilson loops on $S^1\times\Sigma_g$ and the lens spaces, the results match with earlier computations using Feynman diagrams on K3. The second part is motivated by a very curious preprint \cite{Gukov:2020lqm}, where the equivariant index formula for the dimension of the Hilbert space the theory is interpreted as a kind of Verlinde formula. In this interpretation, the fixed points of the target HyperK\"ahler geometry correspond to certain 'states'. In the second half of the paper we extend the formalism of part one to incorporate equivariance on the target geometry. For certain non-compact hyperK\"ahler geometry, we can apply the tilting theory to the derived category of coherent sheaves, whose objects label the Wilson loops, allowing us to pick a 'basis' for the latter. We can then compute the fusion products in this basis and we show that the objects that have diagonal fusion rules are intimately related to the fixed points of the geometry. Using these objects as basis to compute the dimension of the Hilbert space leads back to the Verlinde formula, thus answering the question that motivated the paper.
\end{abstract}

\eject
\normalsize

\tableofcontents
\section{Introduction}
The Rozansky-Witten (RW) theory \cite{Rozansky:1996bq} was introduced as a 3D topological sigma model, which has as source a 3-manifold $M$ and target a hyperk\"ahler manifold $X$. It was understood that the model behaves in some ways like the Chern-Simons theory, including the existence of Wilson loop operators. However the would be gauge field is of degree 1 and hence nilpotent.
This nilpotency gives the model many attractive features, for example, its path integral is captured entirely by perturbation theory. Furthermore the perturbation expansion is finite, being capped off by the dimension of $X$.

These features make the model a very good venue for studying Vasilliev \cite{vasilev1994complements} finite type 3-manifold and knot invariants.
An important point made by Kontsevich \cite{kontsevich_1999} is that certain fields in the theory are better treated as backgrounds fields (what he called 'parameters') rather than integrated over as was done in the original paper \cite{Rozansky:1996bq}, so as to produce finite type invariants valued in the Dolbeault cohomology of $X$. In contrast to the Chern-Simons theory, whose finite type invariants are associated with Lie algebras, the RW invariants are associated with a dg-Lie algebra.
Kapranov \cite{Kapranov99rozansky-witteninvariants} went further
and interpreted the 'structure constant' of the aforementioned dg-Lie algebra as the Atiyah-Class of $X$ and showed that there is in fact an $L_{\infty}$-algebra lurking behind the dg-Lie algebra (especially if one extends from Hyperk\"ahler to the holomorphic symplectic manifolds).

Roberts and Willerton \cite{roberts2010} studied further the RW weight system for finite type invariants in the category $D^b(X)$, the bounded derived category of coherent sheaf on $X$. In \cite{CALDARARU_Ip} and also \cite{CALDARARU_I} (the latter, though superseded by the former, is also inspiring), an extended TFT with $D^b(X)$ as the target category was put forward. The Hilbert space that the TFT assigns to a Riemann surface $\Gs_g$ is
\bea {\cal H}(\Gs_g,X)\simeq H^{0,\sbullet}(X,(\Go_X^{\sbullet,0})^{\otimes g})\label{Hilbert}\eea
originally obtained in \cite{Rozansky:1996bq} via canonically quantising the fermion zero modes in the action.
In \cite{CALDARARU_I} it was pointed out that, in a more categorical approach one is led very naturally to assign the Hochschild homology of $X$ as the Hilbert space on the torus, see the appendix of \cite{CALDARARU_I} for a beautiful pictorial demonstration.
In the case $X$ is smooth, the Hochschild homology is isomorphic via the celebrated HKR theorem to $H^{\sbullet}(X,\Go^{\sbullet}_X)$, which coincides with the $g=1$ case of \eqref{Hilbert}. So the naive canonical quantisation approach turns out correct.

The message from the cited works above is that what irreducible representations to CS theory is what objects in $D^b(X)$ are to the RW theory.
For example in the CS theory, we can insert into the solid torus a Wilson loop labelled by a representation of the Lie algebra. Performing the path integral then produces a vector in the Hilbert space ${\cal H}(T^2,X)$.
In the RW theory, one can write down a similar Wilson-loop operator from the connection and curvature of a holomorphic vector bundle, that is, the vector bundle now takes the role of the representation. One can extend this construction even to sheaves, provided one resolves the sheaf into a complex of vector bundles and write the Wilson-loop using the connection and curvature of the latter.
We find that for each object in $D^b(X)$, its Chern-Character or more precisely its Mukai vector plus some additional framing data of the Wilson loop determines a vector in the Hilbert space. With multiple parallel Wilson loops, their fusion at the classical level is just the tensor product of the vector bundles labelling the Wilson-loops, but it is expected to receive quantum corrections, see \cite{Kapustin:2008sc}, where a perturbative computation of the fusion product has also been attempted.

Thanks to the work of \cite{KaWiYa}, now we can go beyond the perturbation theory in CS theory. The exact partition function as well as the correlation function of Wilson loops are computed much more easily.
These were previously obtained in a roundabout way via relating CS theory to the Wess-Zumino model in the classic work \cite{witten1989}.
With the explicit result, one can obtain the fusion algebra of Wilson loops as well as the $SL(2,\BB{Z})$ that acts on the Hilbert space \cite{Kapustin:2013hpk}.
With this much success, there is no reason not to apply the same technique to the RW theory, who is after all the odd cousin of CS theory. This is only one motivation of the current work.

\smallskip

Another piece of inspiration comes from \cite{Gukov:2020lqm}. The problem is that the HK varieties that arise as the Coulomb branch of 3D $N=4$ susy gauge theories are non-compact and so the Hilbert space is of infinite dimension. Yet these varieties tend to have plenty of $U(1)$ symmetries, with which the Hilbert space \eqref{Hilbert} can be given a graded module structure ${\cal H}(\Gs_g)=\oplus_n{\cal H}_n(\Gs_g)$. It is important that at each grading $\dim {\cal H}_n$ be finite, provided that the $U(1)$ symmetry is chosen well (precise meaning later). By introducing an equivariant parameter $u$ one can investigate the generating function
$\sum_n u^n\opn{sdim}{\cal H}_n$, especially its analytic behaviour.
This idea of using the grading or equivariance has already been implemented successfully in \cite{Gukov:2015sna} to treat the CS theory with non-compact gauge groups and has led to the equivariant Verlinde formula. The latter formula was first derived for CFT in \cite{VERLINDE1988360}, but now it is enriched with the extra equivariant parameters.

The authors of \cite{Gukov:2020lqm} extended this idea to the RW theory, in particular, they saw a striking resemblance between the Verlinde formula for the dimension of ${\cal H}(\Gs_g,X)$
\bea \sum_nu^n\opn{sdim}{\cal H}_n(\Gs_g,X)=\sum_{\gl}(S_{0\gl})^{2-2g}\label{verlinde}\eea
and the equivariant index formula for computing $\dim{\cal H}(\Gs_g,X)$.
More concretely the rhs matrices $S_{\mu\nu}$ are the $S$-duality matrices with the indices $\mu,\nu,\gl\cdots$ labelling the states of ${\cal H}$. On the other hand the equivariant index formula can be computed as a sum over the fixed points of the $U(1)$ actions
\bea \sum_nu^n\opn{sdim}{\cal H}_n(\Gs_g,X)=\sum_{x_{\gl}\in {\rm f.p.}}\frac{\opn{ch}_{eq}(\Go^{\sbullet}(X)^{\otimes g})\opn{Td}_{eq}(TX)}{e_{eq}(T_{\gl}X)}\Big|_{x_\gl}.\nn\eea
Here the rhs sum is over fixed points $x_{\gl}\in X$ of the $U(1)$ actions, assuming of course they are isolated. If one associates the fixed point $x_{\gl}$ with the states labelled by $\gl$ in \eqref{verlinde}, then one gets a formula for $S_{0\gl}^2$. This formula was then checked against alternative derivations via summing over the Bethe vacua of the $N=4$ gauge theory.

It is this interpretation of $U(1)$ fixed points on the HK variety as states of the Hilbert space that first ignited this project. In order to understand this interpretation, we shall follow the sequence of reasoning below. First, the original Verlinde formula was the manifestation of the by now well chanted slogan: $S$-duality diagonalises the fusion rule.
This $S$-duality is an $SL(2,\BB{Z})$ action on the Hilbert space ${\cal H}(T^2)$ associated to the torus. To obtain this $S$-matrix, we can compute the partition function of RW theory on $S^3$. This is because $S^3$ is glued from two solid tori along their common boundary $T^2$, but during the gluing an $S$ transformation is applied to one of the $T^2$'s. Thus the partition function of $S^3$ can be realised as a matrix element of $S$ sandwiched between two states in ${\cal H}(T^2)$. To compute this partition function, we shall reformulate the RW theory in sec.\ref{sec_RoRm} to get two susy's that would work as equivariant differentials. Then we can run the by now well-oiled localisation machinery to finish the calculation on $S^3$ or on more general Seifert manifolds. The main result is that the $S$-matrix is essentially the Todd genus, and this seems vaguely to be behind why the rhs of \eqref{verlinde} should give the equivariant index. Furthermore, the Wilson loops that can appear in a localisation computation are very rigid: they must be placed along certain restricted loci. For such Wilson-loops, we found that they contribute to the correlation functions through their Chern-Characters. More importantly, we found no correction to their fusion product i.e. fusion remains just the (derived) tensor product. It is not yet clear whether this is due to the fact that we work with very specific Wilson-loops.

The next questions are: why or whether $S$ diagonalises the fusion rule, what are the states we are fusing, and what have they got to do with the fixed points? In the CS theory context, the states are identified as the integrable representations at level $k$ of the loop group $LG$ (these are the objects of the so called category ${\cal O}$). The tensor product of two such representations can again be decomposed as direct sums of such, giving the fusion rules. In RW theory we have the derived category of coherent sheaves $D^b(X)$ replacing the category ${\cal O}$. Suppose $D^b(X)$ possesses a tilting object $T$ which furthermore is the direct sum of vector bundles $T=E_1\oplus\cdots\oplus E_n$ \footnote{this immediately forces the HK variety to be non-compact, our main examples come from the construction of \cite{TODA20101}}. It is natural to suggest $\{E_i\}$ as the replacement of the integrable representations earlier.
Just as in CS theory, a state in ${\cal H}(T^2)$ can be realised by inserting a (framed) Wilson loop along the core of the solid torus, whose boundary is the said $T^2$. We will see in sec.\ref{sec_HsWlar} that the Wilson loops descend to the Grothendieck group of $D^b(X)$. Concretely this means that one can simply say that any object in $D^b(X)$ is an integer linear combination $\sum_i n_iE_i$\footnote{a negative $n_i$ means that $E_i$ has its cohomological degree shifted}. The tensor products between two $E_i$'s can also be expanded \bea E_i\otimes^{\BB L} E_j=\sum_kN^k_{ij}E_k\nn\eea
giving us the fusion rules.
We will demonstrate in sec.\ref{sec_TtaVf} (which can be read independently of the rest of the paper) with the simplest examples $T^*\BB{P}^1$, $T^*\BB{P}^2$ that the objects that fuse diagonally have 1-1 correspondence with the $U(1)$ fixed points in the geometry. More precisely, in the examples we treat, they are the structure sheaves of the cotangent fibre sitting above the fixed points. Applying the Verlinde formula \eqref{verlinde} then leads to the correct equivariant index of $H^{\sbullet}(X,\Go^{\sbullet}(X)^{\otimes g})$.

However we also hit some snags along the way. The biggest one is how to add $U(1)$-equivariance to the RW theory that acts holomorphically but not tri-holomorphically, see sec,\ref{sec_Ee}.
Incidentally, these $U(1)$'s are crucial to making sure that at each grading the Hilbert space be of finite dimension.
As a compromise we merely added such equivariance to the reformulation of RW theory as a susy $\gs$-model in sec.\ref{sec_Eea}.
We can compute using this formulation the equivariant partition function on $S^3$, from which we can read off the action of $S\in SL(2,\BB{Z})$.
But it is not clear if we can at all call it \emph{the} $S$-matrix of the Rozansky-Witten theory, as it is computed for a stand-in theory.
But if we trudge on, we observe that the action of $S$ is given in terms of certain double-sine function of the Chern-roots of $TX$.
Based on this and also on the result on $L(p,q)$, we make a precarious proposal for the cohomological kernel of the Fourier-Mukai transform that realises the action of $S$ as well as that of $T^kS\in SL(2,\BB{Z})$ on the category $D^b(X)$, see sec.\ref{sec_TcFMkotSa}. We also argue in sec.\ref{ToaboHs} that the $S$-matrix should not diagonalise the fusion rules in this context.

\bigskip
{\bf Acknowledgements:} It is my pleasure to thank Arkady Vaintrob and Matthew Young for many discussions on equivariant Rozansky-Witten theory over the past year, Wanmin Liu for discussions about exceptional collections on cotangent bundles, Georgios Dimitroglou Rizell for answering my many ignorant questions about surgery, Volodymyr Mazorchuk for his course on category ${\cal O}$, Martin Herschend and Sondre Kvamme for answering my questions about tilting theory, and especially Julian K\"ulshammer for stopping me from making a false claim about the minimal resolution. Finally it will be remiss not to thank Maxim Zabzine for the constant illuminating discussions ranging over many topics over the years.

\section{The Rozansky-Witten theory}
\subsection{Essentials of Hyperk\"ahler geometry}
We recall the essentials of the construction \cite{Rozansky:1996bq}, but our notations are slightly different. Let $M$ be a closed 3-manifold and $(X,g,I,J,K)$ be a Hyperk\"ahler manifold of real dimension $4n$. We will first take $X$ to be compact, but of course the real interesting aspect is when we take it non-compact and with toric symmetries.
Recall that $X$ has three complex structures $I,J,K$, all of which are covariantly constant under the Levi-Civita. To fix the convention, the three complex structures satisfy the relations
\bea I^2=J^2=K^2=-1,~~~I_a^{~c}J_c^{~b}=-J_a^{~c}I_c^{~b}=K_a^{~b},\label{quaternion_alg}\eea
the indices will be raised or lowered freely with the HK metric $g$. We will use $I$ as our main complex structure, i.e. when we talk about complex coordinates, they are defined with respect to $I$.
We have also three K\"ahler forms $(I,J,K)g$, we take the combination
\bea \Go=(J-iK)g\label{hol_symp_form}\eea
as the holomorphic symplectic form. It is non-degenerate as a holomorphic 2-form, and is annihilated by $\bar\partial$ (in fact it is covariantly constant).

That $I,J,K$ are covariantly constant implies certain symmetry properties for the Riemann curvature tensor that will be crucial for the BRST symmetry later. Our notation for the curvature is
\bea ([\nabla_U,\nabla_V]-\nabla_{[U,V]})W=R(U,V)W,~~U,V,W\in TX,\nn\eea
and $R(U,V)\in\opn{End}TX$ satisfies $R(U,V)I=I^*R(U,V)$ and the same for $J,K$. More importantly $R(U,V)$ preserves $\Go$, i.e. $\Go R(U,V)\in\opn{Sym}^2T^*X$. Combined with the first Bianchi identity, this implies that $\Go R(U,-)\in\opn{Sym}^3T^*X$. For the convenience of the reader we summarise these properties explicitly using indices
\bea R_{\bar ii~j}^{~~l}\Go_{lk}=R_{\bar ii~k}^{~~l}\Go_{lj}=R_{\bar ik~j}^{~~l}\Go_{li}\label{total_sym}\eea
as well as the usual identities satisfied by the Riemann tensor on a K\"ahler manifold
\bea R_{ij~?}^{~~?}=R_{\bar i\bar j~?}^{~~?}=R_{??~j}^{~~\bar i}=R_{??~\bar j}^{~~i}=0.\nn\eea

\subsection{Action and BRST symmetry}
Now we come to the action of the RW model, it consists of a topological part $L_0$ and another BRST-exact part $L_1$,
\bea &&\hspace{3cm}S=\int_M L_0+L_1,\label{RW_action}\\
&&L_0=\frac12\Go_{ij}\chi^i\nabla\chi^j-\frac16R(\bar v,\chi,\Go\chi,\chi),~~~~L_1=d\phi*d\bar\phi-\nabla\bar v*\chi.\nn\eea
Here the notations are: $\phi,\bar\phi$ are the complex coordinates of $X$ (again, with respect to $I$), we will also use $\phi$ to denote the map
\bea \phi:\,M\to X.\nn\eea
The fermionic fields $\chi$ are sections of $\Go^1_M\otimes \phi^*T^{1,0}X$, that is, they are 1-forms on $M$ valued in the pullback via $\phi$ of the holomorphic tangent bundle of $X$.
We have another fermionic field $\bar v\in \phi^*T^{0,1}X$, note that deg-$p$ polynomials of $\bar v$ can be regarded as $(0,p)$ forms $\Go^{0,p}_X$ pulled back to $M$ via $\phi$. Finally $R(\bar v,\chi,\Go\chi,\chi)$ is our lazy notation for
\bea R(\bar v,\chi,\Go\chi,\chi)=\bar v^{\bar i}R_{\bar ii~k}^{~~l}\chi^i\wedge (\Go_{lm}\chi^m)\wedge \chi^k.\label{curvature_term}\eea
Note that the $\chi$'s are fermionic but they are also 1-forms, as a result the three $\chi$ above are totally symmetric, matching the symmetry property \eqref{total_sym}.

The model has a BRST symmetry
\bea \gd_0\phi^{\bar i}=\bar v^{\bar i},~~~\gd_0\chi^i=-d\phi^i\label{RW_BRST}\eea
while the variation of the other fields is zero identically. It is clear that $\gd_0^2=0$.
\begin{remark}\label{rmk_cov_susy}
  The second variation in \eqref{RW_BRST} is in fact covariant, even though normally $\gd_0$ of a section of $TX$ is no longer a section of $TX$. Here what made the difference is that on a K\"ahler manifold only $\Gc_{ij}^k$ and $\Gc_{\bar i\bar j}^{\bar k}$ are nonzero. However in general it is wise to incorporate the Levi-Civita connection into the definition of BRST, as we shall see later. Also should one want to check that $\gd_0$ annihilates the action \eqref{RW_action}, one is much better off by defining the BRST covariantly. Again one shall see the example later.
\end{remark}

We speak much of the similarity between RW and CS model in the introduction, so we juxtapose the action of the latter here to accentuate this analogy
\bea S_{CS}=\frac{k}{4\pi}\int_M A\wedge dA+\frac{2}{3}A\wedge A\wedge A+\bra b,d^{\dag}A\ket -\bra\bar c,d^{\dag}d_Ac\ket.\nn\eea
Note the analogy is only complete when the BRST sector of the CS theory is taken along (see sec.2.6 in \cite{Rozansky:1996bq})
\bea A\leftrightarrow\chi,~~c\leftrightarrow \phi,~~b\leftrightarrow\bar v,~~\bar c\leftrightarrow \bar \phi,~~\gd_{BRST}\leftrightarrow \gd_0.\label{analogy}\eea
The CS model has the obvious ghost number symmetry of $+1,-1$ for $c$ and $\bar c$. One could assign the ghost number $+1,-1$ also to $\bar v$ and $\chi$, but then one must assign $+2$ to the holomorphic symplectic form $\Go$ so that the curvature term \eqref{curvature_term} may have ghost number zero. The ghost number restricts the dependence of the partition function on $\Go$.

\smallskip

Back to the RW theory. Both $L_0$ and $L_1$ are annihilated by $\gd_0$, but $L_1$, which depends on the metric on $M$ is $\gd_0$-exact
\bea L_1=\gd_0(-\chi*d\bar\phi).\nn\eea
This is the reason that RW model is a topological sigma model: a change of the metric on $M$ is BRST exact and will not affect the path integral. The role of $L_1$ is merely to give the bosons and fermions a good kinetic term. By giving $L_1$ a large coefficient, the path integral would concentrate on the configuration
\bea d\phi=0\nn\eea
i.e. constant maps $M\to X$, so perturbative computation round the constant maps captures the entire partition function.
One may proceed with the perturbation expansion, where the curvature term becomes a trivalent vertex. The $\bar v$ that comes with the curvature term would then limit the total number of vertices, truncating the perturbation expansion to a finite one.

%

\subsection{Hilbert space, Wilson loops and representations}\label{sec_HsWlar}
The general procedure of semi-classical quantisation says that we solve the equation of motion, then expand the fields to first order in fluctuation round the solutions.
These first order fluctuations are then quantised into raising and lowering operators.
But the topological nature of the RW model says that there are only the zero modes to quantise.

Let $M=\BB{R}\times \Gs_g$ with $\BB{R}$ being the time direction.
We have seen that the equation of motion corresponds to constant maps $\phi:\,M\to X$, so the bosonic zero modes are parameterised by $X$ itself. The fermion zero modes will depend on $\Gs_g$. We can find their commutation relations by looking at the kinetic term of the action. Once this is done we need to pick a polarisation, that is, we pick half of the fermion zero modes as 'coordinates' and the other half as 'momenta'. The Hilbert space assigned to $\Gs_g$ is thus the Fork space of the fermionic coordinates.
To carry this out, we split the 1-form $\chi$ as $\chi_t$ and $\chi_z,\chi_{\bar z}$ then the fermionic kinetic term reads
\bea L_{{\rm ferm\,kin}}=g_{\bar ij}\bar v^{\bar i}\partial_t\chi^j_t+\frac{1}{2}\Go_{ij}\chi^i_{[z}\partial_t\chi^j_{\bar z]}.\nn\eea
This shows that $\bar v$, $\chi_t$ are a pair of dual variables while $\chi_z,\chi_{\bar z}$ form another pair. We pick the zero modes of $\bar v$ and $\chi_{\bar z}$ as the coordinates while the rest as momenta. The zero modes of $\bar v$ have dimension $\dim_{\BB{C}}X$ while those of $\chi_{\bar z}$ have dimension and $g\dim_{\BB{C}}X$ with $g$ being the dimension of $H^{0,1}(\Gs_g)$.
The Fock space built from $\bar v$ and $\chi_{\bar z}$ can clearly be identified as a $(0,\sbullet)$-form on $X$ (grading $\sbullet$ coming from power of $\bar v$) valued in the bundle $(\wedge^{\sbullet} T^{*{1,0}}_X)^{\otimes g}$ (with grading coming from $g$ copies of $\chi_{\bar z}$). Finally we see from \eqref{RW_BRST} that $\gd_0$ acts on the zero modes as the $\bar\partial$-operator on $X$ and we conclude
\bea {\cal H}(\Gs_g)\simeq H^{0,\sbullet}(X,(\Go^{\sbullet,0}_X)^{\otimes g}).\nn\eea

We will be focusing specifically on $\Gs=S^2,T^2$. Recall in the Chern-Simons theory, one can realise the wave function on $T^2$ by inserting Wilson loops inside the solid torus whose boundary is the $T^2$ \cite{Elitzur:1989nr}.
In RW theory we can also write down a Wilson loop operator as in CS theory. Let $C$ be a closed curve in $M$, consider the following parallel transport operator on $\phi^*T^{(1,0)}M$ along $C$
\bea W_C(T^{1,0}X)=\Tr{\bf P}\exp\int_C {\cal A},~~{\rm where}~~{\cal A}^i_{~j}=-d\phi^k\Gc_{kj}^i+\bar v^{\bar i}\chi^kR_{\bar ik~j}^{~~i}.\label{Wilson_adj}\eea
Note we do not merely use the Levi-Civita connection, but also corrected it with a curvature term, so that altogether $\gd_0W=0$.

In CS theory we can label a Wilson loop with different representations of the Lie algebra, here we can do something similar. In fact the choice made above should be regarded as the 'adjoint representation', while the 'fundamental representations' come from holomorphic vector bundles on $X$. Indeed, let $E\to X$ be such a bundle and denote with $A,F$ its connection and curvature respectively. The holomorphicity implies that $F$ is of type (1,1). Now we can write another Wilson loop operator
\bea W_C(E)=\Tr{\bf P}\exp\int_C {\cal A},~~{\rm where}~~{\cal A}=-d\phi^kA_k+\bar v^{\bar i}\chi^kF_{\bar ik},\label{Wilson_E}\eea
here we have assumed that $A^{0,1}=0$ since the bundle is holomorphic. Checking that $\gd_0W=0$ is now slightly more involved.


We make a digression into discussing how to incorporate also coherent sheaves into Wilson loops. Within the derived category $D^b(X)$, one can replace a coherent sheaf with bounded complex of locally free sheaves
\bea \cdots \to E_a\to E_{a+1}\to E_{a+2}\to \cdots\label{perfect_complex}\eea
Equating the locally free sheaves as vector bundles, we can write the connection of this complex of vector bundles as a big block matrix, with the $(a,a)$ block occupied by the connection for $E_a$. There are also lower triangular terms given by the maps $E_a\to E_{a+1}$ (essentially the idea of super connection due to Quillen). In the same way the curvature has diagonal and lower triangular blocks. But seeing that in constructing the Wilson loop \eqref{Wilson_E}, we will eventually take a super-trace, the lower triangular blocks will not matter. Thus as far as Wilson loops are concerned, we can drop the maps in the complex \eqref{perfect_complex} and write it simply as $\sum_a(-1)^aE_a$, where the sign $(-1)^a$ refers to the sign $E_a$ gets in the super-trace. What this is saying is that the construction of Wilson-loops {\bf descends to the Grothendieck group} of $D^b(X)$. In fact during the localisation calculation, the Wilson loops will collapse to the Chern-character and so certainly descends to the Grothendieck group.
When $X$ is non-compact e.g. $T^*\BB{P}^2$ or $T^*Gr(2,4)$, then there is a tilting bundle which is a direct sum of line bundles generating $D^b(X)$ \cite{TODA20101}. The collection of line bundles will play the same role as the integrable representation of Lie algebra for the Chern-Simons theory.

\subsection{Equivariant extension}\label{sec_Ee}
We extend here the original RW-model by adding a tri-holomorphic equivariance to $X$ (see the appendix for some examples of holomorphic, tri-holomorphic actions). We take $X$ to be non-compact, in fact, our go-to example will be the cotangent bundles of K\"ahler manifolds.

Suppose a Lie group $G$ acts tri-holomorphically on $X$, meaning its action preserves the metric and all three complex structures.
For such actions the equivariant RW model has been treated in \cite{Kapustin:2009cd}. In \cite{Kallen:2010ff} the same model has been obtained via the so called AKSZ formalism, which is a very convenient way of constructing TFT's (the original RW model can also be obtained via this method \cite{AKSZ_RW}).
In both works cited above, the $G$-action was gauged and the gauge fields are a part of a gauge multiplet with either Chern-Simons or BF theory interaction.
But we will treat the gauge sector as backgrounds and in the following, we will first turn off everything in the gauge sector except the ghost, leading to an equivariant theory on any 3-manifold. Afterwards we will turn on a background gauge field, but this will necessarily involve further structures on the 3-manifold e.g. a contact structure.

We denote with $e_a,~a=1,\cdots,\dim\FR{g}$ a basis for $\FR{g}$ and let $V_a$ be the fundamental vector field of the action by $e_a$. We denote with $V^i_a$, or $V^{\bar i}_a$
the (anti-)holomorphic component of $V_a$.
The original BRST \eqref{RW_BRST} is modified to be
\bea& \gd_0\bar\phi^{\bar i}=\bar v^{\bar i}+c^aV^{\bar i}_a,~~\gd_0\phi^i=c^aV_a^i,~~\gd_0c^a=-(1/2)f^a_{bc}c^bc^c,\nn\\
&\gd_0\chi^i=-d\phi^i+c^a\chi^j\partial_jV_a^i,~~\gd_0\bar v^{\bar i}=c^a\bar v^{\bar j}\bar\partial_{\bar j}V_a^{\bar i}.\nn\eea
Here $c^a$ is the ghost, it will be a constant on $M$ and be treated as a background. The full BRST variation of the complete gauge sector can be found in \cite{Kallen:2010ff}.
At this point it is useful to briefly recall the equivariant connections that would help understand the $\gd_0$ above. Since $V_a$ is a Killing vector field, one can easily verify the relation
\bea V_a^kR_{k\bar k~?}^{~~~?}=\bar\nabla_{\bar k}X^{?}_{a?},~~{\rm where}~~X^i_{aj}=-\nabla_jV_a^i,~~X^{\bar i}_{a\bar j}=-\bar\nabla_{\bar j}V_a^{\bar i}.\nn\eea
When we regard curvature $R$ as $\opn{End}(TX)$-valued 2-form, we can write the relation above as $\iota_{V_a}R=\nabla X_a$. This means that $R+e^aX_a$ together gives the $G$-equivariant curvature, since it is annihilated by $\nabla-e^a\iota_{V_a}$. We collect some standard formulae involving equivariant curvatures here
\bea \iota_{V_a}R=\nabla X_a,~~\iota_{V_b}\iota_{V_a}R=[X_a,X_b]-f_{ab}^cX_c.\label{equiv_ccurv}\eea
These relations are not restricted to the tangent bundle, they are true for a general equivariant curvature.

With the new notations we will write the $\gd_0$ variation of $\chi,\bar v$ as
\bea \nabla_{\gd_0}\chi=-d\phi-c^aX_a\chi,~~~\nabla_{\gd_0}\bar v=-c^aX_a\bar v\nn\eea
where we have incorporated the Levi-Civita connection on $X$ into the variation e.g.
\bea \nabla_{\gd_0}\chi^i:=\gd_0\chi^i+\Gc^i_{jk}(\gd_0\phi^j)\chi^k\label{covart_delta}.\eea
This way of covariantising $\gd_0$ has the advantage that $\nabla_{\gd_0}$ will go through $g,J,\Go$ as these tensors are covariantly constant.
This is the remark made in \ref{rmk_cov_susy} and we will use such tricks throughout the rest of the paper.

It turns out that the same action
\bea S_{eq\,I}=\int_M \frac12\Go_{ij}\chi^i\nabla\chi^j-\frac16R(\bar v,\chi,\Go\chi,\chi)-\gd_0(d\bar\phi*\chi)\nn\eea
is annihilated by $\gd_0$ thanks to the fact that the $G$ action preserves both the metric and complex structures.
The ghost field does not appear in the Lagrangian, so we should not integrate it in the path integral.
If one tries the same procedure when $G$ only acts holomorphically then one will not get an invariant action, due to the fact that $\Go$ will not be preserved by such actions.

\smallskip

We can also allow more background fields from the gauge sector other than the ghost. The simplest choice is to turn on a flat connection, but as our $G$ is mostly $U(1)$, this would exclude the most important geometry $M=S^3$. To accommodate a non-flat background connection, more auxiliary background fields are needed and this is where the systematic AKSZ approach becomes valuable. We will just state the result here. Let $\hat c^a$ resp. $\rho^a$ be $\FR{g}$-valued odd resp. even constant scalar field with
\bea \gd_0\hat c=\rho-\{c,\hat c\},~~~\gd_0\rho=-[c,\rho].\nn\eea
Let further $\gk$ be a contact 1-form on $M$, so we can write down a background gauge field $A=\rho\gk$ and another 2-form $B=\hat c d\gk$ with $\gd_0B=dA-\{c,B\}$. \footnote{Altogether $c,A,B$ are the fields comprising the gauge multiplet in the AKSZ construction. To compare with \cite{Kallen:2010ff}, we make the identification $c\sim A_{(0)}$, $A\sim A_{(1)}$ and $B\sim A_{(2)}$.}

We have some new BRST rules
\bea \gd_0\bar\phi=\bar v+c^a\bar V_a,~~\gd_0\phi=c^aV_a,~~\nabla_{\gd_0}\chi=-d\phi-c^aX_a\chi-\rho^aV_a,~~\nabla_{\gd_0}\bar v=-c^aX_a\bar v.\label{BRST_eq_II}\eea
The new BRST still closes $\gd_0^2=0$. The action will not be important for later use, but we record it here nonetheless
\bea &&S_{eq\,II}=\int_M \frac12\Go_{ij}\chi^i\nabla\chi^j-\frac16R(\bar v,\chi,\Go\chi,\chi)-\gd_0(d\bar\phi*\chi)\nn\\
&&\hspace{1.6cm}+\frac12\gk\chi\Go (\rho^aX_a)\chi+d\gk(\hat c^aV_a)\Go\chi+\frac12\gk d\gk(\rho-\{\hat c,\hat c\})^a\nu_a\label{action_equiv}\eea
where $\nu_a$ is the holomorphic moment map for $V_a$: $\iota_{V_a}\Go=d\nu_a$.

\section{Reformulation of RW model}\label{sec_RoRm}
The odd symmetry \eqref{curvature_term} is not suited for the localisation computation. A (non-)reason for this is: $\gd_0$ is the counterpart of the BRST symmetry in gauge theory that arises out of gauge invariance. Had one been able to localise using $\gd_0$, then one can also localise any non-supersymmetric gauge theory. This is simply too good to be true. Instead, we will need another set of odd symmetry that resembles equivariant differential (squaring to some Killing vector field), so that we can apply the powerful equivariant localisation in the space of fields.

\subsection{A second set of supersymmetry}\label{sec_Assos}

The original RW action \eqref{RW_action} and the BRST symmetry were obtained from twisting a 6D susy sigma model, with target an HK manifold \cite{Sierra:1983fj}.
The fermions in this model transform as the right handed spinors $S_R$, while the susy parameters are left-handed $S_L$.
Writing the source $\BB{R}^6$ as $\BB{R}^3\times\BB{R}^3$, one reduces the model to 3D by demanding that the fields did not depend on the second $\BB{R}^3$ and treating the $SO(3)$ acting on it as $R$-symmetry. As is by now standard, one performs a topological twist by saying that the new Lorentz group $SO(3)$ acts on the two $\BB{R}^3$ diagonally. This way, the left handed $\opn{spin}(6)$ spinor $S_L$ will have a component that is invariant under the new Lorentz group. This component can be used as the susy parameter, and this susy can be realised on any 3-manifold since its parameter is a scalar. This gives the $\gd_0$ of \eqref{RW_BRST}. The fermions of the original action now becomes either scalars $\bar v$ or 1-forms $\chi$ on $M$, writing down the original susy action leads to \eqref{RW_action}.

The above topological twisting was done without making any assumption on the holonomy of $M$, so the only way to find a supersymmetry is to make its spinor parameter into a scalar.
However, under restricted holonomy, there are more ways to realise supersymmetry on $M$. Next we shall assume that $M$ is one of the following
\begin{enumerate}
\item $S^1\times \Gs_g$, used to get the Hilbert space structure \eqref{Hilbert},
\item $S^3$ with the standard Hopf fibration structure, for getting the $S$-matrix (or rather the combination $TST$),
\item The lens spaces $L(p,q)$ for getting the combinations $T^kS$ that generate $SL(2,\BB{Z})$.
\end{enumerate}
All of our 3D geometry will be of the type of a circle bundle over a Riemann surface or orbifold
\bea S^1\to M\to\Gs.\label{Seifert}\eea
We highlight some aspects of this geometry, focusing mainly on the case $\Gs$ is smooth.
First the structure group of $M$ is reduced to $SO(2)$, and we will have a Killing vector field
\bea \reeb=\frac{\partial}{\partial \gt}\label{reeb}\eea
which rotates the circle fibre. Transverse to $\reeb$, there is also a complex structure, so we can decompose forms transverse to $\reeb$ according to their Hodge type e.g. $\Go_{\perp}^{p,q}$ for the transverse forms of type $(p,q)$. We will use the 1-form $\FR{a}$ to denote the connection of the bundle \eqref{Seifert}, for example in case of $S^3$
\bea\FR{a}=\frac{i}{2}\frac{zd\bar z-\bar zdz}{1+|z|^2} \nn\eea
with $z,\bar z$ being the coordinates of the base $S^2$ for the Hopf fibration. In particular $\gk=d\gt+\FR{a}$ is a well-defined 1-form: it is the contact 1-form of the Hopf contact structure of $S^3$.

Now consider the lens spaces, defined as a $\BB{Z}/p\BB{Z}$ quotient of $S^3$
\bea \{(z_1,z_2)||z_1|^2+|z_2|^2=1\}/\sim,~~[z_1,z_2]\sim [z_1\zeta,z_2\zeta^q],~~\zeta=e^{2\pi i/p},~~p,q>0,~~\gcd(p,q)=1\label{lens_space}\eea
In the $q=1$ case, the $L(p,1)$ is the total space of an ${\cal O}(-p)$ bundle over $S^2$. For general $q$, the base $\Gs$ is an orbifold. But for the supersymmetry, all that matters is the metric and the transverse complex structure. Both structures can be inherited from those of $S^3$, as they are invariant under the $\BB{Z}/p\BB{Z}$ action. In fact for $\gk$ one can take the same expression from $S^3$ above and multiply it by $p$
\bea \gk\to p\gk.\nn\eea

\smallskip

Next we turn to look for the new susy. We will use the susy transformation in \cite{Sierra:1983fj}, written down for a curved target HK manifold.
Though in practice it is more convenient to work with a flat target, find the desired susy transformation written in terms of the fields in \eqref{RW_action}, then manually supply the connection terms to close the algebra on a curved target. We sketch the procedure now.

The 6D gamma matrices will be
\bea &&\Gc_{1,2}=\gs_{1,2}\otimes\gs_3\otimes\gs_3,~~\Gc_{3,4}=1\otimes \gs_{1,2}\otimes\gs_3,~~\Gc_{5,6}=1\otimes 1\otimes \gs_{1,2},\nn\eea
and the $C$ matrix is $C=\Gc_1\Gc_3\Gc_5=\gs_1\otimes\ep\otimes\gs_1$ with which we form spinor bi-linears.
The susy transformation parameters are in $S_L$ with four components
\bea S_L=\opn{span}\{  |\downarrow\downarrow\downarrow\ket,~~|\uparrow\uparrow\downarrow\ket,~~|\uparrow\downarrow\uparrow\ket,~~|\downarrow\uparrow\uparrow\ket\}. \nn\eea
In $\BB{R}^6=\BB{R}^3\times\BB{R}^3$, we let the first $\BB{R}^3$ occupy 1-2-5 direction and the second 3-4-6 direction. Consider the two spinors
\bea \xi^1=|\uparrow\downarrow\uparrow\ket,~~\xi^2=|\downarrow\uparrow\uparrow\ket,\nn\eea
one may check that a particular linear combination of the two will be invariant under the diagonal $SO(3)$.
Using it as the susy parameter leads to the BRST symmetry \eqref{RW_BRST}.

The Lorentz group is now reduced to a smaller $SO(2)$ embedded diagonally into $SO(2)\times SO(2)$ acting on 1-2 and 3-4 directions simultaneously. Now $\xi^{1,2}$ are invariant separately under this diagonal $SO(2)$, and we take them to be the pair of susy parameters.
The remaining procedure is a bit of drudgery, we plug $\xi^{1,2}$ into the susy rules of \cite{Sierra:1983fj}, rewrite the fermions in terms of the scalar $\bar v$ and 1-form $\chi$, we arrive at the following table
\bea
\begin{array}{l|l}
\hline
  \gd_1\phi^i=0, & \gd_2\phi^i=\chi_{\gt}^i-\frac{1}{2}i(J\bar v)^i \\
  \gd_1\bar \phi^{\bar i}=i(J\chi_{\gt})^{\bar i}+\frac{1}{2}\bar v^{\bar i} & \gd_2\bar \phi^{\bar i}=0 \\
  \nabla_{\gd_1}\chi^i_z=(J\bar H_z)^i & \nabla_{\gd_2}\chi^i_z=-i(J\partial^{\FR{a}}_z\bar\phi)^i \\
  \nabla_{\gd_1}\chi_{\bar z}^i=-\partial^{\FR{a}}_{\bar z}\phi^i & \nabla_{\gd_2}\chi^i_{\bar z}=-iH^i_{\bar z} \\
  \nabla_{\gd_1}\chi^i_{\gt}=-\frac12\partial_{\gt}\phi^i & \nabla_{\gd_2}\chi^i_{\gt}=-\frac{i}{2}J(\partial_{\gt}\bar\phi)^i  \\
  \nabla_{\gd_1}\bar v^{\bar i}=i(J\partial_{\gt}\phi)^{\bar i} & \nabla_{\gd_2}\bar v^{\bar i}=-\partial_{\gt}\bar\phi^{\bar i}\\
  \nabla_{\gd_1}H^i_{\bar z}=-i\nabla^{\FR{a}}_{[\gt}\chi_{\bar z]}^i+\frac{1}{2}(J\nabla^{\FR{a}}_{\bar z}\bar v)^i+i(R\chi_{\bar z})^i & \nabla_{\gd_2} H^i_{\bar z}=0 \\
  \nabla_{\gd_1}\bar H^{\bar i}_z=0 & \nabla_{\gd_2}\bar H_z^{\bar i}=(J\nabla^{\FR{a}}_{[\gt}\chi_{z]})^{\bar i}+\frac{i}{2}(\nabla^{\FR{a}}_z\bar v)^{\bar i}-(JR\chi_z)^{\bar i} \\
  \hline
\end{array}.\label{second_susy}\eea
Here the notation is quite a mouthful, let us go through them one by one.
First we have incorporated the Levi-Civita connection into the definition of susy transformation in the same way as in \eqref{covart_delta}. This solves the covariance on the \emph{target} $X$.
Secondly $\chi^i_{\gt,z,\bar z}$ are the components of the original 1-form $\chi^i$ in \eqref{RW_action}, but in order to maintain covariance with respect to the fibration structure \eqref{Seifert} they are defined as
\bea \chi^i_{\gt}:=\bra \partial_{\gt},\chi^i\ket,~~~\chi^i_z:=\bra \partial^{\FR{a}}_z,\chi^i\ket=\bra \partial_z-\FR{a}_z\partial_{\gt},\chi^i\ket,~~~\chi^i_{\bar z}=\bra \partial^{\FR{a}}_{\bar z},\chi^i\ket=\bra \partial_{\bar z}-\FR{a}_{\bar z}\partial_{\gt},\chi^i\ket.\label{covariance_M}\eea
This definition takes care of the covariance on the \emph{source} $M$.
Finally $R\chi_z$ or $R\chi_{\bar z}$ is short for
\bea  (R\chi_z)^i=R(\gd_1\bar\phi,\gd_2\phi,{}^i,\chi_z^k):=R_{\bar ij~k}^{~~i}\gd_1\bar\phi^{\bar i}\gd_2\phi^j\chi_z^k
.\nn\eea
The new set of susy closes as
\bea \gd_1^2=0=\gd_2^2,~~~\{\gd_1,\gd_2\}=-\partial_{\gt}.\label{susy_closure}\eea
The two auxiliary fields (recall that $\Go_{\perp}$ denotes forms transverse to $\reeb$)
\bea H\in \Go_{\perp}^{0,1}\otimes\phi^*T^{1,0}X,~~~\bar H\in\Go_{\perp}^{1,0}\otimes\phi^*T^{0,1}X,\nn\eea
are not part of the fields in \eqref{RW_action}, but are added in order to close the algebra off shell. It is natural to let the original BRST $\gd_0$ act as zero on $H,\bar H$.

One final remark here is that the second set of susy's is valid even when the base $\Gs$ is an orbifold, since in defining \eqref{second_susy} we used only the metric of $M$ and its transverse complex structures. For example to apply \eqref{second_susy} to the lens space $L(p,q)$ \eqref{lens_space},
one can just use the metric, transverse complex structures from $S^3$, while $\gk$ should be $p$ times that of $S^3$.

\subsection{The supersymmetric action}\label{sec_Tsa}
So far we presented the new set of susy transformation using the original set of fields to facilitate comparison with the original BRST $\gd_0$. But to check \eqref{susy_closure}, as well as to check the invariance of the action, it is better to define a new set of fields that will make tab.\ref{second_susy} more canonical looking.
Define
\bea \eta^i=\chi^i_{\gt}-\frac{i}{2}(J\bar v)^i,~&&~\bar\eta^{\bar i}=i(J\chi_t)^{\bar i}+\frac{1}{2}\bar v^{\bar i},\nn\\
\bar\psi_z^{\bar i}=i(J\chi_z)^{\bar i},~&&~\psi_{\bar z}^i=\chi^i_{\bar z}.\nn\eea
The susy \eqref{second_susy} is now much neater looking
\bea
\begin{array}{l|l}
\hline
  \gd_1\phi^i=0, & \gd_2\phi^i=\eta^i \\
  \gd_1\bar \phi^{\bar i}=\bar\eta^{\bar i} & \gd_2\bar \phi^{\bar i}=0 \\
  \nabla_{\gd_1}\bar\psi^{\bar i}_z=-i\bar H^{\bar i}_z & \nabla_{\gd_2}\bar\psi^{\bar i}=-\partial^{\FR{a}}_z\bar\phi^{\bar i} \\
  \nabla_{\gd_1}\psi^i_{\bar z}=-\partial^{\FR{a}}_{\bar z}\phi^i & \nabla_{\gd_2}\psi_{\bar z}^i=-iH^i_{\bar z} \\
  \nabla_{\gd_1}\eta^i=-\partial_{\gt}\phi^i & \nabla_{\gd_2}\eta^i=0  \\
  \nabla_{\gd_1}\bar \eta^{\bar i}=0 & \nabla_{\gd_2}\bar\eta^{\bar i}=-\partial_{\gt}\bar\phi^{\bar i}\\
  i\nabla_{\gd_1}H^i_{\bar z}=\nabla_{\gt}\psi_{\bar z}^i-\nabla^{\FR{a}}_{\bar z}\eta^i-(R\psi_{\bar z})^i & \nabla_{\gd_2} H^i_{\bar z}=0 \\
  \nabla_{\gd_1}\bar H_z^{\bar i}=0 & i\nabla_{\gd_2}\bar H_z^{\bar i}=\nabla_{\gt}\bar\psi^{\bar i}-\nabla^{\FR{a}}_z\bar\eta^{\bar i}-(R\bar\psi_z)^{\bar i} \\
  \hline
\end{array}.\label{second_susy_I}\eea
Here $(R\psi)^i$ means again $R(\gd_1\bar\phi,\gd_2\phi,{}^i,\psi)=R(\bar\eta,\eta,{}^i,\psi)$. It is now also much easier to check the closure \eqref{susy_closure}.

In terms of the new fields the $L_0+L_1$ terms in action \eqref{RW_action} reads
\bea \int_M L_0+L_1&=&\int_M \opn{Vol}\Big(\bra d\phi,d\bar\phi\ket-4\bra\nabla^{\FR{a}}_z\bar\eta ,\psi_{\bar z}\ket-4\bra \nabla^{\FR{a}}_{\bar z}\eta,\bar\psi_z\ket+\bra \bar\eta,\nabla_{\gt}\eta\ket-4\bra \psi_{\bar z},\nabla_{\gt}\bar\psi_z\ket\nn\\
&&+4R_{\bar ii\bar jj}\bar\eta^{\bar i}\eta^i\bra\bar\psi_z^{\bar j},\psi_{\bar z}^j\ket\Big)+\frac12\Go_{ij}\chi^i_{\gt}\chi_{\gt}^j\,d\gk \wedge \gk\nn\eea
where $\bra\,,\ket$ means using the metric on $M$ and/or $X$ to contract indices whenever appropriate.
In the last term $\gk=d\gt+\FR{a}$, its appearance might be surprising, but this is again due to the covariantised definition of $\chi_{z,\bar z}$ in \eqref{covariance_M}.

However this action will {\bf not} be invariant under $\gd_{1,2}$, we need to first add an innocuous $|H|^2$ term
\bea L_{\rm aux}=4\opn{Vol}\bra H_{\bar z},\bar H_z\ket\nn\eea
and another more crucial one
\bea L_{\rm def}=\frac18\bar\Go_{\bar i\bar j}\bar v^{\bar i}\bar v^{\bar j}\,d\gk\wedge \gk.\label{L_3}\eea
For future convenience, we record
\bea \int_M d\gk \wedge \gk=-4\pi^2 \deg\label{cameron}\eea
where $\deg$ denotes the degree of the fibration \eqref{Seifert}, e.g. zero for $S^1\times\Gs$ and $-1$ for $S^1\to S^3\to S^2$.
Also we record that the combination
\bea \frac12\Go_{ij}\chi^i_{\gt}\chi_{\gt}^j+\frac18\bar\Go_{\bar i\bar j}\bar v^{\bar i}\bar v^{\bar j}=i\bra \bar\eta,\eta\ket.\label{L_cb}\eea
\begin{remark}
It is imperative that the modified action should still be annihilated by the original RW BRST symmetry $\gd_0$, otherwise, we are no longer computing the RW theory.
We observe that in the additional term \eqref{L_3}, the $H$'s do not have $\gd_0$ variation nor does the $\bar \Go\bar v\bar v$ term since $\bar\Go$ is a closed 2-form on $X$.
Thus even though we have modified the RW action to suit the new susy's, the new action is a BRST-closed deformation of the old RW action. Furthermore the $\bar \Go$ term depends on the degree of the $S^1$-fibration, which will produce the correct framing dependence of the path integral.

Pushing the slogan of RW theory being odd-Chern-Simons, the combination \eqref{L_cb} becomes the odd version of the term quadratic in the Coulomb branch parameter in the susy Chern-Simons or Yang-Mills.
Such a term also appears in \cite{Pestun:2007rz} (on $S^4$) and \cite{KaWiYa} (on $S^3$) in order to realise susy on the spheres. These terms are suppressed by powers of the radius $r$ (equivalently the curvature) of the sphere, since they are the price to pay for realising susy on curved backgrounds.

In our case, via a simple dimensional analysis we see \eqref{L_cb} is suppressed by a factor $1/r$.
\end{remark}
To summarise, the original BRST of the RW theory arises from twisting a 6D susy sigma model, it makes no demands on the geometry of the source 3-manifold $M$. But if assume $M$ to be a Seifert manifold, a slightly different twisting is possible, giving us with two susy's $\gd_{1,2}$. The old RW action can be made invariant under $\gd_{1,2}$ provided we add to it a BRST closed deformation, giving us the total action
\bea &&S_{\rm susy}=\int_M \sqrt gd^3x\Big(\bra d\phi,d\bar\phi\ket-4\bra\nabla^{\FR{a}}_z\bar\eta ,\psi_{\bar z}\ket-4\bra \nabla^{\FR{a}}_{\bar z}\eta,\bar\psi_z\ket+\bra \bar\eta,\nabla_{\gt}\eta\ket-4\bra \psi_{\bar z},\nabla_{\gt}\bar\psi_z\ket\nn\\
&&\hspace{.9cm}+4R_{\bar ii\bar jj}\bar\eta^{\bar i}\eta^i\bra\bar\psi_z^{\bar j},\psi_{\bar z}^j\ket+4\bra H_{\bar z},\bar H_z\ket\Big)-\gk\wedge d\gk\,\go(\bar\eta,\eta),\label{susy_action}\eea
where $\go$ is the K\"ahler form on $X$.

\subsection{The Wilson loops again}
The old expression for the Wilson loop \eqref{Wilson_adj} \eqref{Wilson_E} will work for any closed loop and holomorphic vector bundle.
Let us rewrite the connection them ${\cal A}$ with the new set of fields
\bea {\cal A}=-\dot\phi^iA_i+F_{\bar ii}\bar v^{\bar i}\chi^i_{\gt}=-\dot\phi^iA_i+\frac12F_{\bar ii}(\bar\eta-iJ\eta)^{\bar i}(\eta+iJ\bar\eta)^i.\nn\eea
If we assume that the holomorphic bundle is in fact {\bf hyperholomorphic} \cite{Verbitsky:430998}, which means that its curvature is type (1,1) with respect to all three complex structures $I,J,K$. Then the expression ${\cal A}$ simplifies
\bea {\cal A}=-\dot\phi^iA_i+F_{\bar ii}\bar\eta^{\bar i}\eta^i.\label{new_start}\eea
Further we see that $\gd_{1,2}$ now involves $\gt$-derivative on the fields while the derivative on $\phi$ in \eqref{new_start} is along the curve supporting the Wilson loop, so to achieve susy-invariance, these two must be aligned, i.e. the Wilson loop must be placed along the closed orbit of the vector field $\partial_{\gt}$. We demonstrate the susy-invariance next.

Parameterise the orbit by $\gt\in[0,2\pi]$, write
\bea W(\gt_1,\gt_2)={\bf P}\exp\int_{\gt_1}^{\gt_2}{\cal A}
={\bf P}\exp\int_{\gt_1}^{\gt_2}d\gt (-\dot\phi^iA_i+F_{\bar ii}\bar\eta^{\bar i}\eta^i),\nn\eea
so the Wilson loop operator is just $\Tr W(0,2\pi)$. The variation of $W$ reads
\bea \gd_1\Tr W(0,2\pi)&=&\Tr\int_0^{2\pi}d\gt W(\gt,2\pi)(\gd_1{\cal A)}W(0,\gt)\nn\\
&=&\Tr\int_0^{2\pi}d\gt W(\gt,2\pi)(-\dot\phi^i\bar\eta^{\bar i}\partial_{\bar i}A_i+F_{\bar ii}\bar\eta^{\bar i}\dot\phi^i+\bar\eta^{\bar j}\partial_{\bar j}F_{\bar ii}\bar\eta^{\bar i}\eta^i) W(0,\gt).\nn\eea
Recall holomorphicity says $A^{0,1}=0$, so the first two terms cancel. For the third term, it vanishes by Bianchi identity satisfied by $F$.
Now for $\gd_2$
\bea \gd_2\Tr W(0,2\pi)&=&\Tr\int_0^{2\pi}d\gt W(\gt,2\pi)(\gd_2{\cal A)}W(0,\gt)\nn\\
&=&\Tr\int_0^{2\pi}d\gt W(\gt,2\pi)(-\nabla_{\gt}\eta^iA_i-\dot\phi^i\eta^j\partial_jA_i-F_{\bar ii}\dot{\bar\phi}^{\bar i}\eta^i+\eta^j\partial_jF_{\bar ii}\bar\eta^{\bar i}\eta^i) W(0,\gt)\nn\\
&=&\Tr\int_0^{2\pi}d\gt W(\gt,2\pi)(-\partial_{\gt}(\eta^iA_i)-\dot\phi^i\eta^j\partial_{[j}A_{i]}-[\eta^jA_j,F_{\bar ii}\bar\eta^{\bar i}\eta^i]) W(0,\gt).\nn\eea
We now use integration by part on the first term and that fact that
\bea \partial_{\gt_2}W(\gt_1,\gt_2)={\cal A}(\gt_2)W(\gt_1,\gt_2),~~~\partial_{\gt_1}W(\gt_1,\gt_2)=-{\cal A}(\gt_1)W(\gt_1,\gt_2)\nn\eea
to get
\bea \gd_2\Tr W(0,2\pi)
&=&\Tr\int_0^{2\pi}d\gt W(\gt,2\pi)(-[{\cal A},\eta^iA_i]-\dot\phi^i\eta^j\partial_{[j}A_{i]}-[\eta^jA_j,F_{\bar ii}\bar\eta^{\bar i}\eta^i]) W(0,\gt)\nn\\
&=&\Tr\int_0^{2\pi}d\gt W(\gt,2\pi)(\dot{\phi}^i\eta^jF_{ij})W(0,\gt)=0,\nn\eea
where we used $F^{2,0}=0$.
\begin{remark}
  If we take instead the susy model with action $L_{\rm susy}$ in \eqref{susy_action} as the starting point and
  \bea W=\Tr{\bf P}\exp\int_0^{2\pi}d\gt (-\dot\phi^iA_i+F_{\bar ii}\bar\eta^{\bar i}\eta^i)\nn\eea
  as the definition of Wilson-loops, then one can again allow any holomorphic bundles. This is perhaps a viable point of view since we would like to involve all objects of $D^b(X)$, some of which presumably can not be resolved using hyper-holomorphic vector bundles. At the time one is left with a quandary as to whether the RW model or susy-model is more fundamental.
\end{remark}
\begin{remark}
  This remark concerns the abundance of hyper-holomorphic bundles. Our most common example for $X$ will be a cotangent bundle $X=T^*Y$ for some K\"ahler manifold $Y$. If one takes any holomorphic vector bundle $E$ on $Y$, then Feix \cite{Feix2002HypercomplexMA} showed that it extends uniquely to $T^*Y$ as a hyper-holomorphic bundle.
  Even when $X$ is not of the form $T^*Y$, but has a holomorphic $U(1)$ with rather generous conditions, one can modify a holomorphic bundle using this $U(1)$ to make it hyper-holomorphic \cite{HAYDYS2008293}.
\end{remark}

\subsection{Equivariant extension of susy}\label{sec_Eea}
As mentioned in sec.\ref{sec_Ee} we failed to incorporate a holomorphic isometry into the RW model. Now we have reformulated the RW model into a susy $\sigma$-model for $M$ Seifert, in particular, the action \eqref{susy_action} retains no memory of its hyperK\"ahler past. This way an isometry, tri-holomorphic or not, is treated entirely the same. The ghost $c$ and also $\hat c$ in sec.\ref{sec_Ee} are recombined into a conjugate pair $c,\bar c$, with the susy rules
\bea \gd_1c=-\frac12\{c,c\},~&&~\gd_2c=\rho-\{c,\bar c\},\nn\\
\gd_1\bar c=\rho-\{c,\bar c\},~&&~\gd_2\bar c=-\frac12\{\bar c,\bar c\},\nn\\
\gd_1\rho=-[c,\rho],~&&~\gd_2\rho=-[\bar c,\rho].\nn\eea
The combination $\varrho=\rho-\{c,\bar c\}$ will appear often in the closure of the susy: $\gd_1^2=\gd_2^2=0$ and $\{\gd_1,\gd_2\}=-Ad_{\varrho}$.
As $\varrho$ has no susy variation, it can be evaluated at any numerical value and will serve as the equivariant parameter.

The equivariant version of the susy rules read
\bea
\begin{array}{l|l}
\hline
  \gd_1\phi^i=c^aV^i_a, & \gd_2\phi^i=\eta^i+\bar c^aV_a^{\bar i} \\
  \gd_1\bar \phi^{\bar i}=\bar\eta^{\bar i}+c^aV_a^{\bar i} & \gd_2\bar \phi^{\bar i}=\bar c^aV_a^{\bar i} \\
  \nabla_{\gd_1}\bar\psi=-i\bar H-(c^aX_a)\bar\psi & \nabla_{\gd_2}\bar\psi=-\partial^{\FR{a}}\bar\phi-(\bar c^aX_a)\bar\psi \\
  \nabla_{\gd_1}\psi=-\bar\partial^{\FR{a}}\phi-(c^aX_a)\psi & \nabla_{\gd_2}\psi=-iH-(\bar c^aX_a)\psi \\
  \nabla_{\gd_1}\eta^i=-\partial_{\gt}\phi^i-(c^aX_a)\eta^i-\rho^aV^i_a & \nabla_{\gd_2}\eta^i=-(\bar c^aX_a)\eta^i  \\
  \nabla_{\gd_1}\bar \eta^{\bar i}=-(c^aX_a)\bar\eta^{\bar i} & \nabla_{\gd_2}\bar\eta^{\bar i}=-\partial_{\gt}\bar\phi^{\bar i}-(\bar c^aX_a)\bar\eta^{\bar i}-\rho^aV_a^{\bar i}\\
  i\nabla_{\gd_1}H=\nabla_{\gt}\psi-\bar\nabla^{\FR{a}}\eta-(c^aX_a)iH-(\rho^aX_a+R)\psi & \nabla_{\gd_2} H^i_{\bar z}=-(\bar c^aX_a)iH \\
  i\nabla_{\gd_1}\bar H=-(c^aX_a)i\bar H & i\nabla_{\gd_2}\bar H=\nabla_{\gt}\bar\psi-\nabla^{\FR{a}}\bar\eta-(\bar c^aX_a)i\bar H-(\rho^aX_a+R)\bar\psi \\
  \hline
\end{array}\label{second_susy_eq_I}\eea
where $(R\psi)^i$ is again short for $R(\eta,\bar\eta,{}^i,\psi)$. The susy rules close to
\bea \gd_1^2=\gd_2^2=0,~~~\{\gd_1,\gd_2\}=-\partial_{\gt}+\varrho{}\circ,\nn\eea
where $\varrho$ acts as $L_{\varrho^aV_a}$, except on $c,\bar c$ where it acts as $-Ad_{\varrho}$.

The following action is invariant under $\gd_{1,2}$
\bea &&S_{\rm susy\;eq}=\int_M \sqrt gd^3x\Big(\bra d\phi,d\bar\phi\ket-4\bra\nabla^{\FR{a}}_z\bar\eta ,\psi_{\bar z}\ket-4\bra \nabla^{\FR{a}}_{\bar z}\eta,\bar\psi_z\ket+\bra \bar\eta,\nabla_{\gt}\eta\ket-4\bra \psi_{\bar z},\nabla_{\gt}\bar\psi_z\ket+4R_{\bar ii\bar jj}\bar\eta^{\bar i}\eta^i\bra\bar\psi_z^{\bar j},\psi_{\bar z}^j\ket\nn\\
&&\hspace{2cm}+4\bra H_{\bar z},\bar H_z\ket+4\bra \bar\psi_z,\rho^aX_a\psi_{\bar z}\ket +\bra\bar\eta,\rho^aX_a\eta\ket-\frac12||\rho^aV_a||^2\Big)-\gk\wedge d\gk(\go(\bar\eta,\eta)+\rho^a\mu_a),\label{complete_action_equiv}\eea
where $\mu_a$ is the moment map of $V_a$ w.r.t the K\"ahler form $\go$. We point out that the combinations
\bea R(\bar\eta,\eta,{}^i{}_j)+\rho^a(X_a)^i_{~j};~~~\go(\bar\eta,\eta)+\rho^a\mu_a\label{equiv_classes}\eea
are respectively the equivariant curvature and K\"ahler class in the Weil-model. We will discuss this point more in sec.\ref{sec_TEea}.
\begin{remark}
  In the previous sec.\ref{sec_Tsa}, the susy action is annihilated by $\gd_{1,2}$ and also by the original $\gd_0$ of the RW theory. This is no longer true of the equivariant action above. But the failure is quite mild, the only offending term comes from the $\gd_0$ hitting the equivariant K\"ahler form, and to fix it one needs to add the last three terms in \eqref{action_equiv}. We choose not to do so since we must allow $V_a$ to be non tri-holomorphic and there will be no moment map w.r.t $\Go$.
  To summarise we will use \eqref{complete_action_equiv} as a {\bf substitute} of the equivariant RW theory with non tri-holomorphic isometries.
\end{remark}

\section{Localisation of RW model}
The localisation technique in susy gauge theory has matured enormously since the ground breaking work of Pestun \cite{Pestun:2007rz}. The reader may consult \cite{review_loc} for a very modern and streamlined presentation on the subject. However, we will try to be self-contained in this paper, whenever possible.

\subsection{A quick review of a toy model}
The gist of equivariant localisation in supersymmetric theories is to organise the susy transformation so that it acts as equivariant differential in the field space. We will take a finite dimensional toy model. Fix $X$ to be a finite dimensional complex (super-)manifold. For definiteness we assume that $X$ is a super-manifold of type
\bea X=\Pi TM\oplus E\oplus \Pi E\nn\eea
with $M$ being a complex manifold and $E$ a holomorphic vector bundles on $M$. The symbol $\Pi$ means that we take the fibre coordinate to be odd. We assume that there is a real Killing vector field $V$ on $M$ preserving complex structure, and that it has been lifted to act holomorphically on $E$ preserving the innerproduct on $E$.
We denote with $x,\bar x$ the coordinate of $M$, $\psi,\bar\psi$ the fibre coordinate of $\Pi TM$, $H,\bar H,\chi,\bar\chi$ the even and odd fibre coordinate of $E\oplus\Pi E$. Let an odd symmetry $\gd_{eq}$ act as
\bea \gd_{eq} x=\psi,~~\gd_{eq}\psi=L_Vx;~~~\gd_{eq}\chi=H,~~\gd_{eq} H=L_V\chi.\label{equivariant_cplx}\eea
In principle we should covariantise the last two transformations as we have done throughout, but for this illustrative purpose, we shall gloss over it.
The variation \eqref{equivariant_cplx} is ubiquitous in localisation computation, it has the basic pattern
\bea \gd_{eq}\,{\rm coordinate}={\rm momentum},~~ \gd_{eq}\,{\rm momentum}={\rm vector\; field}\nn\eea
where 'coordinate' and 'momentum' have opposite parity. The complex \eqref{equivariant_cplx} is what we call the 'equivariant differential complex' and all the work done
in sec.\ref{sec_Assos} was so that we can arrange the susy's into this form.

Let $S(x,\bar x,\psi,\bar\psi,\chi,\bar\chi,H,\bar H)$ be a function that is annihilated by $\gd_{eq}$, in particular $L_VS=0$. Consider the integral
\bea \int dxd\bar xd\psi d\bar\psi d\chi d\bar\chi dH d\bar He^{-S}.\nn\eea
The detailed form of $S$ is not important for our illustration, we just assume that it has enough damping for the $x,H$ fields. One can add to $S$ any term $W$ of the type below without altering the integral
\bea S\to S+tW=S+t\gd_{eq}\Psi,\label{exact_term}\eea
provided
\bea \gd_{eq}^2\Psi=0\label{Killing_condition}.\eea
The invariance comes from an easy application of the Stokes theorem.
Exploiting this $t$-independence, one may send $t\to\infty$ which suppresses the integrand except at the zero of $W=\gd_{eq}\Psi$.
There are many choices of $\Psi$ so that the zero of $W$ has a very small locus. For example
\bea \Psi=(V^i\bar \psi^{\bar i}g_{i\bar i}+\chi^a\bar H^{\bar a}h_{a\bar a}+c.c)\nn\eea
where $g,h$ denote the hermitian metric on $TM$ and $E$. This would satisfy \eqref{Killing_condition} if $V$ preserves both $g$ and $h$.
The bosonic part of $W$ equals $|V|^2+|H|^2$, and so with this choice and $t\to\infty$ we merely need to evaluate $S$ at $V=H=0$ and then perform a Gaussian integral over the infinitesimal fluctuation round these loci.
The result is
\bea \int e^{-S}=\sum_p e^{-S}\frac{\det{\sf F}}{\det {\sf B}}\Big|_p\label{gaussian_I}\eea
where $p$ labels the locus $V=0$ and ${\sf B,F}$ are the matrices
\bea {\sf B}=\left[
         \begin{array}{cc}
           \partial_{x^i}\partial_{\bar x^{\bar j}}W &  \partial_{H^a}\partial_{\bar x^{\bar j}}W\\
            \partial_{x^i}\partial_{\bar H^{\bar b}}W & \partial_{H^a}\partial_{\bar H^{\bar b}}W \\
         \end{array}\right],~~~{\sf F}=\left[
         \begin{array}{cc}
           \partial_{\psi^i}\partial_{\bar\psi^{\bar j}}W &  \partial_{\chi^a}\partial_{\bar\psi^{\bar j}}W\\
            \partial_{\psi^i}\partial_{\bar\chi^{\bar b}}W & \partial_{\chi^a}\partial_{\bar\chi^{\bar b}}W \\
         \end{array}\right].\nn\eea
In fact more simplification can be achieved. At a point $p$ where $V=0$, then $V$ acts on the tangent space round $p$ as a linear transform $V_i^j=\partial_{x^i}V^j|_p$. It will also act as linear transform on the fibre of $E$ over $p$, we write it as $V_a^b$.
Consider the matrix
\bea {\sf V}_x=\left[\begin{array}{cc}
           V_j^i & 0 \\
            & 1 \\
         \end{array}\right],~~{\sf V}_{\chi}=\left[\begin{array}{cc}
           1 & 0 \\
            & V^a_b \\
         \end{array}\right]\nn\eea
Then the fact that $\gd_{eq}W=0$ implies ${\sf F}{\sf V}_x=\bar{\sf V}_{\bar\chi}\bar{\sf B }$, and we can simplify \eqref{gaussian_I} as
\bea \int e^{-S}=\sum_p e^{-S}\frac{\det{\sf \bar V}_{\bar\chi}}{\det {\sf V}_x}\Big|_p=\sum_p e^{-S}\frac{\det(-{\sf V}_{\chi})}{\det {\sf V}_x}\Big|_p\label{gaussian_II}\eea
where the last equality is because $V$ preserves the fibre metric.

In conclusion the original integral is largely insensitive to the choice of $\Psi$, provided that $\sf B$ and $\sf F$ are non-degenerate. This is particularly useful for our path integral application, since we can effectively ignore the details of the action and only focus on the vector field $V$ and its weights close to its zeros.

\subsection{The role of the double complex}
In finite dimension, the formula \eqref{gaussian_II} is a satisfactory result and it recovers the Duistermaat-Heckman formula \cite{Duistermaat1982}.
But for path integral applications, it necessarily happens that the two determinants in \eqref{gaussian_II} are infinite dimension ones and hence ill-defined. However one can make sense of the ratio between the two if there are further cancellations. The double complex structure we highlight here will serve this purpose. This part of the story is quite often left implicit in the literature on localisation, but it is painstakingly expounded in \cite{exotic_instanton} (and is indispensable for the application therein).
The main idea is that there should be a map going from $x$ to $\chi$, which in finite dimension setting typically comes from a section $s:\,M\to E$ but in infinite dimensional setting may take the form of a differential operator.
The goal is to achieve further cancellation between the two determinants, leaving behind a hopefully finite dimensional remanent coming from the kernel and the cokernel of the said differential operator.

To this purpose we observe that we can split the sum of susy $\gd_1+\gd_2$ into two differentials $\gd_{eq}+\gdh$, such that
$(\phi,\eta,\psi,H,\gd_{1,2})$ is equipped with the structure of a double complex
\bea
  \begin{tikzpicture}
  \matrix (m) [matrix of math nodes, row sep=1.7em, column sep=2em]
    {    \phi & \psi  \\
         \vphantom{H}\eta &  H \\ };

  \path[->, font=\scriptsize]
  (m-1-1) edge node[above] {$\gdh$} (m-1-2);
  \path[->, font=\scriptsize]
  (m-2-1) edge node[above] {$\gdh$} (m-2-2);

  \path[->, font=\scriptsize]
  (m-2-1) edge node[left] {$\gd_{eq}$} (m-1-1)
  (m-2-2) edge node[right] {$\gd_{eq}$} (m-1-2);
    \end{tikzpicture}\label{double_cplx}\eea
where $\gd_{eq}$ and $\gdh$ are
\bea &\begin{array}{l|l}
  \hline
                          \gd_{eq} \phi=\eta  &  \gdh \phi=0 \\
         \nabla_{\gd_{eq}}\eta=-\partial_{\gt}\phi & \nabla_{\gdh}\eta=0 \\
         \nabla_{\gd_{eq}}\psi=-iH &  \nabla_{\gdh}\psi=-\partial^{\FR{a}}_{\bar z}\phi \\
         i\nabla_{\gd_{eq}}H=\nabla_{\gt}\psi-R(\psi) & i\nabla_{\gdh}H=-\nabla^{\FR{a}}_{\bar z}\eta \\
         \hline
                         \end{array},\label{second_susy_II}\\
                         &\gd_{eq}+\gdh=\gd_1+\gd_2.\nn\eea
There is of course an identical set of differentials for the barred fields.

The double complex \eqref{double_cplx} has the following significance. The differential $\gd_{eq}$ that acts vertically works exactly as in \eqref{equivariant_cplx}, showing that we can reduce the path integral to the form of \eqref{gaussian_II}, with $x,\chi$ replaced with $\phi,\psi$.
The $\gdh$ operator maps the fields from the first column to the second and so will relate the determinant over the $\phi$ field to that over the $\psi$ field.
As $\gdh$ is the $\bar\partial$-operator transverse to $\reeb$, it has a well understood kernel and cokernel. Thus the determinant \eqref{gaussian_II} will mostly cancel, except a remnant coming from the said kernel and cokernel. We will see next how this all plays out.

\smallskip

We will first need to pick a good $\Psi$ as in \eqref{exact_term}. Let
\bea \Psi=-\bra \eta,\partial_{\gt}\bar\phi\ket-\bra \bar \eta,\partial_{\gt}\phi\ket+\bra\psi,i\bar H-\partial_z^{\FR{a}}\bar\phi\ket+\bra\bar \psi,iH-\partial_{\bar z}^{\FR{a}}\phi\ket
.\nn\eea
We can work out
\bea (\gd_1+\gd_2)\Psi=2|\partial_{\gt}\phi|^2+2\bra \eta,\nabla_{\gt}\bar\eta\ket+2|H|^2+2|\partial_z^{\FR{a}}\bar\phi|^2
-2\bra\psi,\nabla_{\gt}\bar\psi\ket+2\bra\psi,\nabla_z^{\FR{a}}\bar\eta\ket+2\bra\bar\psi,\nabla_{\bar z}^{\FR{a}}\eta\ket+2R(\bar\eta,\eta,\psi,\bar\psi).\nn\eea
Since the bosonic part above is zero at
\bea \partial_{\gt}\phi=\partial_z\bar\phi=\partial_{\bar z}\phi=0,~~~H=0\label{local_locus}.\eea
We will refer to this as the {\bf localisation locus}.
\begin{remark}
The locus \eqref{local_locus} is much bigger than the original RW theory, which as we recall consists of only constant maps $d\phi=0$.
In contrast, if we take $M$ to be an $S^1$ bundle over $\Gs$, then \eqref{local_locus} would allow holomorphic maps $\Gs\to X$, i.e. there are solitons. This is again slightly disconcerting, the best we could hope for is that the original RW theory is the truncation to the zero soliton sector of our new theory.
\end{remark}

We will now ignore solitons and focus on the localisation locus that are just constant maps. We will need to integrate over the infinitesimal deviations from the constant maps, so we separate each field into its 'zero mode part' (meaning it will have vanishing kinetic term) and 'fluctuation part'
\bea \phi=\phi_0+\phi_1,~~~\eta=\eta_0+\eta_1~~\cdots~~{\rm etc}.\label{split}\eea
We can read off the zero modes from the double complex \eqref{double_cplx}.
Excepting the auxiliary field $H$, the zero modes are annihilated by $\partial_{\gt}$ and are in the kernel or cokernel of $\gdh$, since their kinetic terms are constructed out of $\partial_{\gt}$ and $\gdh$ (see sec.1.5 and 2.3 of \cite{exotic_instanton}).
Seeing that $\gdh=\partial^{\FR{a}}$ or $\bar\partial^{\FR{a}}$, the zero modes $\phi_0,\eta_0$ are constants on $M$ while the zero modes $\psi_0,\bar\psi_0$
will depend on the geometry of $M$. Indeed $\psi$ is a section of $\Go_{\perp}^{0,1}\otimes \phi^*T^{1,0}X$, but now the second tensor factor is just a constant vector space $\phi_0^*T^{1,0}X$, which we ignore for the moment. When the geometry of $M$ is $S^1\times\Gs_g$, then requiring $\partial_{\gt}\psi_0=\partial_z^{\FR{a}}\psi_0=0$ tells us that $\psi_0$ is a harmonic section of $\Go^{0,1}(\Gs_g)$, which has complex dimension
$g$.
Being slightly more general, we take $M$ to be a circle bundle of degree $d$ over $\Gs_g$ and enforce first only $\nabla_z^{\FR{a}}\psi=0$. Those $\psi$ satisfying this correspond to the harmonic sections of
\bea \Go^{0,1}(\Gs_g)\otimes{\cal O}(-dn)\nn\eea
where $n$ is the Fourier mode along the circle fibre. Setting $n=0$ i.e. enforcing $\partial_{\gt}\psi=0$ gives back the same result as in $S^1\times\Gs_g$. Therefore in the other two geometries of importance to us, namely the total space of ${\cal O}(-1),{\cal O}(-p)$ over $S^2$, there will be no zero modes for $\psi,\bar\psi$.

Having separated the zero modes, next we will truncate the complexes \eqref{second_susy_I} or \eqref{second_susy_II} to linear order and obtain the corresponding complex for the fluctuations.
As usual there is a bit of technicality related to covariance when we do the split \eqref{split}, the appropriate definition is
\bea &\upphi^i=\phi^i_1,~~~\upeta^i=\eta^i_1+\Gc^i_{jk}\phi_1^j\eta_0^k,~~~\uppsi^i=\psi^i_1+\Gc^i_{jk}\phi_1^j\psi_0^k,\nn\\
&~~~{\sf H}^i=H^i_1+\Gc^i_{jk}\phi_1^jH_0^k-iR(\bar\upphi,\eta_0,{}^i,\psi_0)-iR(\upphi,\bar\eta_0,{}^i,\psi_0).\nn\eea
It is the $\upphi,\upeta,\uppsi,{\sf H}$ fields that will have a nice (and covariant) truncated complex.
Apart from changing the font of the fields, the complex looks mostly the same
\bea \begin{array}{l|l}
  \hline
                          \nabla_{{\gd_{eq}}}\upphi^i=\upeta^i  &  \nabla_{\gdh}\upphi=0 \\
         \nabla_{\gd_{eq}}\upeta^i=-\partial_{\gt}\upphi^i+R(\bar\eta_0,\eta_0,{}^i,\upphi) & \nabla_{\gdh}\upeta=0 \\
         \nabla_{\gd_{eq}}\uppsi^i=-i{\sf H}^i &  \nabla_{\gdh}\uppsi=-\partial^{\FR{a}}_{\bar z}\upphi \\
         i\nabla_{\gd_{eq}}{\sf H}^i=\partial_{\gt}\uppsi-R(\bar\eta_0,\eta_0,{}^i,\uppsi) & i\nabla_{\gdh}{\sf H}=-\partial^{\FR{a}}_{\bar z}\upeta \\
         \hline
\end{array}.\label{second_susy_linear}\eea
What is important for us is that $\gd_{eq}$ now squares to
\bea \nabla_{\gd_{eq}}^2(-)^i=-\partial_{\gt}(-)^i+R(\bar\eta_0,\eta_0,{}^i,-)\label{susy_closure_linear}\eea
on the fluctuations.
We can regard $\eta_0,\bar\eta_0$ as a basis of the 1-forms on $X$, so the second term above is the curvature 2-form of $TX$.
The determinant of \eqref{susy_closure_linear} will produce the right characteristic class out of the path integral.
Indeed, integration over the fluctuations give, according to \eqref{gaussian_II}, a determinant factor
\bea \frac{\det_{\uppsi}(-\partial_{\gt}+R)}{\det_{\upphi}(-\partial_{\gt}+R)}.\nn\eea
We explained above that the double complex leads to further cancellation of the determinants, leaving only
\bea \frac{\det'_{V^1}(-\partial_{\gt}+R)}{\det'_{V^0}(-\partial_{\gt}+R)},~~{\rm where}~~V^0=\ker\gdh,~~V^1=\opn{coker}\gdh.\nn\eea
The prime means that we exclude the modes annihilated by $\partial_{\gt}$, since these modes are by definition the zero modes and are integrated separately.
Since $R$ is the curvature 2-form, taking the determinants above will produce an invariant polynomial of $R$ and hence certain characteristic classes.

\subsection{The equivariant extension again}\label{sec_TEea}
We should write down the truncated complex also for the equivariant version \eqref{second_susy_eq_I}. By now the pattern is perhaps quite predictable,
so to avoid repetition we will not write down the full truncated complex, but only the closure of susy algebra on the truncated fields.

Remember that the fields $c,\bar c,\rho$ are treated as backgrounds, so we choose the background where $c=\bar c$. This is consistent since $\gd_{eq}(c-\bar c)=-c^2+\bar c^2\to0$. Expand the other fields to first order in fluctuation, then $\gd_{eq}$ acting on them closes to
\bea \nabla_{\gd_{eq}}^2(-)^i=-\partial_{\gt}(-)^i+\frac12R(\eta_0+\bar\eta_0+2cV,\eta_0+\bar\eta_0+2cV,{}^i,-)-\varrho^aX_a(-)^i\label{susy_closure_linear_equiv}.\eea
To interpret this result, we first observe that the combinations $\zeta_0^i=\eta^i_0+2cV^i$ and $\bar\zeta_0^{\bar i}=\bar\eta_0+2cV^{\bar i}$ have susy variation
\bea \nabla_{\gd_{eq}}\zeta_0^i=\varrho V^i,~~\nabla_{\gd_{eq}}\bar\zeta_0^{\bar i}=\varrho V^{\bar i}.\label{cartan}\eea
This is actually the {\bf Cartan algebra} if we identify $\zeta^i_0$ as the basis of local 1-forms $dx^i$ on $X$.
The closure \eqref{susy_closure_linear_equiv} can be written as
\bea \nabla_{\gd_{eq}}^2(-)^i=-\partial_{\gt}(-)^i+R(\zeta_0,\bar\zeta_0,{}^i,-)-\varrho^aX_a(-)^i=-\partial_{\gt}(-)^i+R_{eq}(-)^i\label{susy_closure_linear_equiv_I},\eea
where the combination named $R_{eq}$ is nothing but the equivariant curvature, indeed it is annihilated by $d+\varrho^a\iota_{V_a}$, which can be checked using \eqref{equiv_ccurv}.

\smallskip

We end this subsection with a technical remark. The issue is that in the action \eqref{complete_action_equiv} we find the equivariant curvature \eqref{equiv_classes} in the Weil-model. But the closure of $\gd_{eq}$ gives the same equivariant curvature but in the Cartan model. For the readers familiar with the equivariant cohomology, the discrepancy is not serious, it is well-known that the Weil-model (has ghosts) or the Cartan model (has no ghosts) are equivalent for computing the equivariant cohomology. But we recall the argument to be self-contained.
The main idea is to find an interpolation in between so that the change along the interpolation is via $\gd_{eq}$-exact terms.

The discussion is a preparation for later sections, where we will only care about the zero modes. Therefore we will assume here that the fields have no dependence on $M$.
For $t\in[0,1]$ we let \footnote{Note $\rho_t$ will interpolate between $-\rho$ and $\varrho$, we could have fixed it via a flip $\varrho\to -\varrho$, but opted to leave things as they are.}
\bea \eta_t=\eta+2tcV,~~\rho_t=(t-1)\rho+t\varrho+4t(1-t)c^2,~~(R_t)^i_{~j}=R(\bar \eta_t,\eta_t,{}^i,{}_j)-\rho_t X^i_{~j}\nn\eea
where $c^2:=\{c,c\}/2$. Here $R_t$ will extrapolate between the equivariant curvature in the Weil and Cartan model. Omitting some algebra, we can compute the following relations
\bea
&D_tR_t:=(\nabla_{\gd_{eq}}+(1-t)Ad_{2cX})R_t=0,\nn\\
&\dot{R}_t=D_t(2cX).\nn\eea
Recall that one constructs a characteristic class by plugging $R_t$ into an ad-invariant polynomial $f(x)$ with $x\in\FR{g}$ e.g. the determinant or trace of some powers of $x$. We will take $f(x)=\Tr[x^n]$ for concreteness, but the discussion is entirely the same for other choices. Plug $R_t$ into $f$ and differentiate w.r.t $t$
\bea \frac{d}{dt}f(R_t)=n\Tr[D_t(2cX)(R_t)^{n-1}]=n\Tr[D_t(2cXR_t^{n-1})]=n\gd_{eq}\Tr[2cXR_t^{n-1}].\nn\eea
Integrating both sides from 0 to 1, we see that the characteristic classes $f(R_1)$ and $f(R_0)$ differ by something $\gd_{eq}$-exact.

The same argument can be applied to the equivariant K\"ahler class or the Chern-characters arising from the Wilson loops. In what follows, we will just gloss over the distinction between characteristic classes written using $\eta,\rho$ or $\zeta,\varrho$.

\subsection{Warmup example $M=S^1\times\Gs_g$}
This is a more or less trivial example, since physics already tells us what to expect. The circle in $M$ means that the path integral has a Hamiltonian interpretation: it is the super-trace of $e^{-H}$ over the Hilbert space associated with $\Gs$, which is
\bea {\cal H}(\Gs_g,X)=H^{0,\sbullet}(X,(\Go^{\sbullet,0}_X)^{\otimes g}).\nn\eea
As the Hamiltonian is zero ($\gd$-exact), we should just get the dimension of the above vector space.

We can directly use \eqref{gaussian_II} with the localisation locus being constant maps $\phi_0:\;M\to X$. So we should integrate over $\phi_0$, along with the other zero modes $\psi_0,\eta_0$ (the integration over $H$ is straightforward), while the non-zero modes contribute only the determinant factors
\bea {\rm det~factor}=
\frac{\det'_{V^1}(-\partial_{\gt}+R)}{\det'_{V^0}(-\partial_{\gt}+R)},~~{\rm where}~~V^0=\ker\gdh,~~V^1=\opn{coker}\gdh\label{step_det_factor}\eea
where $R$ denotes the matrix $R^i_{~j}=R_{\bar kk~j}^{~~\,i}\bar\eta_0^{\bar k}\eta_0^k$ acting on the tangent space at the point $\phi_0$.

To obtain $V^{0,1}$, we saw earlier that for $M=S^1\times\Gs_g$, $\gdh$ acts as the $\bar\partial$ operator on $\Gs_g$, so \eqref{step_det_factor} gives
\bea {\rm det~factor}=\prod_{\BB{Z}\ni n\neq0}\frac{\det_{V^1}(in+R)}{\det_{V^0}(in+R)},~~~~V^{\sbullet}=H^{0,\sbullet}_{\bar\partial}(\Gs_g)\otimes T^{1,0}_{\phi_0}X.\nn\eea
Here the sum over $n$ comes from the Fourier modes of $\partial_{\gt}$. The $n=0$ mode is excluded because those modes annihilated by $\bar\partial$ and having $n=0$ are by definition the zero modes, which are integrated separately.

Now we need to write the matrix $R$ as the Chern-roots and perform the product over $n$.
As mentioned earlier $\eta_0,\bar\eta_0$ are a basis of the 1-forms on $X$, however the normalisation is tricky, we use the normalisation from \cite{alvarez-gaume1983} (see the discussion there above equation 4.7) and make the association \footnote{thankfully we do not need the normalisation for non-zero modes since we only compute determinants over these modes.}
\bea \eta_0^k\bar\eta_0^{\bar k}\to \frac{i}{(2\pi)^2} dz^k\wedge d\bar z^{\bar k}\label{ad_hoc}\eea
where $z,\bar z$ are the complex coordinates of $X$. Now we let $x_i$ be the eigen-value of the $\opn{End}TX$-valued 2-form $(2\pi)R_{\bar kk~j}^{~~\,i}\bar\eta_0^{\bar k}\eta_0^k$, so that $x$ would conform to the usual normalisation of Chern-roots.
Now with the product formula for sine
\bea \frac{\sin(\pi x)}{\pi x}=\prod_{n=1}^{\infty}(1-\frac{x^2}{n^2}).\nn\eea
we can write the determinant factor as
\bea {\rm det~factor}\sim\prod_i\Big(\frac{\sinh(x_i/2)}{x_i/2}\Big)^{g-1}
.\nn\eea
For hyperk\"ahler manifold, we have an extra simplification. The curvature is valued in $\FR{sp}(2n)$, and so its Chern roots all pair up, say
\bea x_i=-x_{i+n}.\label{eigen_val_paired}\eea
So we may as well write
\bea {\rm det~factor}\sim\prod_i\Big(\frac{1-e^{-x_i}}{x_i}\Big)^{g-1}=\opn{Td}(X)\big(\prod_i\frac{1-e^{-x_i}}{x_i}\big)^g,\nn\eea
where $\opn{Td}(X)$, the Todd genus, is the multiplicative series associated to the polynomial $x/(1-e^{-x})$. Furthermore
\bea \prod_i(1-e^{-x_i})=1-\opn{ch}\Go_X^{1,0}+\opn{ch}\Go_X^{2,0}-\cdots=\opn{ch}(\Go^{\sbullet,0}).\nn\eea
So our determinant factor becomes
\bea
{\rm det~factor}=\opn{Td}(X)\opn{ch}\big((\Go^{\sbullet,0})^{\otimes g}\big)\prod_i(x_i)^{-g}.\nn\eea
This is almost what we need, except the last product over $x_i$ factor (which is proportional to $\det R$). But this will get neatly cancelled after we integrate over the zero modes $\psi_0,\bar\psi_0$.
Indeed, evaluating the action at the localisation locus
\bea S_{\rm susy}\big|_0=\int_M\,R(\bar\eta_0,\eta_0,\bar\psi_0,\psi_0),\nn\eea
where there are actually $g$-copies of $\psi_0$, $\bar\psi_0$. Choosing a basis for $\bar\psi_0,\psi_0$ and integrating, we produce exactly $g$ copies of $\det R$
\bea Z(S^1\times\Gs_g)&\sim& \int d\phi_0d\bar\phi_0d\psi_0d\bar\psi_0d\eta_0d\bar\eta_0e^{-S}\opn{Td}(X)\opn{ch}\big((\Go^{\sbullet,0})^{\otimes g}\big)\prod_i(x_i)^{-g}\label{step_here}\\
&\sim& \int d\phi_0d\bar\phi_0d\eta_0d\bar\eta_0\opn{Td}(X)\opn{ch}\big((\Go^{\sbullet,0})^{\otimes g}\big)\nn\\
&\sim&\int_X\opn{Td}(X)\opn{ch}\big((\Go^{\sbullet,0})^{\otimes g}\big).\nn\eea
It is inherently hard to keep track of the multiplicative factors in a path integral, and hence all the $\sim$ signs along the way.
However we consider this an encouraging result, since the Atiyah Singer index theorem gives
\bea Z(S^1\times\Gs_g,X)=\int_X\opn{Td}(X)\opn{ch}\big((\Go^{\sbullet,0})^{\otimes g}\big)=\dim_{\BB{C}}{\cal H}(\Gs,X)\label{AS_index_Z}.\eea
\begin{remark}
  We have mentioned in an earlier remark that the RW theory is a zero soliton truncation of the theory we are computing here. Let us hazard a guess that by including solitons, the corresponding cohomology group $H^{0,\sbullet}(X,(\Go^{\sbullet,0}_X)^{\otimes g})$ gets some $q$-correction (similar to the quantum cohomology group). But the dimension of the cohomology group remains the same, quantum or not, which probably explains why the zero soliton computation still gives the correct result \eqref{AS_index_Z}.
\end{remark}

Including Wilson loops is also easy. We will take the case $S^2\times S^1$ and insert two Wilson loops labelled with vector bundles $E_1,E_2$ along the circle, one over the north and one over the south pole.
Such insertion does not affect the localisation process until the step of \eqref{step_here}, where one needs to evaluate the $W(E_1)W(E_2)e^{-S}$ altogether at the localisation locus and integrate over the zero modes.
The evaluation of Wilson-loop is easy: one truncates to constant maps
\bea W(E)=\Tr{\bf P}\exp\int_0^{2\pi}d\gt (-\dot\phi^iA_i+F_{\bar ii}\bar\eta^{\bar i}\eta^i)
\stackrel{\rm trunc}{\to} \Tr\exp \big(2\pi F_{\bar ii}\bar\eta_0^{\bar i}\eta_0^i\big)\stackrel{\eqref{ad_hoc}}{=}
\Tr\exp \big(\frac{i}{2\pi}F_{i\bar i}dz^i\wedge dz^{\bar i}\big)=\opn{ch}(E),\nn\eea
where ch denotes its Chern character and in the last step we have used the same normalisation \eqref{ad_hoc} as when we identified the Chern roots earlier. Following the same steps after \eqref{step_here} we get
\bea Z(S^1\times S^2,W(E_1),W(E_2))=\int_X\opn{Td}(X)\opn{ch}(E_1)\opn{ch}(E_2).\label{inner_prod}\eea
We observe that with parallel Wilson loops labelled by $E_1,E_2$, their contribution to the path integral is $\opn{ch}(E_1)\opn{ch}(E_2)=\opn{ch}(E_1\otimes E_2)$.
This suggests that, at least for such special Wilson-loops, the fusion remains the tensor product, devoid of further corrections.

\subsection{$M=S^3$}
The computation itself is not that different from the last example. We can fast forward to the step \eqref{step_det_factor}, up until this point no change has taken place.
The modes of $\uppsi$ and $\upphi$ are related by the $\gdh$ operator as before, but now $\gdh$ is the Dolbeault operator transverse to the vector field $\reeb=\partial_{\gt}$, which we call $\bar\partial_H$. We have therefore
\bea V^{\sbullet}=H^{0,\sbullet}_{\bar\partial_H}(M)\otimes T^{1,0}_{\phi_0}X.\nn\eea

We compute the $\bar\partial_H$-cohomology for $S^3:~S^1\to S^3\to S^2$ first. Since $\gt$ is a good coordinate and $\bar\partial_H$ can be identified with $\bar\partial$ on the base $S^2$.
Given a Fourier mode $-n$ along the $S^1$-fibre, the $\bar\partial_H$-cohomology equals $H^{\sbullet}_{\bar\partial}(S^2,{\cal O}(n))$. Thus for $n\geq 0$ we get $\dim H^0_{\bar\partial_H}=n+1$, $H^1_{\bar\partial_H}=0$; while for $n<-1$ we have $\dim H^0_{\bar\partial_H}=0$, $\dim H^1_{\bar\partial_H}=-1-n$. These modes contribute to \eqref{step_det_factor}
\bea \eqref{step_det_factor}={\det}_{T^{1,0}X}\frac{\prod_{n\geq1}(in+R)^{n-1}}{\prod_{n\geq1}(-in+R)^{n+1}}.\label{det_S^3}\eea
As before we will deal with the determinant over $T^{1,0}X$ by writing $R$ in terms of the Chern roots
\bea \eqref{step_det_factor}=\prod_i\frac{\prod_{n\geq1}(in+x_i/2\pi)^{n-1}}{\prod_{n\geq1}(-in+x_i/2\pi)^{n+1}}.\nn\eea
The last ratio is related to the double-sine function \cite{Multiplesinefunctions} that appeared also in localisation of CS theory \cite{KaWiYa}
\bea S_2(x|\go_1,\go_2)=\frac{\prod_{a,b\in\BB{Z}_{\geq0}^2}(x+a\go_1+b\go_2)}{\prod_{a,b\in\BB{Z}_{\geq1}^2}(-x+a\go_1+b\go_2)}.\label{S_2}\eea
We observe that by setting $\go_1=\go_2$ and $a+b=n$, we will have the multiplicity $n+1$ or $n-1$ as in the previous formula.
With this we can write
\bea \eqref{step_det_factor}=\prod_i\frac{ix_i}{S_2(ix_i|2\pi,2\pi)}.\nn\eea
The $S_2$ function enjoys another reciprocity property
\bea S_2(x|\go_1,\go_2)S_2(-x|\go_1,\go_2)=-4\sin\frac{\pi x}{\go_1}\sin\frac{\pi x}{\go_2}.\nn\eea
From this and \eqref{eigen_val_paired} we get
\bea \eqref{step_det_factor}=\prod_i\frac{x_i/2}{\sinh(x_i/2)}=\prod_i\frac{x_i}{1-e^{-x_i}}=\opn{Td}(X).\nn\eea

What remains is to evaluate the action and possibly the Wilson loops at the localisation locus and integrate over zero modes. Note in contrast to \eqref{step_here} now we have no $\psi$ zero modes due to the new geometry, so the curvature term in \eqref{susy_action} does not survive and the only term left is $-4i\pi^2(-1)\bra\bar\eta_0,\eta_0\ket$ (using \eqref{cameron}). Upon taking into account the normalisation of the zero modes, it turns into the K\"ahler form
\bea 4i\pi^2\bra\bar\eta_0,\eta_0\ket=-i\go.\nn\eea
Thus the path integral gives
\bea Z(S^3,X)= \int d\phi_0d\bar\phi_0d\eta_0d\bar\eta_0\;e^{4i\pi^2\bra \bar \eta_0,\eta_0\ket}\opn{Td}(X)=\int_X\opn{Td}(X)e^{-i\go}.\label{Z_S3}\eea
If we have inserted Wilson loops, then we merely insert in the above integral the Chern-character of the relevant vector bundle, just as in the $\Gs\times S^1$ case
\bea Z(S^3,W(E_1),W(E_2))=\int_X\opn{Td}(X)\opn{ch}(E_1)\opn{ch}(E_2)e^{-i\go}.\label{inner_prod_S^3}\eea
\begin{example}
  If we take $X$ as the K3 surface, then \eqref{Z_S3} can be expanded as
  \bea Z(S^3,X)=\int_{K3}(\frac{c_2}{12}-\go^2)=\int_{K3}(-\frac{p_1}{24}-\go^2)=-\frac{\gs(K3)}{8}-2\opn{Vol}(K3).\nn\eea
  Here $\gs(K3)=-16$ is the signature of K3 and Vol is the volume measured with the K\"ahler class $\go^2/2$. If there were Wilson lines with hyper-holomorphic vector bundle $E$, we insert $\opn{ch}(E)$ into \eqref{Z_S3}. But since by definition $c_1(E)$ is (1,1) w.r.t all three complex structures, it is anti-self-dual and hence orthogonal to $\go$. This means that only $\opn{ch}_2(E)$ would contribute to the integral:
  \bea Z(S^3,X,E)=\opn{ch}_2(E)+2-2\opn{Vol}(K3).\label{Z_S3_K3}\eea

  We compare this result with the explicit one in \cite{Rozansky:1996bq} from combining Feynman diagram computation with Hilbert space arguments.
  If $X$ were the Atiyah-Hitchin manifold $X_{AH}$, then the RW model is the twisted version of the low energy effective theory of an $N=4$ SYM. Its partition function $Z(M,X_{AH})$ would then be the Casson-Walker invariant of $M$. In sec 4 of \cite{Rozansky:1996bq} it was surmised that such low energy effective theory description automatically assigns $M$ the canonical 2-framing.

  How is this relevant for the computation on K3? For any 4-dimension target, the partition function comes from only one 2-loop Feynman diagram, the so called $\theta$-diagram (with two vertices and three propagators in between). Therefore $Z(S^3,K3)$ and $Z(S^3,X_{AH})$ must be proportional, but the latter gives the Casson Walker invariant of $S^3$ which is zero!?

  Actually everything matches. The subtlety comes from precisely the 2-framing. So which framing did we use when localising the RW theory? We observe that the Hopf fibre structure was crucial in writing down the susy complex \eqref{second_susy_I}, the framing associated to the Hopf fibration corresponds to trivialising $TS^3$ with right invariant vector fields\footnote{this depends on convention, we have chosen $\partial_{\gt}$ to be the right multiplication on $SU(2)$ by $U(1)$. A opposite choice would flip the right to left invariant vector field.}
  The canonical 2-framing for $S^3$ trivialises $TS^3\oplus TS^3$ using left/right invariant vector fields for the first/second summand \cite{KirbyMelvin}. The framing we used trivialises both summands with right invariant vector fields and is therefore 2 units off the canonical 2-framing. According to Eq.4.18 of \cite{Rozansky:1996bq}, the partition function is shifted by $\Gd\opn{Fr}=2$. This explains the summand 2 in \eqref{Z_S3_K3}, though we have not been careful with the precise sign.

  What about the $\opn{Vol}(K3)$ term? One can also compute the partition function with a Hilbert space argument. Clearly $S^3$ can be glued from two $B^3$ along their common boundary $S^2$.
  The Hilbert space associated with $S^2$ is 2-dimensional $H^{0,0}(K3)\oplus H^{0,2}(K3)$, we name the bases $|\Psi^0\ket$ and $|\Psi^2\ket$ respectively. Their inner product is worked out in sec.5.1 \cite{Rozansky:1996bq}: $\bra\Psi^0|\Psi^2\ket=-\bra\Psi^2|\Psi^0\ket=1$ and zero otherwise. From these matrix elements $Z(S^3)=\bra \Psi^0|\Psi^0\ket=0$ in accordance with the argument using Casson-Walker invariant above.

  But one can insert local observables ${\cal O}_{\eta}(x_0)$
  \bea {\cal O}_{\eta}:=\frac{1}{2\opn{Vol}(K3)}\bar{\Go}_{\bar i\bar j}\bar v^{\bar i}\bar v^{\bar j}\nn\eea
  with the matrix element
  \bea \bra \Psi^0|{\cal O}_{\eta}|\Psi^0\ket=1\label{inner_prod_Go}\eea
  independent of the point of insertion $x_0$. Here the normalisation $1/\opn{Vol}(K3)$ comes from Eq.5.18 of \cite{Rozansky:1996bq}.

  Recall that the susy action used for localisation is deformed from the original RW action with the term \eqref{L_3}, which is
  \bea \eqref{L_3}\sim \opn{Vol}(K3)\int_{S^3}{\cal O}_{\eta}.\nn\eea
  This term can be expanded down from the exponent only once and would result in the matrix element of the form \eqref{inner_prod_Go}, but now proportional to the volume of K3.
  This extra matrix element corrects the original RW partition function to \eqref{Z_S3_K3}.
  Again we have not kept track of the precise numerical factors.
\end{example}

\subsection{$M=L(p,q)$}\label{sec_M-Lpq}
Start with the simpler $L(p,1)$ case since it can be realised as the total space of the ${\cal O}(\pm p)$ bundle over $\Gs_0=S^2$. We pick $-p$ here.
As there is no $\bar\psi_z,\,\psi_{\bar z}$ zero modes, the action \eqref{susy_action} truncated to the localisation locus is simply $e^{-ip\go}$ i.e. $p$ times the $S^3$ result.
The corresponding determinant \eqref{det_S^3} changes to
\bea {\det}_{T^{1,0}X}\frac{\prod_{n\geq1}(in+R)^{np-1}}{\prod_{n\geq1}(-in+R)^{np+1}}=\prod_i\frac{\prod_{n\geq1}(in+x_i/2\pi)^{np-1}}{\prod_{n\geq1}(-in+x_i/2\pi)^{np+1}}\nn.\eea
where the multiplicity $np\pm 1$ is the dimension of $H^0_{\bar\partial}(S^2,{\cal O}(np))$ and $H^1_{\bar\partial}(S^2,{\cal O}(-np))$ respectively.
We will deal with such infinite products even for $q\neq 1$ shortly, but for the current case we can take a shortcut by combining the product with $x_i$ and $-x_i$, causing big cancellation and get the same result as before
\bea \prod_i\frac{x_i}{1-e^{-x_i}}=\opn{Td}(X).\nn\eea
So our result for $L(p,1)$ is
\bea Z(L(p,1),X)=\int_X\opn{Td}(X)e^{-ip\go}\label{Z_L(p1)}\eea
where adding Wilson loops only inserts the Chern character as before.
\begin{remark}
Comparing the $e^{-ip\go}$ factor with $e^{-i\go}$ from the $S^3$ result \eqref{Z_S3}, it is tempting to write both as $e^{-3i\go+i(3-p)\go}$ (take $p=1$ for $S^3$).
The point is that the number $3-p$ is the difference between the canonical 2-framing of $L(p,1)$ and the one coming from the Hopf bundle structure, see ex.\ref{ex_framing_lens}.
One might therefore attempt to interpret
\bea \int_X\opn{Td}(X)e^{-3i\go}\label{Z_S3_framing}\eea
as the RW partition function computed for the canonical framing. However this does not appear possible, since computing the integral would give us $2-18\opn{Vol}(K3)$ instead of \eqref{Z_S3_K3}, and there is no reason for this to be zero.

We understand the $p\go$ factor as coming from the Hopf fibration, or more precisely, from the transverse holomorphic structure of $M$. But this feature makes the partition function very unfriendly to cutting, gluing or surgeries. In sec.\ref{sec_TcFMkotSa} we will attempt to construct the partition function from a Hilbert space perspective and we must somehow realise the $p\go$ term purely from wave functions and gluing matrices. Our attempt seems to work for $L(p,1)$, but would fail for the $L(p,q)$ space.
\end{remark}

Now we turn to $L(p,q)$, in view of the boring result \eqref{Z_L(p1)} one would hope that by going to the $L(p,q)$ for $q\neq 1$ one would get some more interesting results, but it turns out not the case.
To begin, since the vector field $\partial_{\gt}$, the connection $\FR{a}$ and contact 1-form $\gk$ on $L(p,q)$ are all inherited from $S^3$ through the $\BB{Z}_p$ quotient, we can continue to use these structures to write the susy rules \eqref{second_susy_I}. They clearly remain valid even when we do not have a fibration over a smooth surface, since what really matters is the complex structure transverse to $\partial_{\gt}$. But we will scale the vector field and contact form as
\bea \partial_{\gt}\to \frac{1}{p}\partial_{\gt},~~\gk\to p\gk\label{rescale}\eea
so that when taking $q=1$, the rescaled $\gt$ is the angle coordinate of the $S^1$ fibre.
For example, we will try to evaluate the susy action at the localisation locus, in particular the last term of \eqref{susy_action}. The 3-form $\gk\wedge d\gk$ is rescaled to $p^2\gk\wedge d\gk$ and integrated over $(1/p)^{\rm th}$ of $S^3$, thus giving $p$ times the $S^3$ result $e^{-ip\go}$, coinciding with $L(p,1)$.

Next we need to compute the $\bar\partial_H$ cohomology in order to get \eqref{step_det_factor}. We can mimic the case of $S^3$, that is, we reduce $\bar\partial_H$ to the $\bar\partial$ on the base orbifold, and for each Fourier mode $n$ we compute the corresponding orbifold cohomology. Since we have the orbifold index theorem, there is no essential difficulty.
However there is a neater way that reveals more structure than the orbifold approach.
We note that $\bar\partial_H$ is not elliptic since its symbol clearly has a kernel along the direction of $\partial_{\gt}$. We can still compute its cohomology equivariantly by involving the isometry of $L(p,q)$. Indeed, the Lens space has a $G=2U(1)$ isometry, and $\partial_{\gt}$ is an integer linear combination of the two $U(1)$'s. Transverse to the $G$-action, the symbol of $\bar\partial_H$ is elliptic. Thus one can use the index theorem of transversally elliptic operators \cite{Ellip_Ope_Cpct_Grp} (see also the appendix of \cite{exotic_instanton} for a quick review and a detailed sample calculation).

The result is very easy to describe. The zeroth cohomology $H^0_{\bar\partial_H}(L(p,q))$ can be enumerated as monomials $z_1^mz_2^n$ that is invariant under the $\BB{Z}_p$ action in \eqref{lens_space}. To be explicit
\bea H^0:~~ z_1^{ap-bq}z_2^b,~~b\geq0,~~ap-bq\geq0,\label{lens_p_q_hol}\eea
exhaust all such invariant monomials. As for $H^1_{\bar\partial_H}$ it is enumerated by the same formula with a small change
\bea H^1:~~ z_1^{ap-bq}z_2^b,~~b<0,~~ap-bq<0.\label{lens_p_q_H1}\eea
The perspicacious reader may have recognised such description as the enumeration of the \v{C}ech cohomology of the structure sheaf of a variety
\bea \check{H}^{\sbullet}(V,{\cal O}_V),~~~V=(\BB{C}^2\backslash\{0\})/\BB{Z}_p.\nn\eea
Note $V$ can be covered by two opens $U_1=\{z_1\neq 0\}$ or $U_2=\{z_2\neq 0\}$, then $H^1$ consists of those invariant monomials that are defined on the $U_1\cap U_2$ but not extendable to either $U_1$ or $U_2$. This explains the two inequalities in \eqref{lens_p_q_H1}.

\begin{figure}[h]
\begin{center}
\begin{tikzpicture}[scale=1]
\draw [step=0.3,thin,gray!40] (-2,-1.8) grid (2,1.8);
\draw [->,thick] (-2,0) -- (2,0) node [right] {\scriptsize$b$};
\draw [->,thick] (0,0) -- (0,2) node [left] {\scriptsize$a$};
\draw [-,dashed,thick] (0,-1.8) -- (0,0);

\draw [-,thick] (0,0) -- (1.8,1.2) node[right] {\scriptsize{$(p,q)$}};
\draw [-,dashed,thick] (0,0) -- (-1.8,-1.2);

\node at (.6,1) {I};
\node at (-.6,-1) {II};
\end{tikzpicture}\caption{The $H^{0,1}$ corresponds to the lattice points in region I resp. II. The dashed line means that the boundary lattice points are excluded}\label{fig_lattice_pq}
\end{center}
\end{figure}
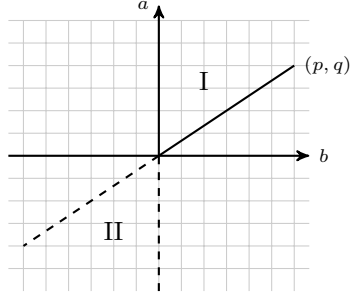

These monomials \eqref{lens_p_q_hol} \eqref{lens_p_q_H1} are the lattice points in fig.\ref{fig_lattice_pq}.
So the determinant factor \eqref{step_det_factor} can be encoded in a way similar to the $S^3$ case, by introducing a generalised double-sine function. Let $C$ be a rational cone based at the origin and $C^{\circ}$ the same cone but excluding the boundary, then
\bea S_2^C(x|\vec\go):=\frac{\prod_{\vec m\in C\cap\BB{Z}^2}(x+\vec m\cdotp\vec\go)}{\prod_{\vec m\in C^{\circ}\cap\BB{Z}^2}(-x+\vec m\cdotp\vec\go)}\label{S_2^C}\eea
taking $C=\BB{R}^2_{\geq0}$ gives back the original double-sine function \eqref{S_2}.
Here in order to be able to regulate the infinite products (typically $\zeta$-function regularisation), one needs to assume that $\vec\go\in C^{\vee}$ with $C^{\vee}$ being the dual cone $C^{\vee}=\{\vec\go\in\BB{R}^2|~\vec \go\cdotp\vec x>0,~\forall\;\vec x\in C\}$.
For a more in depth study of the properties of such generalised multiple-sine functions, see \cite{Winding:2016wpw}.
The determinantal factor is written as
\bea \eqref{step_det_factor}=\prod_i\frac{ix_i}{S^C_2(ix_i|2\pi(1-q)/p,2\pi)},\nn\eea
where the expression for $\vec\go$ comes from the weights of the action of the vector field $\partial_{\gt}$. Indeed $\partial_{\gt}=iz_1\partial_{z_1}+iz_2\partial_{z_2}+c.c.$ acts on a monomial $z_1^{ap-bq}z_2^b$ with weight $ap-b(q-1)$. But remembering the rescaling \eqref{rescale} we divide further by $p$.

To finish the computation, we have discussed what happens when we evaluate the action \eqref{susy_action} at the localisation locus, so the total result is
\bea Z(L(p,q),X)=\int_X\prod_i\frac{ix_i}{S^C_2(ix_i|2\pi(1-q)/p,2\pi)}e^{-ip\go}.\label{Z_Lpq_raw}\eea
If there are Wilson loops present, one just inserts the Chern character into the integral.

If we use the relation of the Chern roots \eqref{eigen_val_paired}, the $S_2^C$ function also collapses to the usual sine function.
What happens is that the product over the interior of the cone $C$ cancels between $S_2^C(x_i)$ and $S^C_2(-x_i)$ leaving only the product over the two bounding rays.
\bea Z(L(p,q),X)=\int_X\prod_{i=1}^{2n}\frac{x_i/2}{\sinh(x_i/2)}e^{-ip\go}=\int_X\prod_{i=1}^{2n}\frac{x_i}{1-e^{x_i}}e^{-ip\go}.\label{Z_Lpq_simple}\eea
The result is disappointingly equal to the $L(p,1)$ result \eqref{Z_L(p1)}, in particular, it does not depend on $q$, meaning the partition function cannot distinguish $L(p,q)$ for different $q$.

\smallskip

\noindent {\bf The equivariant extension ... again}

The culprit for the uninteresting result is of course the HyperK\"ahler property \eqref{eigen_val_paired}. This is perhaps one more reason that one must try to involve equivariance to $X$ (for $X$ non-compact necessarily). If one only introduces a tri-holomorphic $U(1)$ action on $X$, then since the action preserves $\Go$, the equivariant curvature will again have Chern-roots with the same property as \eqref{eigen_val_paired}. No big help here. However if one turns on a $U(1)$ that acts holomorphically i.e. it only preserves $I$ and $g$, then the equivariant curvature will no longer have its Chern-roots paired up. We can hope that the equivariant version of the partition function \eqref{Z_Lpq_raw} would depend non-trivially on $q$ and be a finer invariant.

%
%

\smallskip

The rest of the discussion in the section will lead to an expected outcome: with equivariance incorporated, one needs only take the non-equivariant partition function, replace all cohomology classes e.g. Todd, Chern and K\"ahler class with their equivariant version. The sole purpose of the discussion is to address the issue of Cartan v.s. Weil model, which the reader may safely skip.

It should be clear what the necessary changes are in order to finish the calculation. From the new closure property of the susy algebra \eqref{susy_closure_linear_equiv_I}, we should plug into \eqref{step_det_factor} the equivariant curvature
\bea (R_{eq})^i_{~j}=R(\bar\zeta,\zeta,{}^i,{}_j)-\varrho X^i_{~j}.\nn\eea
The reader might be alarmed by the apparent contradiction: there is no ghosts in the action \eqref{complete_action_equiv}, so there should be no ghost in the final answer. But the determinant of $R_{eq}$ will contain the ghosts hidden in $\zeta$?
This is resolved by our preemptive remark toward the end of
sec.\ref{sec_TEea}: provided the above $R_{eq}$ only appears in ad-invariant polynomials, then the ghosts can be removed at the cost of some $\gd_{eq}$ exact terms, which vanish upon integration.

Next we evaluate the action on the zero modes
\bea S_{\rm susy\,eq}\big|_0=\int_M 4\bra \bar\psi_0,R_{eq}\psi_0\ket+\big(\bra\bar\eta_0,\rho^aX_a\eta_0\ket-\frac12||\rho^aV_a||^2\big)-\gk\wedge d\gk \go_{eq}\to \int_M
4\bra \bar\psi_0,R_{eq}\psi_0\ket-\gk\wedge d\gk \go_{eq}.\nn\eea
The middle term is discarded since it is exact.
Again the argument at the end of sec.\ref{sec_TEea} says that we can treat the current $R_{eq}$ the same as the one above, even though one is in the Cartan model, the other in the Weil-model.

In short if we are to compute, say, the $L(p,q)$, we just take
\eqref{Z_Lpq_raw} and stick the subscript ${}_{\rm eq}$ to $R$ and $\go$.
But now the $S_2^C$ function will no longer collapse into \eqref{Z_Lpq_simple}. To compute this $S_2^C$ with $C$ a $q/p$ rational cone, we need to subdivide the cone by adding more rays such that the neighbouring two rays form an $SL(2,\BB{Z})$ basis.
Incidentally, for obtaining the framing of the lens space, we also need to decompose a rational $q/p$ surgery into a series of integer surgeries. The two procedures i.e. subdivision of a rational cone and decomposition of a rational surgery involve the same technique. We will do so in sec.\ref{sec_RCaRS}.

\section{Hilbert space and $SL(2,\BB{Z})$ action}\label{sec_some_sec}
Here we first interpret our explicit results of the last section, especially from the perspective of Hilbert space and surgery. We do so in the non-equivariant settings, to highlight the problems that will be circumvented when equivariance is added.

\subsection{$S^2\times S^1$}\label{sec_StS}
Consider the special case $S^2\times S^1$, we have seen that the circle in the geometry means that the partition function is a super-trace over the Hilbert space associated to $S^2$, which is ${\cal H}(S^2,X)=H^{0,\sbullet}(X)$. The super-trace is therefore the Euler character $\sum_i(-1)^i\dim H^{0,i}(X)$.
With the insertion of two Wilson loops along $S^1$ but at two arbitrary points of $S^2$, we have worked out in \eqref{inner_prod} that
\bea Z(S^1\times S^2,W(E_1),W(E_2))=\opn{sTr}{\cal H}(S^2,E_1,E_2)=\chi(E_1\otimes E_2).\eea
From this we see that the Hilbert space associated to $S^2$ with two punctures labelled with $E_1,E_2$ has dimension
\bea \dim{\cal H}(S^2,E_1,E_2)=\chi(E_1\otimes E_2).\nn\eea

These statements should be contrasted with the CS theory case.
First ${\cal H}_{CS}(S^2)$ is 1-dimensional, which is responsible for the very simple connected sum formula
\bea \frac{Z_{CS}(M_1\#M_2)}{Z_{CS}(S^3)}=\frac{Z_{CS}(M_1)}{Z_{CS}(S^3)}\frac{Z_{CS}(M_2)}{Z_{CS}(S^3)}.\label{connected_sum}\eea
The Hilbert space ${\cal H}_{RW}$ is not dimensional, e.g. for $X=K3$ its dimension is 2.
The corresponding connected sum formula is much more involved, they are worked out in \cite{Rozansky:1996bq} for $X=K3$ or $X_{AH}$.

Secondly, one can also compute ${\cal H}_{CS}$ associated to $S^2$ with two punctures labelled with two irreducible representations $R_1,R_2$.
Then
\bea \dim{\cal H}_{CS}(S^2,R_1,R_2)=\bigg\{
                                      \begin{array}{cc}
                                        1 & R_1=R_2^{\vee} \\
                                        0 & {\rm otherwise} \\
                                      \end{array}.\label{used_II}\eea
This simple result was crucial for the derivation of the Verlinde formula (see sec.4 of \cite{witten1989} where the derivation was explained admirably).
In the RW setting we will label the punctures with objects in $D^b(X)$, there is no longer the type of orthogonality like \eqref{used_II}.
In fact $\chi(E_1\otimes E_2)$ can certainly be nonzero when $E_1\nsim E_2^{\vee}$. When we go to the equivariant case later, we will again be able to find orthogonal bases.

Alternatively one can also view $S^2\times S^1$ as being glued from two copies of $D^2\times S^1$, where the discs are the northern and southern hemisphere of $S^2$. Associated to the boundary $\partial(D^2\times S^1)=T^2$ there is the Hilbert space
\bea {\cal H}(T^2,X)=H^{\sbullet}(X,\Go^{\sbullet,0})=H^{\sbullet,\sbullet}(X).\nn\eea
We can insert a Wilson loop $W(E)$ along the core $\{0\}\times S^1\in D^2\times S^1$, then performing the path integral on the solid torus produces a vector $\Psi\in {\cal H}$, depending on the choice of $E$ (we put aside for now the framing of the Wilson loop). From this point of view the partition function on $S^2\times S^1$ with Wilson loops inserted is an inner product
\bea Z(S^2\times S^1,W(E_1),W(E^{\vee}_2))=\bra \Psi_2|\Psi_1\ket\nn\eea
where we take $W(E_1)$ (resp. $W(E^{\vee}_2)$) to be inserted along the circle over the north (resp. south) pole of $S^2$. In fact from the explicit formula \eqref{inner_prod}, the inner product is \bea \bra \Psi_1|\Psi_2\ket=\chi(E_2,E_1).\label{inner_prod_I}\eea
We should again contrast this with the CS theory case, where the character $\opn{ch}(R_i)$ of the representations $R_i$ labelling the Wilson loop are naturally orthogonal to each other for $R_1\nsim R_2^{\vee}$. This fails in the category $D^b(X)$.

\subsection{$S^3$ or $L(p,q)$}
It is well-known that $S^3$ can also be glued together from two solid tori, except the a-cycle of one solid torus is glued to the b-cycle of the other and vice versa.
So the partition function has a tentative interpretation as a matrix element
\bea \bra \Psi_1|S|\Psi_2\ket\label{used_I}\eea
where $S$ is the operator that acts on ${\cal H}$ effecting the exchange of a-, b-cycles.
But now we must take into account something we glossed over in previous discussions: the framing of the Wilson loop. This means we push the Wilson loop slightly off the core of the solid torus and specify the combination of a-,b-cycles traversed by the push-off. As a result, the vector $|\Psi(E)\ket\in {\cal H}(T^2,X)$, obtained via the path integral over the solid torus with $W(E)$ inserted, retains not just the information of the Chern-character of $E$, but also the framing.
To emphasise such framing dependence we will write $|\Psi(E,f)\ket$ with $f$ being the framing data.

If we do the path integral via perturbation theory, such a push-off is certainly necessary. Indeed, there can be propagators with two ends on the Wilson loop, in order to regulate the short distance singularity when the two ends collide, one needs to push one end of the propagator off the loop slightly and this push-off gives the framing dependence. We also see in sec.\ref{sec_HsWlar} that when we quantise the theory on $T^2$ we need to choose a polarisation. Concretely, denoting $\chi_a,\chi_b$ the component of $\chi$ along the a,b-cycle of $T^2$, the integer linear combinations $\chi_a$ and $n\chi_a+\chi_b$ will be declared coordinate and momentum respectively. The integer $n$ is the framing of the Wilson loop.

Now that we have computed the RW theory by formulating it as a susy $\sigma$-model, the framing of the Wilson loop is forced. This is because we must insert the Wilson loop along the $\partial_{\gt}$-orbit to preserve supersymmetry. In fact we saw that in our susy $\sigma$-model formulation, we treated the $\gt$-component of the $\chi$ field differently from its other two components \footnote{to head off a potential confusion, the $\chi_{\gt}$ should not be compared to $\chi_t$ in sec.\ref{sec_HsWlar}.}. Moreover the entire Hopf fibration structure was crucial for the reformulation, not only the framing of the Wilson loop, but the framing of the 3-manifold itself is forced.

Now we reconsider \eqref{used_I} and try to get the precise $SL(2,\BB{Z})$ action that should be inserted in the matrix element.
We choose the generators of $SL(2,\BB{Z})$ as
\bea S=\left[
         \begin{array}{cc}
           0 & -1 \\
           1 & 0 \\
         \end{array}\right],~~~T=\left[
         \begin{array}{cc}
           1 & 1 \\
           0 & 1 \\
         \end{array}\right].\nn\eea
The first column of the matrices give the new meridian while the second column is the new longitude.
We see that $T$ keeps the meridian but only adds to the longitude a copy of the meridian, In particular $T$ does not change the topological type of the resulting 3-manifold, only the framing.
We fix the trivialisation of $TS^3$ using the Hopf-bundle structure $S^1\to S^3\to S^2$. That is, the two solid tori to be glued are the northern (resp. southern) hemisphere times the circle fibre. The angle coordinate of the Fubini-Study coordinates for $S^2$ parameterises the $a$-cycle while the Hopf fibre is the b-cycle. Working out the coordinate change from north to south indicates the correct $SL(2,\BB{Z})$ matrix:
\bea S^3:~~\left[
   \begin{array}{c}
     a_{\rm old} \\
     b_{\rm old} \\
   \end{array}\right]= \left[
   \begin{array}{cc}
     1 & 0 \\
     1 & 1 \\
   \end{array}\right]\left[
   \begin{array}{cc}
     a_{\rm new}  \\
     b_{\rm new}  \\
   \end{array}\right].\label{M_S3}\eea
The matrix above is decomposed as $TST$.
More generally for $L(p,q)$, we fix $r,s$ so that $pr-qs=1$ and then the above gluing matrix becomes
\bea L(p,q):~~\left[
   \begin{array}{c}
     a_{\rm old} \\
     b_{\rm old} \\
   \end{array}\right]= \left[
   \begin{array}{cc}
     q & -r \\
     p & -s \\
   \end{array}\right]\left[
   \begin{array}{cc}
     a_{\rm new}  \\
     b_{\rm new}  \\
   \end{array}\right].\label{M_Lpq}\eea
The choices for $r,s$ are not unique, but they only affect the framing of the manifold. The reader can find more details about such surgeries in sec.\ref{sec_RCaRS}.
To summarise, for $M=S^3$ or $L(p,q)$, the partition function, with two Wilson loops inserted along the core of each solid tori, is the matrix element
\bea Z(M, E_1,E_2^{\vee})=\bra \Psi_2(E_2)|A|\Psi_1(E_1)\ket\nn\eea
with $A$ being either \eqref{M_S3} or \eqref{M_Lpq} and the framing of the Wilson loop implicit from the Hopf-fibration structure.

\subsection{The cohomological Fourier-Mukai kernel of the $SL(2,\BB{Z})$ action}\label{sec_TcFMkotSa}
We have seen that the $SL(2,\BB{Z})$ acts on the cohomology group $H^{\sbullet,\sbullet}(X)$. But the latter is also the Hochschild homology of $D^b(X)$ and so is a very basic invariant of $D^b(X)$. We go out on a limb and say that the $SL(2,\BB{Z})$ action actually descends from its action on $D^b(X)$.
As $SL(2,\BB{Z})$ is group, such an action is an auto-equivalence. It was shown by Orlov \cite{Orlov_K3} that for projective varieties, such auto-equivalences can always be realised as a {\bf Fourier-Mukai} transform. Briefly, consider the product $X\times X$ and let $p_{1,2}$ be the projection to the two factors. Let $F$ be an object of $D^b(X\times X)$, define
$\Phi_{F}(-):\,D^b(X)\to D^b(X)$ to act as
\bea \Phi_F(-)=\BB{R}p_{2*}(F\otimes^{\BB{L}}p_1^*(-)).\nn\eea
Here the object $F$ is called the kernel of the Fourier-Mukai transform.

It is easier to study how the Fourier-Mukai transform acts on the Chern-character of the objects of $D^b(X)$. Let $v([E])$ be the Mukai-vector of $E\in\opn{obj}(D^b(X))$ defined as $v([E])=\opn{ch}(E)\sqrt{\opn{Td}(X)}$, then $\Phi_F$ acts on $v([E])$ as
\bea \Phi_{F}(-):~ v([E])\to p_{2*}(v([F])p_1^*v([E])).\nn\eea
Here the class $v([F])$ or simply denoted $v(F)$ is the cohomological Fourier-Mukai kernel and is unique. The reader may consult the survey \cite{FM_survey} for details.

We cannot hope to get what is the kernel $F$, but from the explicit computations of the matrix elements, and some reverse engineering, we put forward a proposal of the cohomological Fourier-Mukai kernel associated to the $SL(2,\BB{Z})$ action. Needless to say, our proposal below is pure unsubstantiated speculation.

Looking back at \eqref{Z_Lpq_raw}, the $S^C_2$ function should correspond to the action of the matrix in \eqref{M_Lpq}, but where are the integers $r,s$ in $S_2^C$? Recall that the $S^C_2$ function \eqref{S_2^C} is a product over points inside a rational $q/p$ cone. To take the product, we need to subdivide the cone by adding more rays. This will be done in sec.\ref{sec_RCaRS}, and the process turns out equivalent to decomposing a $q/p$ surgery into a product of integer surgeries
\bea A=\left[
   \begin{array}{cc}
     q & -r \\
     p & -s \\
   \end{array}\right]=T^{k_0}ST^{k_1}S\cdots T^{k_t}S\label{factor_A}\eea
where the set of integers $k_i$ determine $r,s$.
The product over the lattice points between the $i^{th}$ and $(i+1)^{th}$ ray will correspond to the factor $T^{k_i}S$ in \eqref{factor_A} and
give the usual double-sine function (upon a change of basis). Therefore it seems natural to identity the resulting double-sine function as the cohomological kernel for the action of $T^{k_i}S$. In view of this, we define a new function called $T^k\kern-0.2em S(x|\vec\nu)$, depending on the Chern-roots $x$ of $TX$ and also a 2-vector $\vec \nu$ that encodes the orbit of $\partial_{\gt}$ as a linear combination of the a- and b-cycle on $T^2$.
The same parameter used to be called $\vec\go$ in sec.\ref{sec_M-Lpq}, but we switched notation midway in deference to the K\"ahler form.
This orbit is important as it fixes the framing of the Wilson loops. The function reads
\bea T^k\kern-0.2em S(x|\vec \nu)=\frac{ix}{S_2(\frac{ix}{2\pi}|\vec \nu(T^kS)\vec e_2,\vec \nu\cdotp\vec e_2)}\big(\frac{2}{x}\sinh \frac{x}{2\vec \nu(T^kS)\vec e_2}
\cdotp\frac{2}{x}\sinh \frac{x}{2\vec \nu\cdotp\vec e_2}\big)^{1/2}\label{T^kS_main}\eea
where $\vec e_1=[1;0],\;\vec e_2=[0;1]$.
The idea is that the $S_2$ function accounts for the product over lattice points inside a segment of the $q/p$ rational cone, while the two hanging $\sinh$ factors deal with under/over counting of lattice points on the boundary of the segment. We point out that in the non-equivariant limit the entire $T^k\kern-0.2em S$ function collapses to 1.
The product of two matrices $T^{k_1}S$ and $T^{k_2}S$ simply corresponds to juxtaposing their respective segments of the cone, see fig.\ref{fig_sub_div} and \eqref{composing T^kS} for the pictures.

The framing data also enters the assignment of the wave function. If a Wilson loop labelled with vector bundle $E$ has framing $\vec\nu$, then we associate the wave function
\bea |\Psi(E,\vec\nu)\ket \sim \big(\frac{2}{x}\sinh\frac{x}{2\nu_2}\big)^{-1/2}\opn{ch}(E)e^{i\nu_1\go}.\label{wave_function}\eea
In the special case $\vec\nu=[0,1]$ and the Chern-roots sum to zero, the above expression simplifies to $\sqrt{\opn{Td}_X}\opn{ch}(E)$, i.e. the Mukai vector of $E$.
The idea of the framing factor $e^{i\nu_1\go}$ is that the Wilson loop would go round the meridian $\nu_1$ times as it goes along the longitude once.
Matching this framing factor, we propose that the $T$-matrix should act simply as
\bea T\sim e^{i\go}\nn\eea
with the replacement $\go\to\go_{eq}$ in the equivariant setting.

Now we can assemble the wave functions and the kernel of the $SL(2,\BB{C})$ actions into a partition function.
For the $L(p,q)$ case, suppose we have chosen a sub-division of the $q/p$ rational cone, giving rise to the factorisation of \eqref{M_Lpq} as $A=\prod_{i=0}^tT^{k_i}S$.
We show in \eqref{weight_theta} that the Wilson loops $E_1,E_2$ inserted above the north and south pole have framing 2-vector $\vec\nu_0=[(1-q)/p,1]$ and $\vec\nu_{t+1}=[(1+s)/p,1]$.
The appearance of fraction is due to the rescaling \eqref{rescale} and purely conventional.
For the partition function
\bea Z(L(p,q),E_1,E_2)=\big\bra \Psi(E_1,\vec\nu_0)\big|\prod_{i=0}^te^{i(-3+k_i)\go_{eq}}T^{k_i}S \big|\Psi(E_2,\vec\nu_{t+1})\big\ket\label{used_IX}\eea
where we attached $e^{i(-3+k)\go_{eq}}$ to each $T^kS$ to correct the framing. We provide more detail about the framing correction in appendix \ref{sec_RCaRS}.
The action of $T^kS$ is then via the multiplication with \eqref{T^kS_main}, the product of these factors and the wave function \eqref{wave_function} reproduce by design the $S_2^C$ function
\bea Z(L(p,q),E_1,E_2)=\int_X\prod_i\frac{ix_i}{S^C_2(ix_i/(2\pi)|(1-q)/p,1)}\opn{ch}_{eq}(E_1)\opn{ch}_{eq}(E_2)e^{i\opn{Fr}\go_{eq}}.\nn\eea
The result is independent of the choice of the sub-division. Indeed, the only ingredient that can go wrong is the framing term $i\opn{Fr}\go_{eq}$, so we collect them next.
We have $(1-q)/p$ and $(1+s)/p$ from the wave functions and a further
\bea \sum_{i=0}^t(-3+k_i)=\phi(-A)-3\label{used_VII}\eea
from all the $T^{k_i}S$ factors. Here $\phi$ is the Rademacher function defined on $SL(2,\BB{Z})$ matrices, whose definition is reviewed in the appendix.
For us
\bea \phi(\left[
            \begin{array}{cc}
              -q & r \\
              -p & s \\
            \end{array}\right])=\frac{s-q}{-p}-12s(q,p).\nn\eea
Combining this with the framing contribution from the wave functions
\bea \phi(-A)-3+\frac{2+s-q}{p}=-12s(q,p)-3+\frac{2}{p}.\label{used_VIII}\eea
We see that the choice of $r,s$ disappears. In the special example $L(p,1)$, we have $s(1,p)=p/12+1/(6p)-1/4$ and the sum above simplifies to $-p$. The total framing factor $e^{-ip\go}$ agrees with our earlier result \eqref{Z_L(p1)}.

There is a small caveat. The reason \eqref{used_VII} is true is that $\phi(-A)$ can also be computed from the so called linking matrix $L$ associated to the decomposition $A=\prod T^{k_i}S$. This is reviewed in \eqref{framing_shift}, but briefly $L$ is a matrix with diagonal $0,k_0,k_1,\cdots,k_t$ and 1 on the main off diagonal. The function $\phi(-A)$ equals the trace of $L$ minus 3 times its signature. In general one cannot read the signature directly from the sign of $k_i$'s, but with our assumption $q<p$ and that our continued fraction being what is known as the \emph{negative definite} continued fraction, the signature of $L$ is $t$ and so $\phi(-A)$ can be written as $-3t+\sum k_i$ and hence \eqref{used_VII}. Put differently, our choice of $k_i$ comes from sub-dividing a rational cone. But one could image a subdivision where certain added rays land outside the original cone, thus causing either the $S_2$ function or the square root term in \eqref{T^kS_main} to be ill-defined.

As one last example, we take a different surgery for $L(p,1)$. Indeed the Hopf-fibration structure of $L(p,1)$ dictates that $A=[1,0;p,1]=S^3T^{-p}S$. In this case the framing vector for the Wilson loops are $[(1-q)/p,1]=[0,1]$ and $[(1+s)/p,1]=[0,1]$ while $\phi(-A)=3-p$.
Now we get no framing contribution from the wave functions and $\phi(-A)-3$ gives again $-p$.

There is however a puzzle that we are not able to resolve. When we collect the framing factors from \eqref{used_IX} for $L(p,q)$, we get
\eqref{used_VIII}, while the localisation computation seems to suggest $-p$. It is possible that we have overlooked some other subtleties related to the framings, for example \eqref{used_IX} is actually proportional to certain $\eta$-invariant (see Eq.4.7 \cite{atiyah_patodi_singer_1975}) which is related to the framing and appeared in the perturbative Chern-Simons theory, see sec.2 \cite{witten1989}.
Let us also not forget that the exercises done in this section are entirely futile for the non-equivariant case: the kernels $T^k\kern-0.2em S(x|\vec \go)$ become 1 and the partition functions are given by the pairing of the Mukai vectors of the two Wilson loops inserted, making no distinction for $M=S^2\times S^1$ \eqref{inner_prod} or $M=S^3$
\eqref{inner_prod_S^3}.

\section{Tilting theory and Verlinde formula}\label{sec_TtaVf}
We have highlighted earlier the major deviations of RW theory from the CS theory: 1. $\dim{\cal H}(S^2)\neq 1$ losing us the simple connected sum formula, 2. no \emph{apparent} orthogonality in ${\cal H}(S^2,E_1,E_2)$ that was crucial for the derivation of the Verlinde formula. In this section, we change gear and, rather than working over $\BB{C}$, we shall consider HK varieties relative to another affine variety. This is the setting of \cite{TODA20101} which we follow, the advantage should become clear shortly.

\subsection{Tilting object and basis of Hilbert space}\label{ToaboHs}
Let $X$ be an HK variety and $X\to Y$ be a map to an affine variety. A typical choice for $Y$ is the spectrum of the algebra $H^0(X)$. For example let $X=T^*\BB{P}^1$. The geometry is the total space of ${\cal O}(-2)$ over $\BB{P}^1$, we denote with $A_0=H^0(X)$, which is well-known to be the function ring of the quotient $\BB{C}^2/\BB{Z}_2$ where $\BB{Z}_2$ acts as $-1$ on both coordinates of $\BB{C}^2$. The map $X\to Y$ then simply shrinks the base $\BB{P}^1$ to zero size.

Since the cohomology of any sheaf on $X$ is naturally a module over $A_0$, we will treat them as such (i.e. rather than as mere vector spaces).
Taking our example $X=T^*\BB{P}^1$, and $Y=\BB{C}^2/\BB{Z}_2$, then $X$ and $Y$ would 'agree at infinity', therefore working over $Y$ effectively compactifies $X$.
As perhaps a trivial remark, we saw earlier that the Hilbert space ${\cal H}(S^2,X)=H^{0,\sbullet}(X)$ is not 1-dimensional, but assuming that $H^{0,>0}(X)=0$, then ${\cal H}(S^2,X)$ is tautologically a rank 1 module over $A_0$. That ${\cal H}(S^2)$ is of rank 1 means that a version of the simple connected sum formula is again possible.

We assume that $D^b(X)$ possesses a tilting object $T=\oplus_iE^{\vee}_i$ with $E_i$ locally free i.e. vector bundles. Then we will use $E_i$ as a 'basis' for the Hilbert space ${\cal H}(T^2,X)$, in much the same way the integrable representations of the Lie algebra give a basis for the Hilbert space ${\cal H}_{CS}(T^2)$. The crucial difference is that we must work over $A_0$ rather than $\BB{C}$. We explain the procedure first and then illustrate with some examples.

The tilting bundle $T$ satisfies $\opn{Ext}^i(T,T)=0$ for $i>0$ and we let $\Gl=\opn{Ext}^0(T,T)$ be the endomorphism algebra of $T$.
Let $D^b(\Gl)$ denote the bounded derived category of modules over $\Gl$, then we have a pair of equivalences between $D^b(\Gl)$ and $D^b(X)$
\bea& \Psi:~~D^b(\Gl)\rightleftharpoons D^b(X):~~\Phi\label{equivalence}\\
&\Psi(-)=-\otimes_{\Gl}^{\BB{L}}T^{\vee},~~~\Phi(-)=\BB{R}\Gc(-\otimes^{\BB{L}}_{{\cal O}_X}T).\nn\eea
The composition $\Psi\Phi$ is isomorphic to the identity: $F\simeq \Psi\Phi(F)$ for any $F\in D^b(X)$. But as $T$ is a direct sum of vector bundles $E_i$, by looking at the rhs of $\Psi$, we see that $F\simeq \Psi\Phi(F)$ is now rewritten as a complex consisting of only the $E_i$'s. The key point here is often phrased as saying that $T\otimes^{\BB{L}}_{\Gl}T^{\vee}$ gives a resolution of the diagonal ${\cal O}_{\Gd}$, then applying $\Psi\Phi$ is like applying the Fourier-Mukai transform with ${\cal O}_{\Gd}$ as the kernel, but this is the identity transform.
At the level of Grothendieck group we have then simply
\bea [F]=\sum_i n_i[E_i].\nn\eea
We will constantly omit the square braces in the following. We can also apply $\Psi\Phi$ to the tensor product $E_i\otimes^{\BB{L}}E_j$, which gives us the fusion rules between the $E_i$'s
\bea E_i\otimes^{\BB{L}}E_j=\sum_k N_{ij}^kE_k.\label{fusion_coef}\eea

Moreover we also work equivariantly, this means that we treat all $A_0$-modules as graded modules, with the grading furnished by the action of some $U(1)$.
We typically need a $U(1)$ that acts on $X$ only holomorphically, so that at each grading the cohomology is of finite dimension. Concretely the means that the coefficients $n_i$ and $N_{ij}^k$ will be enriched with monomials of equivariant parameters to keep track of the grading shifts in addition to the multiplicity.

\begin{figure}
\begin{center}
\begin{tikzpicture}[scale=.6]
\draw [blue,fill=blue!40,opacity=.3] (0,0) circle (1.4);
\draw [dashed] (-1.4,0) to [out=90, in=90] (1.4,0);
\draw [-] (-1.4,0) to [out=-90, in=-90] (1.4,0);
\draw [-,blue](0,-1.4cm) to [out=80,in=-80] (0,1.4cm); 
\node at (0,1.4cm) {\small $\times$};
\node at (0,-1.4cm) {\small $\times$};
\node at (-0.1,-.5cm) {\scriptsize{$E$}};
\end{tikzpicture}~~~
\begin{tikzpicture}[scale=.6]
\draw [blue,fill=blue!40,opacity=.3] (0,0) circle (1.4);
\draw [dashed] (-1.4,0) to [out=90, in=90] (1.4,0);
\draw [-] (-1.4,0) to [out=-90, in=-90] (1.4,0);
\draw [-,blue](0,-1.4cm) to [out=80,in=-80] (0,1.4cm); 
\node at (0,1.4cm) {\small $\times$};
\node at (0,-1.4cm) {\small $\times$};
\node at (-0.1,-.5cm) {\scriptsize{$E$}};
\node at (0.8,0cm) {\scriptsize{$E'$}};
\draw [-,blue](0.25,-.2cm) to [out=170, in=-80] (-.3,0cm); 
\draw [-,blue](-.3,0cm) to [out=90, in=-170] (0.05,.3cm); 
\draw [-,blue](0.15,-.2cm) to [out=10, in=-100] (.6,.0cm); 
\draw [-,blue](.6,0cm) to [out=90, in=-10] (0.25,.3cm); 
\end{tikzpicture}
\caption{The blue line is part of a Wilson loop that pierces the 2-sphere at two punctures. The two pictures give us two vectors in the Hilbert space on $S^2$ with two punctures.}\label{fig_one_W_loop}
\end{center}
\end{figure}
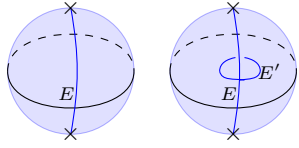

Assume that we have computed the fusion coefficients in \eqref{fusion_coef}, a natural question arises: does the $S$-matrix diagonalise the fusion coefficients? One can find \emph{some} matrix $S^i_{j}$ such that the new basis $F_i=\sum_jE_jS^j_{~i}$ fuses diagonally:
\bea F_i\otimes^{\BB{L}}F_j=\gd_{ij}\gl_iF_i.\label{lambda}\eea
But we will see that the matrix $S^i_{~j}$ does not satisfy any obvious nice relations such as $S^4=1$.
This looks strange at first sight, to understand this one needs to look at the derivation of the Verlinde formula in sec.4 of \cite{witten1989} more closely. We will not review the entire proof, but the crucial catch is that the Hilbert space on $S^2$ with two punctures be either 0 or 1-dimensional, see \eqref{used_II}.
Since two vectors in a 1-dimensional vector space must be proportional, the two vectors in fig.\ref{fig_one_W_loop} must be proportional. The proportionality coefficient is then the eigen-value of the fusion rules. If the Wilson loop labelled $E'$ runs parallel to $E$ in the beginning, then an $S$-surgery makes it wind round $E$ as in the second picture of fig.\ref{fig_one_W_loop}, and this is roughly why $S$-duality diagonalises the fusion rule.
But in the RW theory, the dimension of the Hilbert space does not have this simple dimensionality. We believe this is why the matrices diagonalising $N_{ij}^k$ is not directly related to the $S$-matrix.

On the other hand, we will see that the new basis $F_i=\sum_jE_jS^j_{~i}$ that fuse diagonally are associated to fixed points of the torus actions. In particular, for $T^*\BB{P}^n$, they are the structure sheaf of the cotangent fibre above the fixed points. Even though the matrices $S$ here are invertible, we do not claim that the objects $F_i$ generate the whole category $D^b(X)$ as the $E_i$'s do, only that their Chern-characters can be written in terms of each other when working equivariantly. But this seems a place to explore further, especially if one can repeat the analysis above on some singular non-compact HK variety. In those cases, the number of fixed points will not be equal to the number of $E_i$'s, we are curious as to how the Verlinde formula, if it existed, would fan out.

\begin{figure}
\begin{center}
\begin{tikzpicture}[scale=1]
\centerarc[blue](0,0)(90:270:0.6);
\centerarc[blue,-<](0,0)(90:180:0.6);
\centerarc[blue](3.2,0)(-90:90:0.6);
\centerarc[blue,-<](3.2,0)(-90:0:0.6);
\draw [blue,-] (0,-.6) -- (0.2,.6);
\draw [blue,->] (0,-.6) -- (0.1,0);
\draw [blue,-] (.8,-.6) -- (1,.6);
\draw [blue,->] (.8,-.6) -- (.9,0);
\draw [blue,-] (3.2,-.6) -- (3.4,.55);
\draw [blue,->] (3.2,-.6) -- (3.3,0);
\draw [blue,-] (2.4,-.6) -- (2.6,.6);
\draw [blue,->] (2.4,-.6) -- (2.5,0);

\draw [blue,-] (0,.6) -- (1.2,.6);
\draw [blue,->] (0,.6) -- (0.6,.6);
\draw [blue,-] (3.2,.6) -- (2.0,.6);
\draw [blue,-<] (3.2,.6) -- (2.3,.6);
\draw [blue,-] (0,-.6) -- (1.2,-.6);
\draw [blue,-<] (0,-.6) -- (.5,-.6);
\draw [blue,-] (3.2,-.6) -- (2.0,-.6);
\draw [blue,->] (3.2,-.6) -- (2.8,-.6);

\draw [blue,-,dotted] (1.2,.6) -- (2,.6);
\draw [blue,-,dotted] (1.2,-.6) -- (2,-.6);

\node at (0.2,0.6) [above] {\scriptsize{$\gl_i$}};
\node at (1,0.6) [above] {\scriptsize{$\gl_i$}};
\node at (2.6,0.6) [above] {\scriptsize{$\gl_i$}};
\node at (3.3,0.6) [above] {\scriptsize{$\gl_i$}};

\node at (0,-0.6) [below] {\scriptsize{$\gl^{\vee}_i$}};
\node at (0.8,-0.6) [below] {\scriptsize{$\gl^{\vee}_i$}};
\node at (2.4,-0.6) [below] {\scriptsize{$\gl^{\vee}_i$}};
\node at (3.1,-0.6) [below] {\scriptsize{$\gl^{\vee}_i$}};

\end{tikzpicture}\caption{A Rieman surface of genus $g$ as a trivalent graph.}\label{fig_RS_tri}
\end{center}
\end{figure}
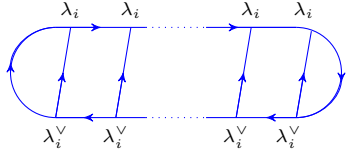

Back to our discussion when $X$ is smooth, we derive now the formula for
\bea \chi_{eq}({\cal H}(\Gs_g,X))=\sum_n u^n \opn{sdim}{\cal H}(\Gs_g,X)_n.\nn\eea
Here we have written only one equivariant parameter $u$, but it is meant to represent all possible equivariant parameters.
The calculation follows \cite{VERLINDE1988360}. We represent a Riemann surface of genus $g$ as a trivalent graph with $2g-2$ vertices as in fig.\ref{fig_RS_tri}, and think of the objects of $D^b(X)$ as states that birth, evolve, merge, split and annihilate following the graph.
Though the objects fuse or split as they go through the vertices, if we use $F_i$ as the basis of objects, there will no mixing of states due to the diagonal fusion of $F_i$. For each tri-valent vertex $F_i$ passes, it only picks up a factor of $\gl_i$ or $\gl_i^{\vee}$, where $\gl_i$ is the fusion coefficient in \eqref{lambda} and $\gl_i^{\vee}$ is the fusion coefficient for the dual object $F_i^{\vee}$ (which is obtained from $\gl_i$ by inverting all equivariant parameters).
This way the graph describes, going from left to right, a series of creations of 'particle anti-particle pair' at each rung, fusions or fissions at each vertex, and finally an annihilation at the last rung. Putting this cartoon into formula we get
\bea \chi_{eq}({\cal H}(\Gs_g,X))=\sum_i\frac{(\gl_i\gl_i^{\vee})^{g-1}}{||F_i||^{2(g-1)}},\label{final_formula}\eea
where the $||F_i||$ factors are explained as follows.
At a rung with pair creation we insert a 'partition of unity'
\bea \sum_i \frac{F_i^{\vee}\otimes F_i}{||F_i||^2}.\nn\eea
Here the norm $||F_i||^2$ is defined in the sense of \eqref{inner_prod_I}. At the last rung with pair annihilation we get $||F_i||^2$. Therefore the total power of $||F_i||$ comes down to $-2g+2$. A more straightforward approach would be to simply normalise each state $F_i$ as
\bea \hat F_i=\frac{F_i}{||F_i||}\nn\eea
and let $\hat \gl_i$ be the normalised fusion coefficient $\hat F_i\otimes^{\BB{L}}\hat F_i=\hat \gl_i\hat F_i$. Then we can write
\bea \chi_{eq}({\cal H}(\Gs_g,X))=\sum_i(\hat\gl_i\hat\gl^{\vee}_i)^{g-1}.\label{verlinde_mine}\eea
We will call this formula the {\bf Verlinde formula} of Rozansky-Witten theory.
We will see from the concrete examples that this formula agrees with the index computation of the lhs. This is one of the major results of this paper, as we have answered the very question that motivated this whole work: why do the fixed points of the geometry have an interpretation as some kind of states, i.e. Eq.(2.2) of \cite{Gukov:2020lqm}:
\begin{center}
The objects in $D^b(X)$ that fuse diagonally are supported above the fixed points.
\end{center}

\subsection{$X=T^*\BB{P}^1$}\label{sec_TP1}
We illustrate the general analysis above with $X=T^*\BB{P}^1$. As we review in sec.\ref{sec_SncHmatcg}, the cohomology groups $A_n=H^{\sbullet}(T^*\BB{P}^1,{\cal O}(n))$ are bi-graded, and we use $s$ for the grading from the tri-holomorphic action while $u$ for the holomorphic action. The $A_n$'s are easily enumerated, see Eqs.\eqref{A_0} \eqref{A_1} \eqref{A_{-2}}.
We have the projection
\bea \begin{tikzpicture}
  \matrix (m) [matrix of math nodes, row sep=2em, column sep=3em]
    { X=T^*\BB{P}^1 \\
    Y=\BB{C}^2/\BB{Z}_2 \\ };
  \path[->, font=\scriptsize]
  (m-1-1) edge (m-2-1);
\end{tikzpicture}\nn\eea
which collapses the $\BB{P}^1$. In fact the tilting theory of $T^*\BB{P}^1$ is quite simple and has already appeared in the context of McKay correspondence, we really do not need the pull power of \cite{TODA20101}.


To begin, the algebra $A_0$ can be presented as
\bea A_0=\BB{C}[a,b,c]/(ab-c^2)\nn\eea
which is also the function ring of $\BB{C}^2/\BB{Z}_2$.\footnote{To make connection with \eqref{A_0}, one should let $z=a/c$, $p=b$ as the base and fibre coordinates. One can also let $z'=b/c$, $p'=a$ and one can check that the change of coordinate matches that of ${\cal O}(-2)$.\label{footnote_notation}}
There are only two non-isomorphic indecomposable $A_0$-modules, one is $A_0$ itself and the other is the ideal $I=(b,c)$. This ideal $I$ corresponds to the ideal ${\cal O}(-[\infty])$ of the structure sheaf, the reader may consult the very well-written lecture notes by Wemyss \cite{NCCR4} for these points.
The direct sum $T=A_0\oplus I$ gives the tilting object, as a result, we can resolve every object in $D^b(X)$ using their duals ${\cal O}$ and ${\cal O}(1)$.

As we discussed below \eqref{equivalence}, the key now is to compute effectively $T\otimes^{\BB{L}}_{\Gl}T^{\vee}$, where ${\Gl}$ is the endomorphism algebra of $T$.
This derived tensor product can be computed by finding first a minimal resolution of ${\Gl}$ as a bi-module over itself, which is treated in \cite{BUTLER1999323}, then tensoring this resolution with $T$ and $T^{\vee}$ from left and right.
Fortunately the first three terms of this minimal resolution is easily described using the Auslander-Reiten (AR) quiver.
If the global dimension of ${\Gl}$ is 2 or less, this will be all we need.
The AR quiver has as nodes the two indecomposables, and as arrows the irreducible maps in between.
There are two such maps from $A_0$ to $I$, i.e. multiplication by $b$ or $c$. There are also two maps going back, given by the inclusion $I\hookrightarrow A_0$ and multiplication with $a/c$. The AR quiver looks like (see the footnote \ref{footnote_notation} for notations)
\bea  \begin{tikzpicture}
  \matrix (m) [matrix of math nodes, row sep=.5em, column sep=5em]
    { A_0 & I \\        };
  \path[->, font=\scriptsize, bend right = 5]
  (m-1-1) edge (m-1-2);
  \path[->, font=\scriptsize, bend left = 5]
  (m-1-1) edge (m-1-2);
  \path[->, blue, font=\scriptsize, bend right = 25]
  (m-1-2) edge (m-1-1);
  \path[->, blue,font=\scriptsize, bend left = 25]
  (m-1-2) edge (m-1-1);

  \node at (0.1,0.2)  {\scriptsize{$pz$}};
  \node at (0.1,-0.15)  {\scriptsize{$p$}};
  \node at (-0.1,0.50)  {\scriptsize{1}};
  \node at (-0.1,-0.47)  {\scriptsize{$z$}};
  \end{tikzpicture}\label{AR_quiver_I}\eea
with only two relations $[\stackrel{p}{\leftarrow}\stackrel{z}{\leftarrow}]=[\stackrel{pz}{\leftarrow}\stackrel{1}{\leftarrow}]$ and $[\stackrel{z}{\leftarrow}\stackrel{p}{\leftarrow}]=[\stackrel{1}{\leftarrow}\stackrel{pz}{\leftarrow}]$ coming from the commutativity.
Note that the blue arrows constitute the Kronecker quiver which is well-known to be the Beilinson quiver for $D^b(\BB{P}^1)$. The black arrows arise because of the cotangent fibre.

For later generalisation to $T^*\BB{P}^n$, we will put the above quiver in a more symmetric form that emphasises the HK geometry and will allow us to use the symmetric polynomial package for computation.
We realise $T^*\BB{P}^1$ as an HK reduction from $\BB{H}^2$ as in sec.\ref{sec_SncHmatcg}. Writing $z_i+jw_i$ as the quaternion coordinates, then the earlier coordinates $z,p$ are $z=z_1/z_2$ and $p=w_1z_2$ in the open $z_2\neq 0$. We also realise the line bundle ${\cal O}(-[\infty])$ as a hyper-holomorphic line bundle (meaning we need to keep track of the weight of the place function for the divisor $-[\infty]$). Then the quiver is rewritten as
\bea  \begin{tikzpicture}
  \matrix (m) [matrix of math nodes, row sep=.5em, column sep=5em]
    { {\cal O} & {\cal O}(-1) \\        };
  \path[->, font=\scriptsize, bend right = 5]
  (m-1-1) edge (m-1-2);
  \path[->, font=\scriptsize, bend left = 5]
  (m-1-1) edge (m-1-2);
  \path[->, blue, font=\scriptsize, bend right = 25]
  (m-1-2) edge (m-1-1);
  \path[->, blue,font=\scriptsize, bend left = 25]
  (m-1-2) edge (m-1-1);

  \node at (0.1,0.2)  {\scriptsize{$w_2$}};
  \node at (0.1,-0.18)  {\scriptsize{$w_1$}};
  \node at (-0.8,0.50)  {\scriptsize{$z_2$}};
  \node at (-0.8,-0.47)  {\scriptsize{$z_1$}};
  \end{tikzpicture}\nn\eea
  where ${\cal O}(-1)$ is realised as the associated bundle \eqref{associated_bdl}.
  All the relations of the quiver stem from the moment map condition $\vec z\cdotp\vec w:=\sum_{i=1}^2z_iw_i$ (see appendix.\ref{sec_SncHmatcg} in particular \eqref{HK_quotient_alg}) and from the commutativity between the two $z$'s and the two $w$'s e.g. the following paths are all zero (we use the letters to denote both the variables and the paths next to avoid clutter)
  \bea \vec z\cdotp \vec w,~~~\vec w\cdotp \vec z,~~~z_iw_kz_j-z_jw_kz_i,~~\sum_iz_iw_kz_lw_i, {\rm etc}.\label{some_zero_paths}\eea

The minimal resolution of ${\Gl}$ as a bi-module over itself reads
\bea 0\to\Gl\otimes_S \SF{T}_2\otimes_S\Gl \stackrel{c}{\to} \Gl\otimes_S\SF{X}\otimes_S\Gl\stackrel{d}{\to} \Gl\otimes_S\Gl\stackrel{e}{\to}\Gl\to0.\label{min_resolution}\eea
We have tried as much as possible to use the same notation as in \cite{BUTLER1999323}: here $S$ is the finite dimensional algebra generated by the idempotents (paths of zero length), $\SF{X}$ is the $S$-module consisting of paths of length 1. The map $e$ is the multiplication map, while $d$ is given by $d(1\otimes_Sx\otimes_S1)=x\otimes_S1-1\otimes_Sx$. The map $c$ is known as the universal derivation, on a path written as a word $x_1\cdots x_r$ we have $c(x_1\cdots x_r)=\sum_{i=1}^rx_1\cdots x_{i-1}\otimes_Sx_i\otimes_Sx_{i+1}\cdots x_r$.
Finally
\bea \SF{T}_2=\frac{R\cap J^2}{JR+RJ}\label{T_2}\eea
with $J$ being ideal of paths of non-zero lengths and $R$ is ideal of relations of $\Gl$ (this is denoted $I$ in \cite{BUTLER1999323}).
Explicitly $\SF{T}_2$ has only two generators
\bea \SF{T}_2=\bra \vec w\cdotp\vec z,\vec z\cdotp\vec w\ket\nn\eea
with the first (resp. second) one starting and ending at the right (resp. left) vertex. The reader might protest that $z_iw_kz_j-[i\leftrightarrow j]$ also belongs to $R\cap J^2$. But this term actually belongs to $JR+RJ$ e.g.
\bea z_{1}w_2z_{2}-z_{2}w_2z_{1}=-z_{1}w_1z_{1}+z_1(\vec w\cdotp \vec z) -z_{2}w_2z_{1}=z_1(\vec w\cdotp \vec z)-(\vec z\cdotp \vec w)z_1\in JR+RJ.\nn\eea
So this term is zero in $\SF{T}_2$. Similarly one can show that the last relation in \eqref{some_zero_paths} is also zero in $\SF{T}_2$.
Further, the higher terms $\SF{T}_n$ for $n>2$ all vanish since $\Gl$ has global dimension 2 (and one can also check it explicitly).

Now one takes \eqref{min_resolution} and tensors $T$ to its left and $T^{\vee}$ to its right, and gets a complex
\bea 0\to T\otimes_S \SF{T}_2\otimes_ST^{\vee} \stackrel{c}{\to} T\otimes_S\SF{X}\otimes_ST^{\vee}\stackrel{d}{\to} T\otimes_ST^{\vee}\to0\label{min_resolution_I}\eea
quasi-isomorphic to ${\cal O}_{\Gd}$ as we have sketched earlier.
Given any object $F$ of $D^b(T^*\BB{P}^1)$, we tensor it over ${\cal O}_{T^*\BB{P}^1}$ to the left of the complex, then take derived section. This leads to $\Psi\Phi(F)\simeq F$, except now $F$ is rewritten as a complex involving only ${\cal O}_{T^*\BB{P}^1}$ and ${\cal O}_{T^*\BB{P}^1}(1)$. We will only write ${\cal O}(n)$ next, with the space $T^*\BB{P}^1$ understood.

Spelling out the complex \eqref{min_resolution_I}, and reading off all the maps from \eqref{AR_quiver_I} leads to a complex
\bea
\begin{tikzcd}[row sep=normal, column sep=large]
R\Gc(F)\otimes{\cal O} \arrow[rr,"{\left[\displaystyle{\mathop{\vphantom{x}}^{\;\textrm{-}w_2}_{\textrm{-}w_1}}\right]\otimes 1}"] \arrow[rrdd,"{1\otimes\left[\displaystyle{\mathop{\vphantom{x}}^{z_2}_{\textrm{-}z_1}}\right]}",near end] & & 2\times {\color{blue}R\Gc(F(-1))\otimes{\cal O}}  \arrow[rr,blue,"{[z_2,z_1]\otimes 1}"] \arrow[ddrr,blue,"{(-1)\otimes[z_2,z_1]}",very near end] & & {\color{blue}R\Gc(F)\otimes{\cal O}} \\
\vphantom{E} & & & \vphantom{E} \\
R\Gc(F(-1))\otimes{\cal O}(1) \arrow[rr,"{\left[\displaystyle{\mathop{\vphantom{x}}^{z_2}_{\textrm{-}z_1}}\right]\otimes 1}"] \arrow[rruu,crossing over, "{1\otimes\left[\displaystyle{\mathop{\vphantom{x}}^{\;\textrm{-}w_2}_{\textrm{-}w_1}}\right]}", very near start] & & 2\times R\Gc(F)\otimes{\cal O}(1) \arrow[rr,"{[\textrm{-}w_2,w_1]\otimes 1}"] \arrow[rruu, crossing over,"{1\otimes[w_2,\textrm{-}w_1]}",very near start] & & {\color{blue}R\Gc(F(-1))\otimes{\cal O}(1)} \\
\end{tikzcd}\label{fusion_complex}\eea
that is isomorphic to ${\cal O}_{\Gd}$. Here $F(-1)=F\otimes_{{\cal O}}{\cal O}(-1)$ and all the tensor products are now over $\BB{C}$.
Again, the three blue terms are what one gets in the case of $D^b(\BB{P}^1)$.
In fact, a lot of the work here is unnecessary, since we are working at the level of Grothendieck group and we could have ignored the maps in the complex.

Now we can plug in any concrete $F$ and we will get their expansion in terms of ${\cal O}$ and ${\cal O}(1)$.
It is perhaps prudent to first check that if we plug in $F={\cal O}(0)$ or ${\cal O}(1)$, we should get the same thing back.
For $F={\cal O}(0)$ the complex reads
\bea 0\to A_0\otimes{\cal O}\oplus A_{-1}\otimes{\cal O}(1)\to 2A_{-1}\otimes{\cal O}\oplus 2A_0\otimes{\cal O}(1)\to A_0\otimes{\cal O}\oplus A_{-1}\otimes{\cal O}(1)\to0.\nn\eea
Taking an alternating sum, but keeping track of the weights, we get
\bea{\cal O}\stackrel{?}{\simeq} (u^2A_0-u(s+1)A_{-1}+A_0){\cal O}+(u^2A_{-1}-u(s^{-1}+1)A_0+A_{-1}){\cal O}(1).\nn\eea
By taking a look at the weights of $A_n$'s in \eqref{pictures_P^1}, the first term gives ${\cal O}$ and the second term vanishes, as expected. One can check the case of $F={\cal O}(1)$ the same way. Now take $F={\cal O}(2)$ and we get 
\bea {\cal O}(2) &\simeq& (u^2A_2-u(s+1)A_1+A_2){\cal O}+(u^2A_1-u(s^{-1}+1)A_2+A_1){\cal O}(1)\nn\\
&=&-su^2{\cal O}+u(1+s){\cal O}(1).\label{O2}\eea
We record some more examples
\bea {\cal O}(-1)&\simeq &(u^2A_{-1}-u(s+1)A_{-2}+A_{-1}){\cal O}+(u^2A_{-2}-u(s^{-1}+1)A_{-1}+A_{-2}){\cal O}(1)\nn\\
&=&(u^{-1}+s^{-1}u^{-1}){\cal O}-u^{-2}s^{-1}{\cal O}(1).\label{O(-1)}\eea
\begin{remark}
  Recall that on $T^*\BB{P}^1$ the bundles ${\cal O}(n)$ were obtained as an associated bundle as in \eqref{associated_bdl}, rather than the naive pullback from $\BB{P}^1$. So the sections of ${\cal O}(n)$ have a grading shift $u^n$ w.r.t  the ones in $\pi^*{\cal O}(n)$. We abuse notation and write simply
  \bea {\cal O}_{T^*\BB{P}^1}(n)= u^n\pi^*{\cal O}_{\BB{P}^1}(n).\label{abuse}\eea
  Taking into account the grading shifts, we realise that \eqref{O2} and \eqref{O(-1)} are actually identical to the their counterparts in $D^b(\BB{P}^1)$. The latter can be computed from the blue part of the complex \eqref{fusion_complex}, i.e. the Beilinson complex
  \bea {\cal O}_{\BB{P}^1}(n)&=&H^{\sbullet}({\cal O}_{\BB{P}^1}(n-1))\otimes{\cal O}_{\BB{P}^1}(1)+H^{\sbullet}({\cal O}_{\BB{P}^1}(n))\otimes{\cal O}_{\BB{P}^1}
   -(s+1)H^{\sbullet}({\cal O}_{\BB{P}^1}(n-1))\otimes{\cal O}_{\BB{P}^1}\nn\\
   &=&(1+s+\cdots s^{n-1}){\cal O}_{\BB{P}^1}(1)-(s+s^2+\cdots s^{n-1}){\cal O}_{\BB{P}^1},~~{\rm for}~n\geq0,\nn\\
   &=&-(s^{-1}+\cdots s^n){\cal O}_{\BB{P}^1}(1)+(1+s^{-1}+\cdots+s^n){\cal O}_{\BB{P}^1},~~{\rm for}~n<0. \nn\eea
   Comparing this with \eqref{O2} and \eqref{O(-1)}, we see the agreement once the grading shift is included.
\end{remark}

Now we can use \eqref{O2} to write the fusion matrix. First ${\cal O}$ acts as the identity under the tensor product with any other sheaf, while the tensor product
\bea {\cal O}(1)\otimes_{\cal O}{\cal O}(1)\simeq-su^2{\cal O}+u(1+s){\cal O}(1).\label{used_V}\eea
By expanding any object $F$ into ${\cal O}$, ${\cal O}(1)$, we may write $F$ as a column vector $F=[F_1;F_2]$ i.e. $F\simeq F_1{\cal} O +F_2{\cal O}(1)$, with $F_{\sbullet}$ valued in the equivariant cohomology ring $\BB{Z}[s,u,s^{-1},u^{-1}]$.
Then the action of tensoring $F$ with ${\cal O},\,{\cal O}(1)$ can be presented as $2\times 2$ matrices,
\bea &({\cal O}(n)\otimes F)_{\sbullet}=(N_nF)_{\sbullet},\nn\\
 &N_{0\sbullet}^{\sbullet}={\bf 1}_{2\times 2},~~~N_{1\sbullet}^{\sbullet}=\left[
                    \begin{array}{cc}
                      0 & -su^2 \\
                      1 & u(1+s) \\
                    \end{array}\right],\nn\eea
where the second column of $N_1$ comes from the expansion \eqref{used_V}.
We call these matrices the {\bf fusion matrices}.

\smallskip

Next we will identify the objects in $D^b(T^*\BB{P}^1)$ that fuse diagonally.
First it is clear that $N_{1\sbullet}^{\sbullet}$ has eigen-values $u$ and $us$ with eigen-vectors
\bea E_0=-us{\cal O}+{\cal O}(1),~~~E_{\infty}=-u{\cal O}+{\cal O}(1).\label{E_0,E_infty}\eea
It is now quite easy to compute the fusion of $E_0$ and $E_{\infty}$, as well as their duals
\bea &E_0\otimes^{\BB{L}}E_0=u(1-s)E_0,~~E_{\infty}\otimes^{\BB{L}}E_{\infty}=-u(1-s)E_{\infty},~~E_0\otimes^{\BB{L}}E_{\infty}=0,\nn\\
&E^{\vee}_0\otimes^{\BB{L}}E^{\vee}_0=u^{-1}(1-s^{-1})E^{\vee}_0,~~E^{\vee}_{\infty}\otimes^{\BB{L}}E^{\vee}_{\infty}=(us)^{-1}(1-s)E^{\vee}_{\infty},~~E^{\vee}_0\otimes^{\BB{L}}E^{\vee}_{\infty}=0.\nn\eea
This means in particular the orthogonality
\bea{\cal H}(S^2,E_0,E_{\infty})=0\nn\eea
when the two Wilson loops are labelled with $E_{0,\infty}$.
We already discussed in sec.\ref{sec_StS} the importance of the orthogonality for obtaining any thing remotely like the Verlinde formula.
As the change of basis from ${\cal O}$, ${\cal O}(1)$ to $E_{0,\infty}$ is invertible, we opt to use the $E$'s as a basis for $D^b(T^*\BB{P}^1)$.
It needs to be emphasised that we are by no means saying that $E_0$, $E_{\infty}$ generate the whole derived category, only that when working over the equivariant cohomology ring, the Chern-character of ${\cal O},{\cal O}(1)$ can be written in terms of those for the $E$'s.
It is curious here that in the Rozansky-Witten theory, the basis in which we have the orthogonality is also the basis that fuses diagonally.

So what are the objects $E_0$, $E_{\infty}$?
Consider first the object in $D^b(\BB{P}^1)$ given as the complex ${\cal O}_{\BB{P}^1}(-1)\stackrel{z}{\to}{\cal O}_{\BB{P}^1}$. As the complex is exact everywhere except $z=0$, the object it represents is supported only at $z=0$, in fact it is the skyscraper sheaf at $z=0$.
In the Grothendieck group the skyscraper reads ${\cal O}_{\BB{P}^1}-s{\cal O}_{\BB{P}^1}(-1)$ with $s$ being the $U(1)$ weight of $z$.
Pulling this relation to $T^*\BB{P}^1$, comparing with $E_0$ of \eqref{E_0,E_infty} and taking into account the grading shift mentioned earlier, we realise that $E_0(-1)$ is the structure sheaf of the cotangent fibre above $z=0$. Similarly $E_{\infty}(-1)$ is that of the cotangent fibre above $z=\infty$.
We can also check their section (or rather the generating function thereof)
\bea \chi_{eq}(\Gc(E_0))=\frac{u}{1-u^2s^{-1}}=u(1+u^2/s+\cdots),~~~\chi_{eq}(\Gc(E_{\infty}))=\frac{us}{1-u^2s}=us(1+u^2s+\cdots)\nn\eea
which up to a shift is the Hilbert series of the functions on the affine line.
From this identification of the two $E$'s, it is perhaps natural that they should fuse the way they do.

We come now to the Verlinde formula itself. For objects $F,F'$ in $D^b(T^*\BB{P}^1)$, we can define the inner product as $\bra F|F'\ket=\chi_{eq}(F^{\vee}\otimes^{\BB{L}}F')$. Computing this for $E_0,E_{\infty}$ gives
\bea ||E_0||^2=\frac{1-s^{-1}}{1-u^2s^{-1}},~~~||E_{\infty}||^2=\frac{1-s}{1-su^2},~~~\bra E_0|E_{\infty}\ket=0.\nn\eea
Following the plan laid out at the end of sec.\ref{ToaboHs}, we need the normalised fusion coefficient
\bea \frac{\gl_i\gl_i^{\vee}}{||E_i||^2}=\bigg\{
                                                    \begin{array}{cc}
                                                      (1-s)(1-u^2s^{-1}) & i=0 \\
                                                      (1-s^{-1})(1-u^2s) & i=\infty \\
                                                    \end{array}\nn\eea
The final formula \eqref{final_formula} for the dimension of the Hilbert space on $\Gs_g$ and $X=T^*\BB{P}^1$ is
\bea \chi_{eq}({\cal H}(\Gs_g,X))&=&\sum_i\Big(\frac{\gl_i\gl_i^{\vee}}{||E_i||^2}\Big)^{g-1}=\frac{((1-s)(1-u^2s^{-1}))^g}{(1-s)(1-u^2s^{-1})}+\frac{((1-s^{-1})(1-u^2s))^g}{(1-s^{-1})(1-u^2s)}
\label{used_VI}\\
&=&\sum_i\frac{\opn{ch}_{eq}((\Go^{\sbullet,0}_X)^{\otimes g})\opn{Td}_{eq}(TX)}{e_{eq}(T^*X)}\Big|_i.\nn\eea
where we have recognised the rhs of \eqref{used_VI} as the localisation formula for the index of $(\Go^{\sbullet,0}_X)^{\otimes g}$.

\smallskip

To summarise this section, thanks to the tilting theory, we were able to rewrite any object in $D^b(T^*\BB{P}^1)$ as linear combinations of only two objects ${\cal O},{\cal O}(1)$ (at the level of Grothendieck group). Then the fusion, which is just the derived tensor product, is given in terms of some $2\times 2$ matrices. Diagonalising these matrices gives us two objects $E_i$ that fuse diagonally and are related to the cotangent fibre above the two torus fixed points of the geometry $T^*\BB{P}^1$. Finally the Verlinde formula of the Hilbert space comes from a simple minded diagrammatic calculation where the objects $E_i$ are treated as particles that propagate along the graph representing the Riemann surface.
Next we do the same for $T^*\BB{P}^n$.

\subsection{$X=T^*\BB{P}^n$, $n\geq2$}\label{sec_TPn}
We deal with $T^*\BB{P}^2$ first, the general pattern should become clear after this. In this case the authors of \cite{TODA20101} provide us with the tilting bundle
\bea T={\cal O}\oplus{\cal O}(-1)\oplus{\cal O}(-2).\nn\eea
As usual ${\cal O}(n)$ means ${\cal O}_{T^*\BB{P}^2}(n)$, and in the same way ${\cal O}(n)$ differs from $\pi^*{\cal O}_{\BB{P}^2}(n)$ by a grading shift $u^n$ as in \eqref{abuse}. The irreducible maps between the three line bundles have a similar look
\bea  \begin{tikzpicture}
  \matrix (m) [matrix of math nodes, row sep=.5em, column sep=5em]
    { {\cal O} & {\cal O}(-1) & {\cal O}(-2) \\        };
  \path[->, font=\scriptsize, bend right = 10] (m-1-1) edge node [below] {\scriptsize{$\vec w$}} (m-1-2);
  \path[->, font=\scriptsize, bend right = 10] (m-1-2) edge node [below] {\scriptsize{$\vec w$}} (m-1-3);
  \path[->, blue,font=\scriptsize, bend right = 10] (m-1-2) edge node [above] {\scriptsize{$\vec z$}} (m-1-1);
  \path[->, blue,font=\scriptsize, bend right = 10] (m-1-3) edge node [above] {\scriptsize{$\vec z$}} (m-1-2);
\end{tikzpicture}\nn\eea
where each arrow represents three arrows, since we now have three $z$'s and three $w$'s. The blue part is as before the Beilinson quiver for $D^b(\BB{P}^2)$.
The relations of the quiver still come from the commutativity of the $z,\,w$'s plus the moment map condition $\vec z\cdotp\vec w=0=\vec w\cdotp\vec z$.

Following thm.7.2. of \cite{BUTLER1999323}, the resolution \eqref{min_resolution} will stop at the $\SF{T}_4$ term
\bea 0\to \Gl\otimes_S \SF{T}_4\otimes_S\Gl\to \Gl\otimes_S \SF{T}_3\otimes_S\Gl\to\Gl\otimes_S \SF{T}_2\otimes_S\Gl\to\Gl\otimes_S\SF{X}\otimes_S\Gl\to \Gl\otimes_S\Gl\to\Gl\to0\label{hope_less_I}.\eea
This is because the algebra $\Gl$ has global dimension 4 and the resolution is minimal. The two new terms $\SF{T}_{3,4}$ are as follows
\bea \SF{T}_3=\frac{JR\cap RJ}{R^2+JRJ},~~~\SF{T}_4=\frac{R^2\cap JRJ}{JR^2+R^2J}.\nn\eea
The complex \eqref{min_resolution_I} also extends to level 4
\bea 0\to T\otimes_S \SF{T}_4\otimes_ST^{\vee}\to T\otimes_S \SF{T}_3\otimes_ST^{\vee}\to T\otimes_S \SF{T}_2\otimes_ST^{\vee}\to T\otimes_S\SF{X}\otimes_ST^{\vee}\to T\otimes_ST^{\vee}\to0\label{hope_less_II}\eea
There is also a proliferation of terms in $\SF{T}_i$ as $n$ increases, we will give the list below but we do not have any neat systematics in how to exhaust all possible terms in $\SF{T}_i$.
\bea &&\SF{T}_2=\left[
           \begin{array}{c}
             e_1(z\cdotp w\oplus w\cdotp z)e_1 \\
             e_2(z\cdotp w)e_2 \\
             e_0(w\cdotp z)e_0 \\
             e_1([w_iz^j]_{\rm t.f.})e_1 \\
             e_2(z^{[i}z^{j]})e_0 \\
             e_0(w_{[i}w_{j]})e_2 \\
           \end{array}\right]~~~\SF{T}_3=\left[
           \begin{array}{c}
             e_2(z^kw_kz^j-z^kz^jw_k+z^jz^kw_k)e_1 \\
             e_0(w_kz^kw_j-w_kw_jz^k+w_jw_kz^k)e_1 \\
             e_1(z^jw_kz^k-w_kz^jz^k+w_kz^kz^j)e_0 \\
             e_1(w_jz^kw_k-z^kw_jw_k+z^kw_kw_j)e_2
           \end{array}\right],~\nn\\
           &&
           ~~\SF{T}_4=\left[
           \begin{array}{c}
             e_2(z^{[k}z^{l]}w_{[k}w_{l]}-2z^kw_kz^lw_l)e_2 \\
             e_0(w_{[k}w_{l]}z^{[k}z^{l]}-2w_kz^kw_lz^l)e_0 \\
             e_1(\left[z^i,w_j\right]\left[w_i,z^j\right]-z^iw_iw_jz^j-w_iz^iz^jw_j)e_1
\end{array}\right].\label{T_1234}\eea
The notations are as follows, the $e_{0,1,2}$ are the idempotents at the node $0,1,2$, each letter represents an arrow going left e.g. $e_2z^{[i}z^{j]}e_0$ represents a path $[2\stackrel{z^i}{\leftarrow}1\stackrel{z^j}{\leftarrow}0]-[2\stackrel{z^j}{\leftarrow}1\stackrel{z^i}{\leftarrow}0]$. The subscript in $[w_iz^j]_{\rm t.f.}$ means trace free, which will be handy once we organise these paths into irreducible representations of $GL(3,\BB{C})$.
In an earlier version of the draft we had a few more terms for each $\SF{T}_i$. We will explain in the appendix sec.\ref{sec_Smdatmr} why these and other seemingly valid terms in $\SF{T}_i$ should be excluded, as well as the maps between the $\SF{T}_i$'s.

To effectively keep track of the relations and prepare for later computations, we will label the relations using the $GL(3,\BB{C})$ representation under which they transform. For example $z^i$ and $w_i$ transform respectively in
\bea z^i\in \tiny{\yng(1)},~~~~w_i\in\gl^{-1}\tiny{\yng(1,1)},~~~{\rm where}~\gl{\bf 1}\simeq\tiny{\yng(1,1,1)}\nn\eea
Here $\gl$ is a singlet under $GL(3,\BB{C})$, we kept it only because it carries weights under the $U(1)$'s (one can also discard them during the calculation as they can easily be recovered in the end)
\bea \SF{T}_2:&& z^iw_i\in u^2{\bf 1},~~~w_iz^i\in u^2{\bf 1},~~~z^{[i}z^{j]}\in u^2\tiny{\yng(1,1)},
~~~w_{[i}w_{j]}\in u^2\gl^{-1}\tiny{\yng(1)},~~~[z^i,w_j]_{\rm t.f.}\in u^2\gl^{-1}\tiny{\yng(2,1)},\nn\eea
All terms in $\SF{T}_3$ transform either as $u{\tiny\yng(1)}$ or $u\gl^{-1}{\tiny\yng(1,1)}$ and all terms in $\SF{T}_4$ transform as singlets $u^4{\bf 1}$.

From the minimal resolution \eqref{hope_less_II}, we can get the resolution of the diagonal using the bundles ${\cal O}(0,\pm1,\pm2)$ as we already sketched in sec.\ref{sec_TP1}.
The complex for $T^*\BB{P}^2$ is a lot bigger than the complex \eqref{fusion_complex} for $T^*\BB{P}^1$ and is therefore given in the appendix, see Eq.\ref{tilting_complex_P2}. We also spell out the maps there.
%
In a separate work, we will compute these complexes for more cases such as $T^*Gr(2,4)$. This exercise may be interesting on its own as it provides some non-trivial examples of the minimal resolutions of \cite{BUTLER1999323}.

The main purpose of obtaining a resolution of the diagonal is of course to find the expansion of any sheaf in terms of ${\cal O}_{T^*\BB{P}^2}(0,1,2)$, and in particular $F={\cal O}_{T^*\BB{P}^2}(3),{\cal O}_{T^*\BB{P}^2}(-1)$. The calculations and some checks are again done in the appendix, we found
\bea {\cal O}(n)=u^n\gl \opn{YD}(n-3){\cal O}-u^{n-1}\opn{YD}(n-2,1){\cal O}(1)+u^{n-2}\opn{YD}(n-2){\cal O}(2),~~~n\geq3 \label{expansion_P2_pos}\\
     {\cal O}(-n)=\frac{1}{u^n\gl^n}\opn{YD}(n,n){\cal O}-\frac{1}{u^{n+1}\gl^n}\opn{YD}(n,n-1){\cal O}(1)+\frac{1}{u^{n+2}\gl^n}\opn{YD}(n-1,n-1){\cal O}(2),~~n\geq1\label{expansion_P2_neg}\eea
where the Young-diagrams are now the corresponding Schur polynomials of the equivariant parameters, which we name as $s_1,s_2,s_3$ (which we eventually will rename to $s_1=s$, $s_2=t$ and $s_3=1$).
These expansions agree with the case of $\BB{P}^2$, where we only need to compute the much easier Beilinson quiver. The only difference is that one must not forget the identification ${\cal O}(n)=u^n{\cal O}_{\BB{P}^2}(n)$ as we did in \eqref{abuse} to account for the grading shift.

We need only the expansion \eqref{expansion_P2_pos} for $n=3,4$ to compute the fusion matrix for $D^b(T^*\BB{P}^2)$. Again, we arrange ${\cal O}(0,1,2)$ into a column vector, then tensoring with ${\cal O}(n)$ is represented by the matrix $N_{n\sbullet}^{\sbullet}$
\bea N_{0\sbullet}^{\sbullet}={\bf 1},~~~ N_{1\sbullet}^{\sbullet}=\left[
                                \begin{array}{ccc}
                                  0 & 0 & u^3\gl \\
                                  1 & 0 & -u^2\tiny{\yng(1,1)} \\
                                  0 & 1 & u\tiny{\yng(1)} \\
                                \end{array}\right],~~~N_{2\sbullet}^{\sbullet}=\left[
                                \begin{array}{ccc}
                                  0 &  u^3\gl & u^4\tiny{\yng(1)}\\
                                  0 & -u^2\tiny{\yng(1,1)} & -u^3\tiny{\yng(2,1)}\\
                                  1 & u\tiny{\yng(1)} & u^2\tiny{\yng(2)} \\
                                \end{array}\right].\label{fusion_matrix}\eea
There are some easy sanity checks one can do. First $N_2=(N_1)^2$ is quite straightforward. Secondly $(N_1)^{-1}$ should be the matrix of tensoring with ${\cal O}(-1)$, so its first column should agree with \eqref{expansion_P2_neg}, and it does.

Let us now diagonalise the $N_1$ matrix, the three eigen-vectors are
\bea E_i=\frac{\partial}{\partial s_i}\left[
                                        \begin{array}{c}
                                          u^2\gl \\
                                          -u\tiny{\yng(1,1)} \\
                                          \tiny{\yng(1)} \\
                                        \end{array}\right],~~~N_1E_i=us_iE_i.\nn\eea
For example the $E_1$ is represented by a complex
\bea
\begin{tikzcd}[row sep=normal, column sep=large]
{\cal O} \arrow[r,"{\left[\displaystyle{\mathop{\vphantom{x}}^{\;z_2}_{\textrm{-}z_1}}\right]}"] & 2\times {\cal O}(1)  \arrow[r,"{[z_1,z_2]}"] & {\cal O}(2) \\
\end{tikzcd}\nn\eea
which is up to a shift of 2 the structure sheaf of the cotangent fibre above the point $z_2=z_3=0$.

The remainder of the steps are the same as in the $T^*\BB{P}^1$ case. The fusion of these objects read
\bea E_i\otimes^{\BB{L}}E_j=\gl_i\gd_{ij}E_i,~~~\gl_i=u^2(s_i - s_{i-1})(s_i-s_{i+1}),\nn\eea
the fusion of the dual objects are again given by the same formula with all equivariant parameters inverted.
The inner products of these objects read
\bea ||E_i||^2=\frac{(s_i - s_{i+1})(s_i - s_{i-1})}{(s_{i+1}- s_iu^2)(s_{i-1}-s_iu^2)},~~\bra E_i|E_j\ket=0,~~i\neq j\nn\eea
and hence the ratio
\bea \frac{\gl_i\gl_i^{\vee}}{||E_i||^2}&=&(s_i^{-1}-s_{i-1}^{-1})(s_i^{-1}-s_{i+1}^{-1})(s_{i+1}- s_iu^2)(s_{i-1}-s_iu^2)\nn\\
&=&(1-\frac{s_{i+1}}{s_i})(1-\frac{s_{i-1}}{s_i})(1-u^2\frac{s_i}{s_{i+1}})(1-u^2\frac{s_i}{s_{i-1}}).\nn\eea
We recognise here e.g. $s_{i\pm1}/s_i$ as the weights of the inhomogeneous coordinates $z_{i\pm1}/z_i$, and $u^2s_i/s_{i\pm1}$ as the weights of the fibre coordinates $w_{i\pm1}z_i$, all valid in the open set $z_i\neq 0$. Thus raising this ratio to the $(g-1)^{th}$ power gives simply
\bea \Big(\frac{\gl_i\gl_i^{\vee}}{||E_i||^2}\Big)^{g-1}=\frac{((\Go^{\sbullet,0}_X)^{\otimes g})\opn{Td}_{eq}(TX)}{e_{eq}(T^*X)}\Big|_i\nn\eea
exactly the same as in the last section. The remaining steps are also identical to the $X=T^*\BB{P}^1$ case.

It is not hard to stretch our result to the $T^*\BB{P}^n$ case, where the objects comprising the tilting object is ${\cal O},{\cal O}(n),\cdots,{\cal O}(n)$ and the crucial fusion matrix is obtained from \eqref{fusion_matrix} by adding one more box, e.g. for $n=4$
\bea N_{1\sbullet}^{\sbullet}=\left[
                                \begin{array}{cccc}
                                  0 & 0 & 0 & -u^4\gl \\
                                  1 & 0 & 0 & u^3\tiny{\yng(1,1,1)} \\
                                  0 & 1 & 0 & -u^2\tiny{\yng(1,1)} \\
                                  0 & 0 & 1 & u\tiny{\yng(1)} \\
                                \end{array}\right].\nn\eea
Admittedly our stock of examples is a bit thin, but in a future publication we will provide more non-trivial ones.

\appendix
\section{Some non-compact HK manifolds and their cohomology groups}\label{sec_SncHmatcg}
In this section we set down some notations of some non-compact HK manifolds that will be used in the main text, as well as recording some cohomology groups.

\subsection{HK quotient}
The cotangent bundle of a K\"ahler manifold is naturally HK. A typical example is $T^*\BB{P}^n$, one can present it as the HK reduction of $\BB{H}^{n+1}$ by a $U(1)$ action.
Write the quaternion coordinates of $\BB{H}^{n+1}$ as $q_i,\;i=0,1,\cdots,n$, with these we can write a triplet of K\"ahler forms
\bea &&\hspace{3.6cm}\vec\go=\frac12\sum_idq_i[i,j,k]d\bar q_i,\nn\\
&&\go_1=\frac{i}{2}(dz d\bar z+dwd\bar w),~~~\go_2=\frac12(dzdw+d\bar z d\bar w),~~~\go_3=\frac{i}{2}(dzdw-d\bar zd\bar w)\label{3-symp_form}\eea
where $q_i=z_i+jw_i$.
The simultaneous left-multiplication by $e^{i\ga}$ obviously preserves all three $\go$'s and has moment maps
\bea &&\hspace{2.6cm}\vec\mu=-\frac12\bar qiq,\nn\\
&&\mu_1=\frac{i}{2}\sum_i(-|z_i|^2+|w_i|^2),~~\mu_2-i\mu_3=i\vec z\cdotp\vec w.\label{mement_quat}\eea
We take the HK quotient \cite{Hitchin1987}
\bea \BB{H}^{n+1}///U(1)=Z/U(1)=\{\mu_1=-ic/2,~\mu_{2,3}=0\}/U(1)\label{HK_quotient}\eea
with $c$ a constant. When $c>0$ the $z_i$'s cannot be zero altogether and $[z_0,z_1,\cdots z_n]$ give the homogeneous coordinates for $\BB{P}^n$. On an open set, say, $z_n\neq 0$ the combinations $w_0z_n,w_1z_n,\cdots,w_{n-1}z_n$ give the coordinates of the cotangent fibre.
One can also express \eqref{HK_quotient} as
\bea T^*\BB{P}^n=\{\mu_2-i\mu_3=i\vec z\cdotp\vec w=0\}^{ss}/\BB{C}^*\label{HK_quotient_alg}\eea
where $ss$ means the semi-stable locus $\{\vec z\neq0\}$. Eq.\ref{HK_quotient_alg} is more algebraic and more useful for us.

With this description we have $n$ hyper-holomorphic isometries denoted $U(1)_i,\,i=0,\cdots ,n-1$ namely $U(1)_i$ left multiplies $q_i$ by a phase. They clearly preserve \eqref{3-symp_form}. The \emph{simultaneous right} multiplication by $e^{i\ga}$ preserves only $\go_1$ while rotates $\Go=\go_2-i\go_3$ by weight 2. This is the holomorphic isometry, denoted $U(1)_r$. We see that $U(1)_r$ leaves the inhomogeneous coordinates of $\BB{P}^n$ unchanged, but rotates the fibre coordinate $T^*\BB{P}^n$ by weight 2.

The distinction between these two types of isometries are crucial. Recall that the aim for introducing equivariance is so that the infinite dimensional cohomology group $H^{\sbullet}(X,(\Go^{\sbullet})^{\otimes g})$ may obtain a grading, and at each grading it should be finite dimensional.
Take the $T^*\BB{P}^2$ example and consider the simplest case
\bea H^{\sbullet}(T^*\BB{P}^2,{\cal O}_{T^*\BB{P}^2})\simeq \oplus_{n\geq0} H^{\sbullet}(\BB{P}^2,\opn{Sym}^nT\BB{P}^2).\label{used_III}\eea
Since $U(1)_r$ only rotates the fibre of $T^*\BB{P}^2$, the grading it provides is essentially the grading $n$ in $\opn{Sym}^nT\BB{P}^2$. It is clear that for given $n$, the cohomology group is finite dimensional. The other tri-holomorphic $U(1)$'s will not have this property. 
There is a very easy and heuristic way to check if a certain $U(1)$ ensures finite dimensionality. Consider its moment map $\mu_1$ w.r.t $\go_1$, if its level set is compact, then it will do the job. For example the moment map $\mu_1$ for $U(1)_i$ is the restriction of $|z_i|^2-|w|_i^2$ (see \eqref{mement_quat}) and so has no compact level set. In contrast, the moment map for $U(1)_r$ is $\sum_i(|z_i|^2+|w_i|^2)$ with a compact level set.

\subsection{Some cohomology groups}
We record here come cohomology groups of some line bundles over $T^*\BB{P}^n$, which will be used in the main text.
For $T^*\BB{P}^1$, we realise it as the HK quotient of $\BB{H}^2$ by a $U(1)$ that acts as the simultaneous multiplication from the left. Let $z_a+jw_a$, $a=1,2$ be the coordinates of the two $\BB{H}$, then in the open $z_2\neq0$, we can use $z=z_1/z_2$ as the holomorphic coordinate of $\BB{P}^1$ and $p=w_1z_2$ the fibre coordinate.
Then function ring $A_0=H^0(T^*\BB{P}^1,{\cal O})$ can be enumerated as (note $H^{>0}(T^*\BB{P}^1,{\cal O})=0$)
\bea A_0=\BB{C}\bra p^nz^i,~~n\geq0,~~0\leq i\leq 2n\ket.\label{A_0}\eea
We will also need the cohomology of ${\cal O}(n)$. Amongst these ${\cal O}(n)$, $n\geq-1$ has no higher cohomology and we denote with $A_{\pm1}=H^0(T^*\BB{P}^1,{\cal O}(\pm1))$
\bea A_1=\BB{C}\bra p^nz^i,~~n\geq0,~~0\leq i\leq 2n+1\ket,~~~A_{-1}=pA_1.\label{A_1}\eea
In contrast ${\cal O}(-2)$ will have $H^1\neq 0$
\bea &&H^0(T^*\BB{P}^1,{\cal O}(-2))=\BB{C}\bra p^nz^i,~~p\geq0,~~0\leq i\leq 2n-1\ket,\nn\\
&&H^1(T^*\BB{P}^1,{\cal O}(-2))=\BB{C}\bra z^{-1}\ket.\label{A_{-2}}\eea

The way we have enumerated the cohomology glosses over the graded module structure e.g. writing $A_{-1}$ as $pA_1$ misses the grading completely. On the quotient we have a tri-holomorphic $U(1)$ that acts as left multiplication on $z_1+jw_1$ by a phase. We use $s$ as the equivariant parameter for this action.
The holomorphic $U(1)$ acts as right multiplication on both $\BB{H}$ factor by a phase. We use $u$ as its equivariant parameter. Thus $z$ has weight $s$ and $p$ has weight $u^2s^{-1}$. There is however another small subtlety. We stated earlier that in order for a vector bundle $E$ to be used in a Wilson loop, it has to be hyper-holomorphic. When we naively pull the line bundle ${\cal O}(1)$ from $\BB{P}^1$ to $T^*\BB{P}^1$, it is not hyper-holomorphic. But it is always possible to modify the connection of the bundle to make it so \cite{Feix2002HypercomplexMA}. In the current case, the procedure is much simpler and is given in \cite{Hitchin_hyper} sec.2.4. If we present the ${\cal O}(n)$ as a bundle of degree $n$ associated to the quotient \eqref{HK_quotient}
\bea \begin{tikzpicture}
  \matrix (m) [matrix of math nodes, row sep=1.5em, column sep=.2em]
    { Z\times_{U(1)} \BB{C} \\
       X \\ };
  \node at(-1.3,0) {\small{${\cal O}(n)\sim$}};
  \path[->, font=\scriptsize]
  (m-1-1) edge node[right] {$\pi$} (m-2-1);
\end{tikzpicture}\label{associated_bdl}\eea
with $U(1)$ acting on $\BB{C}$ with weight $n$, then it will be hyper-holomorphic. Apart from the grading, this subtlety has very little impact otherwise. The result is $z^ap^b$ has weight $s^{a-b}u^{2b+n}$ for ${\cal O}(n)$ rather than the expected $s^{a-b}u^{2b}$.
Let us plot the weights of some of the cohomology groups above, which will be a great visual help.
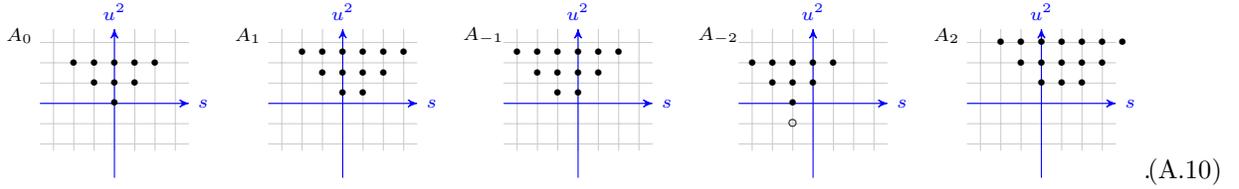
\bea \begin{tikzpicture}[scale=.9]
\draw [step=0.3,thin,gray!40] (-1.1,-0.7) grid (1.1,1.1);

\draw [->,blue] (-1.1,0) -- (1.1,0) node[right] {\scriptsize{$s$}};
\draw [->,blue] (0,-1.1) -- (0,1.1) node[above] {\scriptsize{$u^2$}};

\node at (0,0) {\scriptsize{$\sbullet$}};
\foreach \x in {0,1,...,2}
\node at (0.3*\x-0.3,0.3) {\scriptsize{$\sbullet$}};
\foreach \x in {0,1,...,4}
\node at (0.3*\x-0.6,0.6) {\scriptsize{$\sbullet$}};

\node at (-1.4,1) {\scriptsize{$A_0$}};
\end{tikzpicture}~
\begin{tikzpicture}[scale=.9]
\draw [step=0.3,thin,gray!40] (-1.1,-0.7) grid (1.1,1.1);

\draw [->,blue] (-1.1,0) -- (1.1,0) node[right] {\scriptsize{$s$}};
\draw [->,blue] (0,-1.1) -- (0,1.1) node[above] {\scriptsize{$u^2$}};

\foreach \x in {0,1}
\node at (0.3*\x,0.15) {\scriptsize{$\sbullet$}};
\foreach \x in {0,1,...,3}
\node at (0.3*\x-0.3,0.45) {\scriptsize{$\sbullet$}};
\foreach \x in {0,1,...,5}
\node at (0.3*\x-0.6,0.75) {\scriptsize{$\sbullet$}};

\node at (-1.4,1) {\scriptsize{$A_1$}};
\end{tikzpicture}~
\begin{tikzpicture}[scale=.9]
\draw [step=0.3,thin,gray!40] (-1.1,-0.7) grid (1.1,1.1);

\draw [->,blue] (-1.1,0) -- (1.1,0) node[right] {\scriptsize{$s$}};
\draw [->,blue] (0,-1.1) -- (0,1.1) node[above] {\scriptsize{$u^2$}};

\foreach \x in {0,1}
\node at (0.3*\x-0.3,0.15) {\scriptsize{$\sbullet$}};
\foreach \x in {0,1,...,3}
\node at (0.3*\x-0.6,0.45) {\scriptsize{$\sbullet$}};
\foreach \x in {0,1,...,5}
\node at (0.3*\x-0.9,0.75) {\scriptsize{$\sbullet$}};

\node at (-1.4,1) {\scriptsize{$A_{-1}$}};
\end{tikzpicture}~
\begin{tikzpicture}[scale=.9]
\draw [step=0.3,thin,gray!40] (-1.1,-0.7) grid (1.1,1.1);

\draw [->,blue] (-1.1,0) -- (1.1,0) node[right] {\scriptsize{$s$}};
\draw [->,blue] (0,-1.1) -- (0,1.1) node[above] {\scriptsize{$u^2$}};

\node at (-0.3,-0.3) {\scriptsize{$\circ$}};
\node at (-0.3,0) {\scriptsize{$\sbullet$}};
\foreach \x in {0,1,...,2}
\node at (0.3*\x-0.6,0.3) {\scriptsize{$\sbullet$}};
\foreach \x in {0,1,...,4}
\node at (0.3*\x-0.9,0.6) {\scriptsize{$\sbullet$}};

\node at (-1.4,1) {\scriptsize{$A_{-2}$}};
\end{tikzpicture}~
\begin{tikzpicture}[scale=.9]
\draw [step=0.3,thin,gray!40] (-1.1,-0.7) grid (1.1,1.1);

\draw [->,blue] (-1.1,0) -- (1.1,0) node[right] {\scriptsize{$s$}};
\draw [->,blue] (0,-1.1) -- (0,1.1) node[above] {\scriptsize{$u^2$}};

\foreach \x in {0,1,...,2}
\node at (0.3*\x,0.3) {\scriptsize{$\sbullet$}};
\foreach \x in {0,1,...,4}
\node at (0.3*\x-0.3,0.6) {\scriptsize{$\sbullet$}};
\foreach \x in {0,1,...,6}
\node at (0.3*\x-0.6,0.9) {\scriptsize{$\sbullet$}};

\node at (-1.4,1) {\scriptsize{$A_2$}};
\end{tikzpicture}\label{pictures_P^1}.\eea
%
%
In the above pictures, the solid dots are in $H^0$ and the hollow ones are in $H^1$.

The reason we make such cumbersome plots above is to demonstrate the fact that we can use ${\cal O}(0)$ or ${\cal O}(1)$ to resolve any sheaf in $T^*\BB{P}^1$. The real reason comes from the tilting theory as in the main text, but at a heuristic level, we can see how this plays out from the pictures. Consider e.g. $A_2$, by comparing its plot with that of $A_0,A_1$ we see immediately
\bea 
\begin{tikzcd}[row sep=normal, column sep=normal]
A_1 \arrow[r,tail,"{\left[\displaystyle{\mathop{\vphantom{x}}^{w_2z_1}_{w_1}}\right]}"] & A_1\oplus A_0 \arrow[r,two heads, "{\left[{z_2,z_1^2}\right]}"] & A_2 \\
\end{tikzcd}\label{res_A2}\eea
is exact as $A_0$-modules. At the level of Grothendieck group we will simply write
\bea A_2=A_1u+s^2u^2A_0-u^3sA_1.\label{rel_A2}\eea
We can do the same for $A_1$
\bea
\begin{tikzcd}[row sep=normal, column sep=normal]
A_1 \arrow[r,tail,"{\left[\displaystyle{\mathop{\vphantom{x}}^{w_2}_{w_1}}\right]}"] & A_0\oplus A_0 \arrow[r,two heads, "{\left[{z_2,z_1}\right]}"] & A_1 \\
\end{tikzcd}\label{res_A1}\eea
and get the relation
\bea A_1=u(s+1)A_0-u^2A_1.\label{rel_A1}\eea
The same is of course true for the other sheaves, not just $A_n$'s.
Here the fact that we have a finite length complex consisting of $A_0,A_1$ resolving the other $A_0$-modules is simply due to $A_0$ having global dimension 2.
But it is also convenient for computations to have resolutions using only free modules i.e. complexes consisting of $A_0$ alone. The resolutions will be of infinite length but quite easily expressed, thanks to the graded structure.
For example, the exact sequence \eqref{res_A1} is periodic, so we can splice the same two term complex into an infinite complex where each term is $A_0\oplus A_0$.
Heuristically we can 'solve' \eqref{rel_A1} and get
\bea A_1=\frac{u(1+s)}{1+u^2}A_0.\label{inf_free_res}\eea
The way to read this equation is to expand $u(1+s)(1+u^2)^{-1}A_0=uA_0+usA_0-u^3A_0-u^2sA_0+\cdots$, where $-$ means the particular summand is placed at an odd cohomological degree.
Similarly for $A_2$
\bea A_2=\frac{u^2(1+s+s^2-u^2s)}{1+u^2}A_0.\nn\eea
One can check the consistency of the relations above by using the explicit the generating function for $A_0$
\bea \chi_{eq}(A_0)=\frac{1+u^2}{(1-u^2s)(1-u^2s^{-1})},~~~\chi_{eq}(A_n)=\frac{(1-s^{n+1})-u^2s(1-s^{n-1})}{(1-s)(1-u^2s)(1-u^2/s)}u^n.\nn\eea
It would not have escaped the reader that this formula comes from $H^{\sbullet}(T^*\BB{P}^1,{\cal O}_{T^*\BB{P}^1})\simeq \oplus_{n\geq0} H^{\sbullet}(\BB{P}^1,\opn{Sym}^nT\BB{P}^1)=\oplus_{n\geq0} H^{\sbullet}(\BB{P}^1,{\cal O}(2n))$, which one can use to get similar formulae for $A_n$.


\smallskip

Going up to $T^*\BB{P}^2$, one can use \eqref{used_III} to reduce the computation of cohomology to $\BB{P}^2$, where one can use the equivariant index theorem.
The equivariant index of a bundle on $\BB{P}^2$ can be computed using the localisation formula
\bea \chi_{eq}(E)=\sum_{x\in {\rm f.p.}}\frac{\opn{ch}_{eq}(E)\opn{Td}_{eq}(T\BB{P}^2)}{e_{eq}(T\BB{P}^2)}\Big|_{x}.\nn\eea
Spelling the above formula out makes it a lot less formidable.
We assign weights $us,ut,u$ to the homogeneous coordinates $[z_1,z_2,z_3]$ of $\BB{P}^2$. In the neighbourhood of the fixed point, say, $x=\{z_1=z_2=0\}$, we use $z_1/z_3$ and $z_2/z_3$ as the local coordinates, with weights $s,t$, then
\bea \opn{Td}_{eq}\big|_x=\frac{st}{(1-s)(1-t)},~~~e_{eq}(T\BB{P}^2)\big|_x=st,\nn\eea
As for $\opn{ch}_{eq}(\opn{Sym}^nT\BB{P}^2)$, at $x$, the tangent fibre coordinates have weights $u^2s^{-1}$ and $u^2t^{-1}$ and so
\bea \oplus_{n\geq0}\opn{ch}_{eq}(\opn{Sym}^nT\BB{P}^2)\big|_x=\frac{1}{1-u^2s^{-1}}\frac{1}{1-u^2t^{-1}}.\nn\eea
Making the appropriate substitutions for the various weights, we can get the contribution from all fixed points.
Summing together these contributions, we get the formula of the index. Note that when combined, the index will have the denominator
\bea D=(1- u^2t/s)(1- u^2s/t)(1- u^2/s)(1- u^2s)(1- u^2/t)(1- u^2t).\nn\eea
The numerator are organised as representations of $GL(3,\BB{C})$
\bea &&D\chi_{eq}(A_0)=(1+u^4)(1+u^2)^2\gl^2-u^4\gl\tiny{\yng(2,1)},\nn\\
&&D\chi_{eq}(A_1)=(u+u^5)(1+u^2)\gl^2\tiny{\yng(1)}-u^3(1+u^2)\gl\tiny{\yng(2,2)},\nn\\
&&D\chi_{eq}(A_2)=u^2(1+u^2+u^4)\gl^2\tiny{\yng(2)}-u^4\gl\tiny{\yng(3,2)}-u^4\gl^2\tiny{\yng(1,1)},\nn\\
&&D\chi_{eq}(A_3)=-u^5\gl\tiny{\yng(4,2)}+u^3\gl^2(1+u^2+u^4)\tiny{\yng(3)}+u^7\gl\tiny{\yng(3,3)}-u^5\gl^2(1+u^4)\tiny{\yng(2,1)}+u^7\gl^3(1+u^2+u^4),\nn\\
&&D\chi_{eq}(A_{-1})=u(1+u^2+u^4+u^6)\gl\tiny{\yng(1,1)}-u^3(1+u^2)\gl\tiny{\yng(2)},\label{massive}\\
&&D\chi_{eq}(A_{-2})=u^2(1+u^2+u^4)\tiny{\yng(2,2)}-u^4\tiny{\yng(3,1)}-u^4\gl\tiny{\yng(1)},\nn\\
&&D\chi_{eq}(A_{-3})=-u^3\gl^{-1}\tiny{\yng(4,2)}+u\tiny{\yng(3)}+\gl^{-1}(u+u^3+u^5)\tiny{\yng(3,3)}-(u^{-1}+u^3)\tiny{\yng(2,1)}+(u^{-3}+u^{-1}+u^1)\gl,\nn\eea
where each Young-diagram represents the corresponding Schur polynomial in the variables $s,t,1$. These relations will be sufficient for the computation of fusion product.

It is perhaps also helpful to plot the cohomologies as a cone in the weight space, albeit this time, there will be non-trivial multiplicities at each given weight.
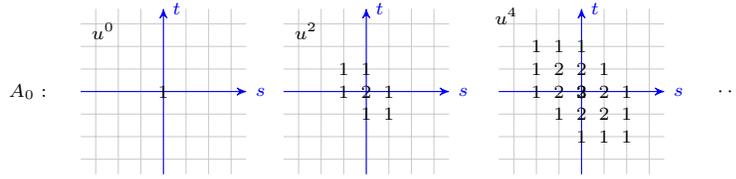
\bea \begin{tikzpicture}[scale=1]
\draw [step=0.3,thin,gray!40] (-1.1,-1.1) grid (1.1,1.1);

\draw [->,blue] (-1.1,0) -- (1.1,0) node[right] {\scriptsize{$s$}};
\draw [->,blue] (0,-1.1) -- (0,1.1) node[right] {\scriptsize{$t$}};

\node at (0,0) {\scriptsize{1}};

\node at (-.8,.8) {\scriptsize{$u^0$}};
\node at (-1.8,0) {\scriptsize{$A_0:$}};
\end{tikzpicture}~
\begin{tikzpicture}[scale=1]
\draw [step=0.3,thin,gray!40] (-1.1,-1.1) grid (1.1,1.1);

\draw [->,blue] (-1.1,0) -- (1.1,0) node[right] {\scriptsize{$s$}};
\draw [->,blue] (0,-1.1) -- (0,1.1) node[right] {\scriptsize{$t$}};

\node at (0,0) {\scriptsize{2}};
\node at (-0.3,0.3) {\scriptsize{$1$}};
\node at (0,0.3) {\scriptsize{$1$}};
\node at (-0.3,0) {\scriptsize{$1$}};
\node at (0.3,0) {\scriptsize{$1$}};
\node at (0,-0.3) {\scriptsize{$1$}};
\node at (0.3,-0.3) {\scriptsize{$1$}};

\node at (-.8,.8) {\scriptsize{$u^2$}};
\end{tikzpicture}~
\begin{tikzpicture}[scale=1]
\draw [step=0.3,thin,gray!40] (-1.1,-1.1) grid (1.1,1.1);

\draw [->,blue] (-1.1,0) -- (1.1,0) node[right] {\scriptsize{$s$}};
\draw [->,blue] (0,-1.1) -- (0,1.1) node[right] {\scriptsize{$t$}};

\node at (0,0) {\scriptsize{2}};
\node at (-0.6,0.6) {\scriptsize{$1$}};
\node at (-0.3,0.6) {\scriptsize{$1$}};
\node at (0,0.6) {\scriptsize{$1$}};
\node at (-0.6,0.3) {\scriptsize{$1$}};
\node at (-0.3,0.3) {\scriptsize{$2$}};
\node at (0,0.3) {\scriptsize{$2$}};
\node at (0.3,0.3) {\scriptsize{$1$}};
\node at (-0.6,0) {\scriptsize{$1$}};
\node at (-0.3,0) {\scriptsize{$2$}};
\node at (0,0) {\scriptsize{$3$}};
\node at (0.3,0) {\scriptsize{$2$}};
\node at (0.6,0) {\scriptsize{$1$}};
\node at (-0.3,-0.3) {\scriptsize{$1$}};
\node at (0,-0.3) {\scriptsize{$2$}};
\node at (0.3,-0.3) {\scriptsize{$2$}};
\node at (0.6,-0.3) {\scriptsize{$1$}};
\node at (0,-0.6) {\scriptsize{$1$}};
\node at (0.3,-0.6) {\scriptsize{$1$}};
\node at (0.6,-0.6) {\scriptsize{$1$}};

\node at (-1,1) {\scriptsize{$u^4$}};
\node at (2,0) {\scriptsize{$\cdots$}};
\end{tikzpicture}.\nn\eea
It should be quite clear what is the monomial at each of the weights. Here the growth pattern as we increase the $u$ power reflects precisely \eqref{used_III}.
If one stacks these pictures together according to the power of $u$, one gets a 3D cone and the multiplicity is controlled by the distance of a lattice point to the boundary of the cone. For more examples, including singular HK varieties, see \cite{Iakovidis:2020znp}.
We will plot the cohomology of ${\cal O}(1)$ and ${\cal O}(2)$, their higher cohomology vanishes too.
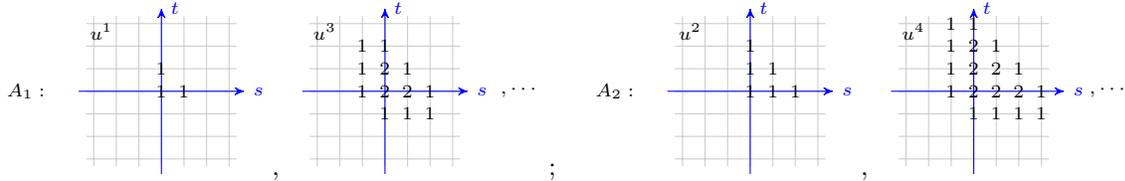
\bea \begin{tikzpicture}[scale=1]
\draw [step=0.3,thin,gray!40] (-1,-1) grid (1,1);

\draw [->,blue] (-1.1,0) -- (1.1,0) node[right] {\scriptsize{$s$}};
\draw [->,blue] (0,-1.1) -- (0,1.1) node[right] {\scriptsize{$t$}};

\node at (0,0) {\scriptsize{1}};
\node at (0.3,0) {\scriptsize{1}};
\node at (0,0.3) {\scriptsize{1}};

\node at (-.8,.8) {\scriptsize{$u^1$}};
\node at (-1.8,0) {\scriptsize{$A_1:$}};
\end{tikzpicture},~~
\begin{tikzpicture}[scale=1]
\draw [step=0.3,thin,gray!40] (-1,-1) grid (1,1);

\draw [->,blue] (-1.1,0) -- (1.1,0) node[right] {\scriptsize{$s$}};
\draw [->,blue] (0,-1.1) -- (0,1.1) node[right] {\scriptsize{$t$}};

\node at (-0.3,0.6) {\scriptsize{1}};
\node at (0,0.6) {\scriptsize{1}};
\node at (-0.3,0.3) {\scriptsize{1}};
\node at (0,0.3) {\scriptsize{2}};
\node at (0.3,0.3) {\scriptsize{1}};
\node at (-0.3,0) {\scriptsize{1}};
\node at (0,0) {\scriptsize{2}};
\node at (0.3,0) {\scriptsize{2}};
\node at (0.6,0) {\scriptsize{1}};
\node at (0,-0.3) {\scriptsize{1}};
\node at (0.3,-0.3) {\scriptsize{1}};
\node at (0.6,-0.3) {\scriptsize{1}};

\node at (-.8,.8) {\scriptsize{$u^3$}};
\node at (1.8,0) {\scriptsize{$,\cdots$}};
\end{tikzpicture};~~~
\begin{tikzpicture}[scale=1]
\draw [step=0.3,thin,gray!40] (-1,-1) grid (1,1);

\draw [->,blue] (-1.1,0) -- (1.1,0) node[right] {\scriptsize{$s$}};
\draw [->,blue] (0,-1.1) -- (0,1.1) node[right] {\scriptsize{$t$}};

\node at (0,0) {\scriptsize{1}};
\node at (0.3,0) {\scriptsize{1}};
\node at (0,0.3) {\scriptsize{1}};
\node at (0.6,0) {\scriptsize{1}};
\node at (0,0.6) {\scriptsize{1}};
\node at (0.3,0.3) {\scriptsize{1}};

\node at (-.8,.8) {\scriptsize{$u^2$}};
\node at (-1.8,0) {\scriptsize{$A_2:$}};
\end{tikzpicture},~~
\begin{tikzpicture}[scale=1]
\draw [step=0.3,thin,gray!40] (-1,-1) grid (1,1);

\draw [->,blue] (-1.1,0) -- (1.2,0) node[right] {\scriptsize{$s$}};
\draw [->,blue] (0,-1.1) -- (0,1.1) node[right] {\scriptsize{$t$}};

\node at (-0.3,0.9) {\scriptsize{1}};
\node at (0,0.9) {\scriptsize{1}};
\node at (-0.3,0.6) {\scriptsize{1}};
\node at (0,0.6) {\scriptsize{2}};
\node at (0.3,0.6) {\scriptsize{1}};
\node at (-0.3,0.3) {\scriptsize{1}};
\node at (0,0.3) {\scriptsize{2}};
\node at (0.3,0.3) {\scriptsize{2}};
\node at (0.6,0.3) {\scriptsize{1}};
\node at (-0.3,0) {\scriptsize{1}};
\node at (0,0) {\scriptsize{2}};
\node at (0.3,0) {\scriptsize{2}};
\node at (0.6,0) {\scriptsize{2}};
\node at (0.9,0) {\scriptsize{1}};
\node at (0,-0.3) {\scriptsize{1}};
\node at (0.3,-0.3) {\scriptsize{1}};
\node at (0.6,-0.3) {\scriptsize{1}};
\node at (0.9,-0.3) {\scriptsize{1}};

\node at (-.8,.8) {\scriptsize{$u^4$}};
\node at (1.8,0) {\scriptsize{$,\cdots$}};
\end{tikzpicture}.\nn\eea
We hope that the growth pattern for $A_1$ and $A_2$ as the power of $u$ increases should be clear.

There are also relations amongst $A_n$ similar to those of \eqref{rel_A2} \eqref{rel_A1} that will be useful for computing the minimal resolution
\bea &A_{-2}(u^3-u^{-3})\gl=A_{-1}(u^2-u^{-2})\tiny{\yng(1,1)}-A_0(u-u^{-1})\tiny{\yng(1)},\label{rel_I}\\
&A_0(u^2-u^{-2})=A_{-1}(u^3-u^{-3})\tiny{\yng(2,1)}-A_{-2}(u^4-u^{-4})\gl\tiny{\yng(1)},\label{rel_II}\\
&A_{-1}(1+u^2+u^4)(1+u^2)\gl=A_{-2}u^3\gl\tiny{\yng(1)}+A_0(-u^3\tiny{\yng(2)}+u(1+u^2+u^4)\tiny{\yng(1,1)}).\label{rel_III}\eea
The last relation can be derived from the first two but is handy to have.
To check these relations, one may multiply both sides by $D$ and use \eqref{massive}. This results in a finite polynomial of $u$ and the relations can be checked order by order in $u$. One can also solve these relations as we did earlier and get
\bea &&A_{-1}=-A_0\frac{u^3(1+u^2)\gl\tiny{\yng(2)}-u\gl(1+u^2)(1+u^4)\yng(1,1)}{(1+u^4)(1+u^2)^2\gl^2-u^4\tiny{\yng(2,1)}\gl},\nn\\
&&A_{-2}=-A_0\frac{u^4\tiny{\yng(3,1)}-u^2(1+u^2+u^4)\tiny{\yng(2,2)}+u^4\gl\tiny{\yng(1)}}{(1+u^4)(1+u^2)^2\gl^2-u^4\gl\tiny{\yng(2,1)}}.\nn\eea
Again this result should be read as saying that one can resolve $A_{-1,-2}$ using only free $A_0$ modules, but one gets an infinite resolution by expanding the rhs in powers of $u$. One can do the same for $A_{1,2}$ by relating them to $A_{-1,-2}$
\bea &A_1=u^{-1}\tiny{\yng(1)}A_0-u^{-2}\tiny{\yng(1,1)}A_{-1}+u^{-3}\gl A_{-2},\nn\\
&A_2=u^{-2}\tiny{\yng(2)}A_0-u^{-3}\tiny{\yng(2,1)}A_{-1}+u^{-4}\gl\tiny{\yng(1)} A_{-2}.\nn\eea


\section{Some more details about the minimal resolution}\label{sec_Smdatmr}
The setting of thm.7.2. of \cite{BUTLER1999323} is: $S$ a separable algebra over $\BB{C}$ (any $S$-module or bi-module is projective), $\SF{X}$ an $S$-bi-module, $\Gc$ the tensor algebra over $S$ of $\SF{X}$.
We will be in the graded setting where $\Gc$ gets its grading from the tensor power. Let $R$ be a homogeneous ideal of $\Gc$ and let $\Gl=\Gc/R$, we also denote with $J$ the kernel of the augmentation $\Gc\to S$.
The minimal resolution below is a resolution of $\Gl$ as a $\Gl$-bi-module. It has terms
\bea\cdots\to \Gl\otimes_S\SF{T}_3\otimes_S\Gl\to \Gl\otimes_S\SF{T}_2\otimes_S\Gl\to \Gl\otimes_S\SF{T}_1\otimes_S\Gl\to \Gl\otimes_S\Gl\to \Gl\to0.\label{min_resolution_app}\eea

We apply this to our quiver algebra
\bea  \begin{tikzpicture}
  \matrix (m) [matrix of math nodes, row sep=.5em, column sep=5em]
    { 2 & 1 & 0 \\        };
  \path[->, font=\scriptsize, bend right = 10] (m-1-1) edge node [below] {\scriptsize{$\vec w\times 3$}} (m-1-2);
  \path[->, font=\scriptsize, bend right = 10] (m-1-2) edge node [below] {\scriptsize{$\vec w\times 3$}} (m-1-3);
  \path[->, blue,font=\scriptsize, bend right = 10] (m-1-2) edge node [above] {\scriptsize{$\vec z\times 3$}} (m-1-1);
  \path[->, blue,font=\scriptsize, bend right = 10] (m-1-3) edge node [above] {\scriptsize{$\vec z\times 3$}} (m-1-2);
\end{tikzpicture}\nn\eea
where each arrow is actually three arrows labelled by $z^i$ or $w_i$. We should really have named the two sets of blue (resp. black) arrows using different letters, but from the starting and ending point of a path one can tell which $z$ (resp. $w$) it is. Another reason is that we will soon imply commutativity relations and so there is no need for extra notations.
The separable algebra $S$ is the $\BB{C}$-algebra generated by the idempotents $e_0,e_1,e_2$, the zero length paths. The $S$-bi-module $\SF{X}$ is generated by the twelve length 1 paths and so $\Gc$ is the path algebra but with no relations. Now we come to the ideal of relations $R$. One part of $R$ is from commutativity: in a path labelled with $z$'s and $w$'s, then one can switch $z,w$ around \emph{provided} it remains a valid path, e.g. $[2\stackrel{z^i}{\leftarrow}\stackrel{z^j}{\leftarrow}0]=[2\stackrel{z^j}{\leftarrow}\stackrel{z^i}{\leftarrow}0]$, $[1\stackrel{z^i}{\leftarrow}\stackrel{w_j}{\leftarrow}1]=[1\stackrel{w_j}{\leftarrow}\stackrel{z^i}{\leftarrow}1]$ and all the commutativity relations generated by them.
The second part is from the moment map condition \eqref{HK_quotient_alg}:
\bea \sum_i[a\stackrel{z^i}{\leftarrow}\stackrel{w_i}{\leftarrow}a]=0~~{\rm for}~a=1,2; ~~~~\sum_i[a\stackrel{w_i}{\leftarrow}\stackrel{z^i}{\leftarrow}a]=0~~{\rm for}~a=0,1.\nn\eea
After imposing the relations $R$, it is not hard to see that $e_a\Gl e_a$ is isomorphic to the algebra $A_0=H^0(T^*\BB{P}^2,{\cal O}_{T^*\BB{P}^2})$ for any $a$, while
$e_{a+n}\Gl e_a$ is isomorphic to the $A_0$ module $A_n=H^0(T^*\BB{P}^2,{\cal O}_{T^*\BB{P}^2}(n))$ for $n=\pm2,\pm1$ and the appropriate $a$. Recall that for these values of $n$ there are no higher cohomologies.

Now we deal with the terms in the resolution \eqref{min_resolution_app}
\bea\SF{T}_1=\SF{X},~~\SF{T}_2=\frac{R\cap J^2}{JR+RJ},~~\SF{T}_3=\frac{JR\cap RJ}{R^2+JRJ},~~\SF{T}_4=\frac{R^2\cap JRJ}{JR^2+R^2J}.\nn\eea
We stop at $\SF{T}_4$ due to the global dimension of $\Gl$ being four.
Also strictly speaking $\SF{T}_3$ is the $S$-bi-module complement of $R^2+JRJ$ in $JR\cap RJ$ and $\SF{T}_4$ is the $S$-bi-module complement of $JR^2+R^2J$ in $R^2\cap JRJ$. Such splits are possible since the algebra $S$ is separable. But we will stick to our lazy notation above.

We have listed in \eqref{T_1234} in the main text all generators of $\SF{T}_{1,2,3,4}$.
Here we will recall the maps between the terms of \eqref{min_resolution_app}. Eventually we will only take alternating sums of the complex and the maps are not essential, but we include them here for completeness. The maps from $\SF{T}_{1,2}$ are standard
\bea \Gl\otimes_S\SF{T}_1\otimes_S\Gl \ni 1\otimes_Sx\otimes_S1 \stackrel{\partial_1}{\to} x\otimes_S1-1\otimes_Sx\in \Gl\otimes_S\Gl,\nn\eea
\bea \Gl\otimes_S\SF{T}_2\otimes_S\Gl \ni 1\otimes_Sxy\otimes_S1 \stackrel{\partial_2}{\to} x\otimes_Sy\otimes_S1+1\otimes_Sx\otimes_Sy\in \Gl\otimes_S\SF{T}_1\otimes_S\Gl\nn\eea
and extended to all of $\Gl\otimes_S\SF{T}_{1,2}\otimes_S\Gl$ as a $\Gl$-bi-module map.
The other maps are formulated in \cite{BUTLER1999323} using bi-module derivations. We will not go into the general theory but use examples to illustrate the maps.
By definition, a length 3 path in $\SF{T}_3$ can be written in two ways $p=xr=r'x'$ where $x,x'$ are two length 1 paths and $r,r'$ are two length 2 relations. Then $\partial_3(1\otimes_Sp\otimes_S1)$ reads
\bea \Gl\otimes_S\SF{T}_3\otimes_S\Gl \ni 1\otimes_Sp\otimes_S1 \stackrel{\partial_3}{\to} x\otimes_Sr\otimes_S1-1\otimes_Sr'\otimes_Sx'\in \Gl\otimes_S\SF{T}_2\otimes_S\Gl,\nn\eea
and to see that $\partial_2\partial_3=0$, one needs only remember that product in $\Gc$ is also just tensor product over $S$.
Take for example $p=e_1(z^jw_kz^k-w_kz^jz^k+w_kz^kz^j)e_0$, then its image in will be
\bea \partial_3(1\otimes_S p\otimes_S 1)=z^j\otimes_S(w_kz^k)\otimes_S1-w_k\otimes_S[z^j,z^k]\otimes_S1-1\otimes_S[z^j,w_k]\otimes_Sz^k-1\otimes_Sw_kz^k\otimes_Sz^j.\nn\eea
Finally a path $q$ of length 4 in $\SF{T}_4$ is written in two ways $q=rr'=xr''y$ with $x,y$ of length 1 and $r,r',r''$ being relations. We have the map
\bea \Gl\otimes_S\SF{T}_4\otimes_S\Gl \ni 1\otimes_Sq\otimes_S1 \stackrel{\partial_4}{\to} x\otimes_Sr''y\otimes_S1+1\otimes_Sxr''\otimes_S y\in \Gl\otimes_S\SF{T}_3\otimes_S\Gl.\nn\eea
Though it is not obvious that $r''y$ or $xr''$ belongs to $\SF{T}_3$, it is guaranteed thanks to prop.A.1 (b) of \cite{BUTLER1999323}.
Let us look at an example $q=e_2(z^{[k}z^{l]}w_{[k}w_{l]}-2z^kw_kz^lw_l)e_2$, then
\bea \partial_4(1\otimes_Sq\otimes_S1)=2z^k\otimes_S([z^l,w_k]w_l-z^lw_lw_k)\otimes_S1+2\otimes_S(z^k[z^l,w_k]-z^lz^kw_k)\otimes_Sw_l\nn\eea
and each parenthesis is manifestly in $JR\cap RJ$ and so the rhs is in $\SF{T}_3$.

\smallskip

We will explain next why some of the apparently valid terms are excluded from the $\SF{T}_i$'s. For example $e_2z^{[i}w_kz^{j]}e_1$ does belong to $R$, but it can also be written as
\bea e_2z^{[i}w_kz^{j]}e_1=e_2z^{[i}[w_k,z^{j]}]e_1+e_2z^{[i}z^{j]}w_ke_1\in JR+RJ\nn\eea
and hence projected out in the quotient $\SF{T}_2=R/(JR+RJ)$. In fact one can show that all generators of $\SF{T}_2$ are of length 2.

Another less obvious one is $e_2z^{[i}w_p z^j z^{k]}e_0$, the total anti-symmetrisation on $i,j,k$ makes it a member of $RJ\cap JR$ and so potentially a member of $\SF{T}_3$. But note that
\bea \ep_{ijk}z^{[i}w_p z^j z^{k]}=\ep_{pij}z^iw_k[z^j,z^k]+\underbrace{\ep_{jkp}(z\cdotp w)z^jz^k}_{R^2}.\nn\eea
We investigate the first term
\bea e_2(z^{i}w_kz^{j}z^k-z^{i}w_kz^kz^{j})e_0-[i\leftrightarrow j]=e_2z^{[i}w_kz^{j]}z^ke_0-\underbrace{e_2(z^{i}w_kz^kz^{j}-z^{j}w_kz^kz^{i})e_0}_{JRJ}.\nn\eea
In the first term one can switch the middle $z,w$ mod $JRJ$ and get $e_2z^{[i}z^{j]}w_kz^ke_0\in R^2$.
So the entire term is projected out in the quotient $\SF{T}_3=(JR\cap RJ)/(R^2+JRJ)$.

The most subtle term is perhaps $e_1z^{[i}w_{[k}z^{j]}w_{l]}e_1$ which is potentially a member of $\SF{T}_3$. We first show that $e_1z^{[p}w_{[p}z^{i]}w_{j]}e_1$ is in $R^2+JRJ$
\bea e_1z^{[p}w_{[p}z^{i]}w_{j]}e_1=e_1(z^pw_pz^iw_j-\underbrace{z^iw_pz^pw_j}_{JRJ}-z^pw_jz^iw_p+z^iw_jz^pw_p)e_1.\nn\eea
We add to the above $e_1[z^p,w_j][z^i,w_p]e_1\in R^2$ (we use $\sim$ to mean equal modulo $R^2+JRJ$)
\bea e_1z^{[p}w_{[p}z^{i]}w_{j]}e_1&\sim&e_1(z^pw_pz^iw_j-w_jz^pz^iw_p-z^pw_jw_pz^i+z^iw_jz^pw_p)e_1\nn\\
&\sim& e_1(z^pw_pz^iw_j-w_jz^iz^pw_p-z^pw_pw_jz^i+z^iw_jz^pw_p)e_1\nn\\
&=&e_1(z^pw_p[z^i,w_j]-[w_j,z^i]z^pw_p)e_1\sim0.\nn\eea
This result shows that $e_1z^{[i}w_{[k}z^{j]}w_{l]}e_1\sim 0$ if the cardinality of the index set $\{i,j,k,l\}$ is 3, e.g. $i=k$ but $i,j,l$ are distinct. Indeed $e_1z^{[1}w_{[1}z^{2]}w_{3]}e_1=e_1z^{[p}w_{[p}z^{2]}w_{3]}e_1\sim0$ from what we have shown. When the cardinality of $\{i,j,k,l\}$ is two, say $i=k=1$ and $j=l=3$, then
\bea 0&\sim& \sum_pe_1z^{[p}w_{[p}z^{2]}w_{2]}e_1=e_1\big(z^{[1}w_{[1}z^{2]}w_{2]}+z^{[3}w_{[3}z^{2]}w_{2]}\big)e_1\nn\\
&\sim& e_1\big(-z^{[1}w_{[1}z^{3]}w_{3]}-z^{[3}w_{[3}z^{1]}w_{1]}\big)e_1=-2e_1z^{[1}w_{[1}z^{3]}w_{3]}e_1,\nn\eea
exactly what we want to show. Therefore we proved that $e_1z^{[i}w_{[k}z^{j]}w_{l]}e_1\in R^2+JRJ$ and so excluded from $\SF{T}_3$.
In contrast the closely related term $e_2z^{[i}w_{[k}z^{j]}w_{l]}e_2$ is much easily excluded
\bea e_2z^{[i}w_{[k}z^{j]}w_{l]}e_2=e_2z^{[i}[w_{[k},z^{j]}]w_{l]}e_2+e_2z^{[i}z^{j]}w_{[k}w_{l]}e_2\in JRJ+R^2.\nn\eea
In the end the minimal resolution complex is homogeneous in the sense that $\SF{T}_i$ has terms of length $i$ only. It was misstated in an earlier version of the draft that there are terms of length greater than $i$ in $\SF{T}_i$. It is possible that by using an induction, one can show that this is a general feature for the minimal resolution even for the case of $T^*\BB{P}^n$.

\smallskip

From the minimal resolution, we immediately get a resolution of the diagonal in the sense reviewed in the main text by tensoring the tilting bundle and its dual to the left and right of the minimal resolution \eqref{hope_less_II}
\bea 0\to T\otimes_S \SF{T}_4\otimes_ST^{\vee}\to T\otimes_S \SF{T}_3\otimes_ST^{\vee}\to T\otimes_S \SF{T}_2\otimes_ST^{\vee}\to T\otimes_S\SF{X}\otimes_ST^{\vee}\to T\otimes_ST^{\vee}\to0.\nn\eea
We can spell out the complex a bit. To avoid clutter, we will denote each term in $\SF{T}_i$ by its representation under $GL(3,\BB{C})$. We lose of course all information about the maps of the complex in this way, but as we will only care about the Grothendieck group, this is justified. We get a complex resolving the diagonal
\bea &&\to u^2\left[
           \begin{array}{c}
             {\cal O}(-1)\otimes 2 \otimes {\cal O}(1) \\
             {\cal O}(-2)\otimes 1 \otimes {\cal O}(2) \\
             {\cal O}(0)\otimes 1 \otimes {\cal O}(0) \\
             {\cal O}(-1)\otimes \gl^{-1}{\tiny\yng(2,1)} \otimes {\cal O}(1) \\
                  {\cal O}(-2)\otimes {\tiny\yng(1,1)}\otimes {\cal O}(0) \\
             {\cal O}(0)\otimes\gl^{-1} {\tiny\yng(1)}\otimes {\cal O}(2) \\
           \end{array}\right] \to u\left[
           \begin{array}{c}
             {\cal O}(-1)\otimes{\tiny \yng(1)}\otimes {\cal O}(0) \\
             {\cal O}(-2)\otimes {\tiny\yng(1)} \otimes {\cal O}(1) \\
             {\cal O}(0)\otimes \gl^{-1}{\tiny\yng(1,1)}\otimes {\cal O}(1) \\
             {\cal O}(-1)\otimes \gl^{-1}{\tiny\yng(1,1)}\otimes {\cal O}(2) \\
           \end{array}\right]\to \left[
           \begin{array}{c}
             {\cal O}(0)\otimes {\cal O}(0) \\
             {\cal O}(-1)\otimes {\cal O}(1) \\
             {\cal O}(-2)\otimes {\cal O}(2) \\
           \end{array}\right]\to0\nn\\
&&0\to u^4\left[
           \begin{array}{c}
             {\cal O}(-2)\otimes 1\otimes {\cal O}(2) \\
             {\cal O}(0)\otimes 1\otimes {\cal O}(0) \\
             {\cal O}(-1)\otimes 1 \otimes {\cal O}(1)
                        \end{array}\right]\to u^3\left[
           \begin{array}{c}
             {\cal O}(-2)\otimes {\tiny\yng(1)}\otimes {\cal O}(1) \\
             {\cal O}(0)\otimes \gl^{-1}{\tiny\yng(1,1)} \otimes {\cal O}(1) \\
             {\cal O}(-1)\otimes{\tiny\yng(1)}\otimes {\cal O}(0) \\
             {\cal O}(-1)\otimes \gl^{-1}{\tiny\yng(1,1)} \otimes {\cal O}(2)
           \end{array}\right]\to\nn\\
            \label{tilting_complex_P2}\eea
where the power of $u$ keeps track of the path lengths.
If we tensor any sheaf $F$ to the left of the complex and take derived section, we will get a complex involving ${\cal O}_{T^*\BB{P}^2}(0,1,2)$ that is quasi-isomorphic to $F$. We then take alternating sum and get $F$ as a linear combination of ${\cal O}(0,1,2)$, with coefficients in the polynomials in the equivariant parameters.
We will only need $F={\cal O}(n)$ for our purpose. We will abuse notation and write $A_n$ for the derived section $\BB{R}\Gc({\cal O}_{T^*\BB{P}^2}(n))$ as well as its equivariant character
\bea {\cal O}(n)=\Big(A_n(1+u^2+u^4)-A_{n-1}(u+u^3){\tiny\yng(1)}+A_{n-2}u^2{\tiny\yng(1,1)}\Big){\cal O}(0),\nn\\
+\Big(A_{n-1}(1+2u^2+u^2\gl^{-1}{\tiny\yng(2,1)}+u^4)-A_{n-2}(u+u^3){\tiny\yng(1)}-A_n\gl^{-1}{\tiny\yng(1,1)}(u+u^3)\Big){\cal O}(1)\nn\\
+\Big(A_{n-2}(1+u^2+u^4)-A_{n-1}(u+u^3)\gl^{-1}{\tiny\yng(1,1)}+A_nu^2\gl^{-1}{\tiny\yng(1)}\Big){\cal O}(2)\nn\eea
One can drastically simplify the parentheses above. For example setting $n=0$, the coefficient of ${\cal O}(2)$ on the rhs is
$A_{-2}(1+u^2+u^4)-A_{-1}(u+u^3)\gl^{-1}{\tiny\yng(1,1)}+A_0u^2\gl^{-1}{\tiny\yng(1)}$ which vanishes thanks to the relation \eqref{rel_II}.
On can likewise check using those relations that by plugging in $n=0,1,2$, one gets the same thing back.
Finally we set $n=3,-1$ and use \eqref{massive} for the $A_n$'s the above formula yields
\bea &{\cal O}(3)=u^3\gl{\cal O}(0)-u^2{\tiny\yng(1,1)}{\cal O}(1)+u{\tiny\yng(1)}{\cal O}(2)\nn\\
&{\cal O}(-1)=(u\gl)^{-1}{\tiny\yng(1,1)}{\cal O}(0)-(u^2\gl)^{-1}{\tiny\yng(1)}{\cal O}(1)+(u^3\gl)^{-1}{\cal O}(2).\nn\eea

\section{Rational cones and rational surgeries}\label{sec_RCaRS}
We will need some basics of surgery on 3,4-manifolds, a general reference would be \cite{gompf19994}.

The lens spaces are defined as a $\BB{Z}/p\BB{Z}$ quotient of $S^3$
\bea \{(z_1,z_2)||z_1|^2+|z_2|^2=1\}/\sim,~~[z_1,z_2]\sim [z_1\zeta,z_2\zeta^q],~~\zeta=e^{2\pi i/p},~~p,q>0,~~\gcd(p,q)=1.\nn\eea
We will review how this space is obtained from surgery on $S^1\times S^2$ or $S^3$. First we divide the geometry into two solid tori:
\bea {\rm solid\;torus\;I}=\{z_1\neq 0\},~~~{\rm solid\;torus\;II}=\{z_2\neq 0\}.\nn\eea
On $I,II$ we pick $\BB{Z}_p$-invariant coordinates for $D^2\times S^1$ as:
\bea I:~\psi=\arg z_1^p,~~u=z_1^{-q}z_2;~~~~~II:~\phi=\arg z_2^p,~~v=z_1z_2^s,\nn\eea
where we picked integers $r,s$ so that $pr-qs=1$. The coordinates $u,v$ parameterise the disc while $\psi,\phi$ the $S^1$. It will be important for us to know the weight of the action of $\partial_{\gt}$ (which rotates $z_1,z_2$ simultaneous by a phase:
\bea I:~~\partial_{\gt}=p\partial_{\psi}+(1-q)\partial_{\arg u};~~~~II:~\partial_{\gt}=p\partial_{\phi}+(1+s)\partial_{\arg v}.\label{weight_theta}\eea
These weights are important for two reasons. First,
since we will insert Wilson loops along the $\partial_{\gt}$ orbit, knowing how the orbit winds round the two 1-cycles gives the framing of the Wilson loop. Second, in the quantisation procedure outlined in sec.\ref{sec_HsWlar}, we need to pick a polarisation for the 1-form $\chi$ field, this polarisation is fixed by the $\partial_{\gt}$ orbit.

From the change of coordinates $(\psi,u)\to (\phi,v)$, one can easily see how the 1-cycles are mapped as one glues together the two solid tori. The result of the simple analysis says: the lens space can be obtained from $S^2\times S^1$ with a $q/p$ rational surgery along one $S^1$ fibre or from a $-p/q$ surgery along an unknot in $S^3$.
For readers unfamiliar with the rational surgeries, a $q/p$ surgery means that we remove a solid torus (which is the tubular neighbourhood of a knot) from a 3-manifold $M$ and glue back a new one, so that $q$ times the old meridian plus $p$ times the old longitude becomes the new meridian. The new longitude is left unfixed here, that is, giving the ratio $q/p$ leaves the framing of $L(p,q)$ ambiguous.
We recall that a framing is the trivialisation of the tangent bundle, but oftentimes it is more convenient to work with 2-framing, which is the trivialisation of $TM\oplus TM$.  In Chern-Simons theory, the framing is responsible for the famous shift of the Chern-Simons level $k\to k+h$. We will see here that in RW theory the framing would shift the last term in \eqref{susy_action}. We mentioned that this term has the interpretation of an odd Coulomb branch parameter, and so this shift is in agreement with the Chern-Simons theory.

The new meridian $[q;p]$ after the surgery is the first column of an $SL(2,\BB{Z})$ matrix. Specifying the second column means specifying what is the new longitude in terms of the old ones.
This amounts to picking $r,s$ as we did above, which we emphasise is not unique.
However for an integer surgery $k/1$, we can pick the second column as $[-1;0]$ giving us the $SL(2,\BB{Z})$ matrix $T^kS$
\bea S=\left[
         \begin{array}{cc}
           0 & -1 \\
           1 & 0 \\
         \end{array}\right],~~~T=\left[
         \begin{array}{cc}
           1 & 1 \\
           0 & 1 \\
         \end{array}\right],~~~T^kS=\left[
         \begin{array}{cc}
           k & -1 \\
           1 & 0 \\
         \end{array}\right].\nn\eea
Thus with an integer surgery $k/1$, we can remove the framing ambiguity.
A deeper reason for this is that: a closed 3-manifold bounds a 4-manifold and we can build the latter from $D^4$ by attaching handles. When one only cares about the resulting 3-manifold boundary, then one needs only the 2-handles (see chapter 5 \cite{gompf19994}). The $T^kS$ surgery corresponds to a 2-handle with framing $k$ and the resulting 3-manifold would inherit a framing from before the handle attachment.
Since any $SL(2,\BB{Z})$ matrix can be decomposed as products of $T^kS$, any rational surgery can be decomposed as a series of integer ones.
The framing change after these surgeries has been worked out in \cite{freed1991}. It is intimately related to the Dedekind function as we shall see later.

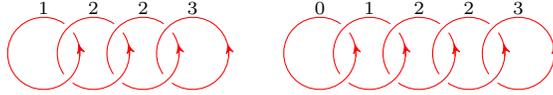
\begin{figure}[h]
\begin{center}
\begin{tikzpicture}[scale=0.6]
\centerarc[red](3.2,0)(235:575:0.8);\centerarc[red,->](3.2,0)(-5:15:0.8);
\centerarc[red](2.1,0)(55:215:0.8); \centerarc[red](2.1,0)(235:395:0.8); \centerarc[red,->](2.1,0)(-5:15:0.8);
\centerarc[red](1,0)(55:215:0.8); \centerarc[red](1,0)(235:395:0.8); \centerarc[red,->](1,0)(-5:15:0.8);
\centerarc[red](-0.1,0)(55:395:0.8); \centerarc[red,->](-0.1,0)(-5:15:0.8);
\node at (3.2,1) {\scriptsize{3}};
\node at (2.1,1) {\scriptsize{2}};
\node at (1,1) {\scriptsize{2}};
\node at (-0.1,1) {\scriptsize{1}};
\end{tikzpicture}
~~~~\begin{tikzpicture}[scale=0.6]
\centerarc[red](3.2,0)(235:575:0.8); \centerarc[red,->](3.2,0)(-5:15:0.8);
\centerarc[red](2.1,0)(55:215:0.8); \centerarc[red](2.1,0)(235:395:0.8); \centerarc[red,->](2.1,0)(-5:15:0.8);
\centerarc[red](1,0)(55:215:0.8); \centerarc[red](1,0)(235:395:0.8); \centerarc[red,->](1,0)(-5:15:0.8);
\centerarc[red](-0.1,0)(55:215:0.8); \centerarc[red](-0.1,0)(235:395:0.8); \centerarc[red,->](-0.1,0)(-5:15:0.8);
\centerarc[red](-1.2,0)(55:395:0.8); \centerarc[red,->](-1.2,0)(-5:15:0.8);
\node at (3.2,1) {\scriptsize{3}};
\node at (2.1,1) {\scriptsize{2}};
\node at (1,1) {\scriptsize{2}};
\node at (-0.1,1) {\scriptsize{1}};
\node at (-1.2,1) {\scriptsize{0}};
\end{tikzpicture}
\caption{Left: Decomposition of 2/7 surgery along $S^1$ fibre in $S^1\times S^2$, right: decomposition of $-7/2$ surgery along an unknot in $S^3$.}\label{fig_surgery_link}
\end{center}
\end{figure}

We first deal with how to decompose the $q/p$ surgery via continued fraction. Write first $q=k_0p-r_1$ with $0<r_1<p$ i.e. $q/p=k_0-1/(p/r_1)$. We now call $p=r_0$ and write $p=r_0=k_1r_1-r_2$ and continue
\bea \frac{q}{p}=k_0-\frac{r_1}{p}=k_0-\frac{1}{\frac{p}{r_1}}=k_0-\frac{1}{\displaystyle{k_1-\frac{1}{\frac{r_1}{r_2}}}}=\cdots=k_0-\frac{1}{\displaystyle{\cdots -\frac{1}{k_{t-1}-\frac{1}{k_t}}}}.\label{continued_frac}\eea
With this continued fraction, the $q/p$ surgery is decomposed as
\bea
T^{k_0}S\,T^{k_1}S\,\cdots T^{k_t}S\label{decomp_SL2}\eea
In fact reading from right to left starting with $T^{k_t}S$, the ratio of the two entries of the left column is $k_t/1$, then successively multiplying with $T^{k_{t-1}}S$, the ratio becomes $k_{t-1}-1/k_t$ and so on, exactly reflecting the continued fraction.
For example decomposing $2/7$ gives a continued fraction with $(k_0,k_1,k_2,k_3)=(1,2,2,3)$.
One can visualise such a decomposition as a chain of linked circles as in fig.\ref{fig_surgery_link}.
That the circles are linked in this way reflects simply the fact that in \eqref{decomp_SL2}, there is an $S$ in between two $T^{k_i}$.

\begin{figure}[h]
\begin{center}
\begin{tikzpicture}[scale=1]
\draw [step=0.3,thin,gray!40] (-.4,-.4) grid (2,1.8);
\draw [->,thick] (-.5,0) -- (2,0) node [right] {\scriptsize$b$};
\draw [->,thick] (0,-0.4) -- (0,2) node [left] {\scriptsize$0$};
\draw [-,dashed,blue] (0,0) -- (.3,1.8) node [above] {\scriptsize$1$};
\draw [-,dashed,blue] (0,0) -- (.6,1.2) node [above] {\scriptsize$2$};
\draw [-,dashed,blue] (0,0) -- (1.2,1.5) node [above] {\scriptsize$3$};

\draw [-,thick] (0,0) -- (1.8,1.2) node[right] {\scriptsize{$(p,q)$}};

\end{tikzpicture}\caption{Subdivision by adding new rays, the slopes of the rays are $k_0/1$, $k_0-1/k_1$, $k_0-1/(k_1-1/k_2)$, ... corresponding to the continued fraction, and $q/p$ is equal to the rhs of \eqref{continued_frac}.}\label{fig_sub_div}
\end{center}
\end{figure}
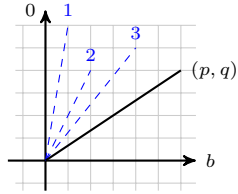

Now we subdivide the cone in fig.\ref{fig_lattice_pq}. We denote with $\vec v_0=[0;1]$ the vertical ray in fig.\ref{fig_sub_div}, we then write $q=k_0p-r_1$ with $0<r_1<p$, and add a ray
$\vec v_1=[1;k_0]=(T^{k_0}S)^{-1}\vec v_0$, which clearly form an $SL(2,\BB{Z})$ basis with $\vec v_0$.
Due to our choice $[p;q]\times\vec v_1<[p;q]\times\vec v_0$.
Continuing let $p=k_1r_1-r_2$ with $0<r_2<r_1$, we add $\vec v_2=(T^{k_1}ST^{k_0}S)^{-1}\vec v_0$ so that $[p;q]\times\vec v_2<[p;q]\times\vec v_1$, the process stops when the cross product becomes 1.
If in an intermediate step one chooses $r_{i-1}=k_ir_i-r_{i+1}$ but with $r_{i+1}>r_i$, that means the newly added ray 'doubles back' toward $\vec v_0$, but the formalism is such that such doubling back will not affect the result.
The process stops at step $t$, so that the last ray $\vec v_{t+1}$ added
\bea \vec v_{t+1}=(T^{k_t}S\cdots T^{k_0}S)^{-1}\vec v_0=[p;q]\nn\eea
is in fact the second ray $[p;q]$ of the original cone.

With the newly added rays, we can perform the product over the lattice points within successive segments bounded by the rays $\vec v_{s+1},\vec v_s$. This will give an $S_2$ function per segment. The lattice points precisely along the rays $\vec v_1,\vec v_2,\cdots \vec v_t$ are either over- or under-counted, their product will produce an $S_1$ function which happens to be the same as the usual sine function.

To write down the product, we denote with
\bea \vec \go_0=[(1-q)/p,1]\nn\eea
the $1/p^{th}$ of the weight of the $\partial_{\gt}$ action, see the left one in \eqref{weight_theta}.
Since $\vec v_{s+1}\times\vec v_s=1$, the product over lattice points between $\vec v_{s+1},\vec v_s$ is the standard $S_2$ function after an $SL(2,\BB{Z})$ change of basis
\bea S_2(\frac{ix}{2\pi}|\vec \go_0\cdotp\vec v_s,\vec \go_0\cdotp \vec v_{s+1}).\nn\eea
Rather than keeping $\vec\go_0$ fixed and letting the $T^kS$ act on $\vec v_0$, we can instead keep $\vec v_0$ and let $T^kS$ act on $\vec \go_0$
\bea \vec \go_0\cdotp\vec v_s=\vec \go_s\cdotp\vec v_0,~~\vec \go_0\cdotp\vec v_{s+1}=\vec \go_{s+1}\cdotp\vec v_0,~~{\rm where}~~\vec \go_s=\vec \go_0(T^{k_0}S)^{-1}\cdots(T^{k_{s-1}}S)^{-1}\nn\eea
This has the following interpretation. Start from the solid torus I: $z_1\neq 0$, where $\partial_{\gt}$ goes along $p\vec \go_0=[1-q,p]$ with the second entry denoting the longitude, we apply the $k_0/1$ surgery $T^{k_0}S$, then from the perspective of the new solid torus $\partial_{\gt}$ would go along $p\vec \go_1$. Keep applying the surgeries $k_1/1,k_2/1$, we eventually reach $p\vec \go_{t+1}=p\vec \go_0[-s,p;-r,q]=[-1-s,p]$. This solid torus is going to be glued back-to-back to the solid torus II: $z_2\neq 0$. The back-to-back gluing flips the sign of the meridian: $[-1-s,p]\to [1+s,p]$, and this matches the right one of \eqref{weight_theta} valid in the solid torus II.

When we juxtapose two neighbouring segments together, the lattice points along the shared ray are over/under counted.
The over/under counting is encoded by the $S_1$ function
\bea {\rm along}\;\vec v_{s+1}:~~\frac{1}{ix}S_1(\frac{ix}{2\pi}|\vec \go_{s+1}\cdotp\vec v_0)=\frac{2}{x}\sinh \frac{x}{2\vec \go_{s+1}\cdotp\vec v_0},~~~~{\rm along}\;\vec v_s:~~\frac{1}{ix}S_1(\frac{ix}{2\pi}|\vec \go_s\cdotp\vec v_0)=\frac{2}{x}\sinh\frac{x}{2\vec \go_s\cdotp \vec v_0}.\nn\eea
To eliminate the over/under counting we can consistently assign the lattice points on a ray to the segment to its left or right. As there is no clear advantage for either choice, we take the Solomonic option, that is, we split each lattice point on the ray in half, giving one half to each side. Concretely this means for each segment we get
\bea \frac{1}{S_2(\frac{ix}{2\pi}|\vec \go_s\cdotp\vec v_0,\vec \go_{s+1}\cdotp\vec v_0)}\big(\frac{2}{x}\sinh \frac{x}{2\vec \go_s\cdotp\vec v_0}\cdotp
\frac{2}{x}\sinh \frac{x}{2\vec \go_{s+1}\cdotp\vec v_0}\big)^{1/2}.\label{used_IV}\eea
Note since $x$ is the Chern-root and is of degree 2, what is in the square root is thus a polynomial with constant term $(\vec\go_s\cdotp\vec v_0)^{-1}(\vec\go_{s+1}\cdotp\vec v_0)^{-1}>0$. That the constant term of a polynomial is positive means we can take the square root without ambiguity.

In view of the similarity between subdividing a rational cone and decomposing a rational surgery, we propose that the factor \eqref{used_IV} be interpreted as $T^{k_s}S$.
That is, we introduce a new function suggestively denoted as $T^k\kern-0.2em S(x|\vec \go)$
\bea T^k\kern-0.2em S(x|\vec \go)=\frac{ix}{S_2(\frac{ix}{2\pi}|\vec \go (T^kS)\vec v_0,\vec \go\cdotp\vec v_0)}\big(\frac{2}{x}\sinh \frac{x}{2\vec \go(T^kS)\vec v_0}
\cdotp\frac{2}{x}\sinh \frac{x}{2\vec \go\cdotp\vec v_0}\big)^{1/2}\label{T^kS}\eea
where $\vec\go\cdotp\vec v_0$ simply picks out the second entry of $\vec \go$.

With the notation \eqref{T^kS}, we can now write the $S_2^C$ function as simply the product
\bea \frac{ix}{S^C_2(\frac{ix}{2\pi}|(1-q)/p,1)}=\big(\prod_{s=0}^tT^{k_s}\kern-0.2em S(x|\vec\go_{s+1})\big)\,\big(\frac{2}{x}\sinh \frac{x}{2\vec \go_0\cdotp\vec v_0}
\cdotp\frac{2}{x}\sinh \frac{x}{2\vec \go_{t+1}\cdotp\vec v_0}\big)^{-1/2}.\nn\eea
The two hanging square root factors come from the first ray and last ray and will combine neatly with the Chern-Character to form the Mukai-vector. Indeed, from the explicit $\vec\go_0$ and $\vec\go_{t+1}$, the two inner products are 1 and so the two factors go to the $\hat A$ genus
\bea \hat A(x)=\frac{x/2}{\sinh{x/2}}.\nn\eea
In particular when the sum of Chern-roots are zero, the $\hat A$ collapses to $\opn{Td}(x)$ and we combine $\sqrt{\opn{Td}_X}$ with $\opn{ch}(E)$ to give the Mukai vector $v([E])$.

We propose that to a Wilson loop labelled with vector bundle $E$ and a general framing $\vec\nu=[\nu_1,\nu_2]$, one should assign
\bea |\Psi(E,\vec\nu)\ket \to \big(\frac{2}{x}\sinh\frac{x}{2\nu_2}\big)^{-1/2}\opn{ch}(E)e^{i\nu_1\go}\nn\eea
where $\go$ is now the K\"ahler form of the HyperK\"ahler target and this is why we switched from $\vec\go$ to $\vec \nu$ to denote the framing.
Furthermore for integer surgeries we assign
\bea &&\begin{tikzpicture}[scale=0.6]
\centerarc[red](3.2,0)(0:360:0.8); \centerarc[red,->](3.2,0)(-5:15:0.8);
\node at (3.2,1) {\scriptsize{$k$}};
\end{tikzpicture}\sim T^k\kern-0.2em S(x|\vec\nu),\nn\\
&&
\begin{tikzpicture}[scale=0.6]
\centerarc[red](3.2,0)(235:575:0.8); \centerarc[red,->](3.2,0)(-5:15:0.8);
\centerarc[red](2.1,0)(55:395:0.8); \centerarc[red,->](2.1,0)(-5:15:0.8);
\node at (3.2,1) {\scriptsize{$k_1$}};
\node at (2.1,1) {\scriptsize{$k_0$}};
\end{tikzpicture}
\sim T^{k_0}\kern-0.2em S(x|\vec \nu T^{k_1}S)T^{k_1}\kern-0.2em S(x|\vec \nu).\label{composing T^kS}\eea
One has to be a bit cautious in applying the formulae when the cone $C$ is degenerate, e.g. composing $S$ with itself or $TS$ with itself twice would result in a cone $C$ that spans $180^{\circ}$. For such cones, the $S_2^C$ is ill defined. So one cannot directly check that $S^2=(TS)^3=-1$, unfortunately.
Another unsavoury aspect of the definition of $T^k\kern-0.2em S$ is that if we assume $x_i=-x_{i+n}$ as in \eqref{eigen_val_paired} due to the HK property, then $T^k\kern-0.2em S$ collapses to 1. All
that is left is the Mukai vector from the Wilson loops and we recover the boring result of \eqref{Z_Lpq_simple}.

\smallskip

Finally we deal with the 2-framing. When a 3-manifold $M$ is obtained from $S^3$ (with canonical 2-framing) via a surgery along a link, Freed and Gompf gave a formula for the difference between the 2-framing of $M$ and its canonical 2-framing \cite{freed1991}. We will not need the most general formula since our link is always going to look like the right panel of fig.\ref{fig_surgery_link}. Suppose that the components of the link are marked with $a_0,a_1,\cdots, a_n$, then we let $L_{\{a\}}$ be an integer matrix whose diagonal is $a_0,a_1,\cdots a_n$ and whose first off diagonals are all 1. Then the 2-framing shift is
\bea {}^2\opn{Fr}=-3\gs(L_{\{a\}})+\Tr[L_{\{a\}}]\label{framing_shift}\eea
with $\gs(L_{\{a\}})$ being the signature of $L_{\{a\}}$. According to thm.1.12 of \cite{Framing_Dedekind}, \eqref{framing_shift} can be written using the Rademacher $\phi$ function, which involves the Dedekind sum
\bea &&\hspace{1.5cm}{}^2\opn{Fr}=\phi(B),~~~B=ST^{a_0}S\cdots T^{a_n}S,\nn\\
&&\phi\big(\left[
       \begin{array}{cc}
         a & b \\
         c & d \\
       \end{array}\right]\big)=\bigg\{
                                        \begin{array}{cc}
                                          b/d & c=0 \\
                                          (a+d)/c-12\opn{sgn}(c)s(a,c) & c\neq 0 \\
                                        \end{array}.\nn\eea
\begin{example}\label{ex_framing_lens}
First the $L(7,2)$ has 2-framing shift $-1$ using the right panel of fig.\ref{fig_surgery_link}.

For $L(p,1)$, we can choose different surgeries. We can naturally write $1/p$ as $(1,2,\cdots,2)$ where there are $p-1$ $2$'s.
This means we have the integer surgery $(0,1,2,\cdots,2)$ and $B=SS(TS)(T^2S)^{p-1}=-[1,-1;p,1-p]$.
Computing the framing using \eqref{framing_shift} we get ${}^2\opn{Fr}=-3(p-1)+(p-1)\times 2+1=2-p$. Or if we use the $\phi$ function we get
\bea {}^2\opn{Fr}=\frac{p-2}{-p}+12s(-1,-p)=\frac{p-2}{-p}-12s(1,p)=\frac{p-2}{-p}-12(\frac{p}{12}+\frac{1}{6p}-\frac14)=2-p\nn\eea
where we have used the reciprocity formula for $s(a,c)$.

But the Hopf fibration structure of $L(p,1)$ naturally entails a different surgery matrix
\bea B=S^2\left[
          \begin{array}{cc}
            1 & 0 \\
            p & 1 \\
          \end{array}\right]=ST^{-p}S,~~~\phi(B)=3-p.\nn\eea
See sec.3 of \cite{KirbyMelvin} for the explicit description of this framing. In particular setting $p=1$ for $S^3$ and its 2-framing associated with the Hopf fibration is off by 2 units from the canonical one.

\end{example}
\bibliographystyle{alpha}
\bibliography{RW_localisation}
\end{document}